\documentclass[a4paper,onecolumn,11pt, accepted=2026-03-16]{quantumarticle}
\pdfoutput=1
\usepackage[utf8]{inputenc}
\usepackage[english]{babel}
\usepackage[T1]{fontenc}
\usepackage{amsmath}
\usepackage{amssymb}
\usepackage{hyperref}
\usepackage{doi}
\usepackage{mathtools}
\usepackage{braket}
\usepackage{physics}
\usepackage{dsfont}
\usepackage{empheq} 
\usepackage{float}
\usepackage{nomencl}
\usepackage{cite}
\usepackage{enumitem} 
\usepackage{orcidlink} 
\usepackage{tabularx}
\usepackage{amsthm}
\usepackage[final]{changes}
\definechangesauthor[name={Authors}, color=blue]{A}

\usepackage{tikz}
\usetikzlibrary{arrows.meta}
\usetikzlibrary{trees} 
\usetikzlibrary{shapes.misc}
\usepackage{lipsum}

\newtheorem{theorem}{Theorem}
\newtheorem{lemma}{Lemma}
\newtheorem{definition}{Definition}
\newtheorem{remark}{Remark}
\newtheorem{example}{Example}

\usepackage{footmisc}
\usepackage{graphicx}

\definecolor{equation_box}{rgb}{1, 1, 1} 

\renewcommand{\Pr}{\mathrm{Pr}}
\makenomenclature

\begin{document}

\tikzstyle{level 1}=[level distance=5.5cm, sibling distance=2.5cm]
\tikzstyle{level 2}=[level distance=4cm, sibling distance=1.5cm]

\tikzstyle{bag} = [text width=8em, text centered]
\tikzstyle{end} = [circle, minimum width=3pt,fill, inner sep=0pt]

\title{A consolidated and accessible security proof for finite-size decoy-state quantum key distribution}

\author{Jerome Wiesemann\,\orcidlink{0009-0009-1509-6133}}
\affiliation{Fraunhofer Institute for Telecommunications, Heinrich-Hertz-Institut, HHI, 10587 Berlin, Germany}
\orcid{0009-0009-1509-6133}
\email{jwiesemann@uwaterloo.ca}

\author{Jan Krause\,\orcidlink{0000-0002-3428-7025}}
\affiliation{Fraunhofer Institute for Telecommunications, Heinrich-Hertz-Institut, HHI, 10587 Berlin, Germany}
\orcid{0000-0002-3428-7025}

\author{Devashish Tupkary\,\orcidlink{0000-0002-7586-3738}}
\affiliation{Institute for Quantum Computing and Department of Physics and Astronomy, University of Waterloo, N2L 3G1, Canada}
\orcid{0000-0002-7586-3738}

\author{Norbert L\"utkenhaus\,\orcidlink{0000-0002-4897-3376}}
\affiliation{Institute for Quantum Computing and Department of Physics and Astronomy, University of Waterloo, N2L 3G1, Canada}
\orcid{0000-0002-4897-3376}

\author{\\Davide Rusca\,\orcidlink{0000-0002-3319-6893}}
\affiliation{Vigo Quantum Communication Center, University of Vigo, Vigo E-36310, Spain}
\affiliation{Escuela de Ingeniería de Telecomunicación, Department of Signal Theory and Communications, University of Vigo, Vigo E-36310, Spain}
\affiliation{AtlanTTic Research Center, University of Vigo, Vigo E-36310, Spain}
\orcid{0000-0002-3319-6893}

\author{Nino Walenta\,\orcidlink{0000-0001-7243-0454}}
\affiliation{Fraunhofer Institute for Telecommunications, Heinrich-Hertz-Institut, HHI, 10587 Berlin, Germany}
\orcid{0000-0001-7243-0454}
\maketitle

\begin{abstract}
    In recent years, quantum key distribution (QKD) has evolved from a scientific research field to a commercially available security solution, supported by mathematically formulated security proofs. However, since the knowledge required for a full understanding of a security proof is scattered across numerous publications, it has proven difficult to gain a comprehensive understanding of all steps involved in the process and their limitations without considerable effort and attention to detail. Our paper aims to address this issue by providing a rigorous and comprehensive security proof for the finite-size 1-decoy and 2-decoy BB84 protocols against coherent attacks within Renner's entropic uncertainty relation framework. We resolve important technical flaws found in previous works regarding the fixed-length treatment of protocols and the careful handling of acceptance testing. To this end, we provide various technical arguments, including an analysis accounting for the important distinction of the 1-decoy protocol where statistics are computed after error correction, along with a slight improvement of the secure-key length. We also explicitly clarify the aspect of conditioning on events, addressing a technical detail often overlooked and essential for rigorous proofs. We extensively consolidate and unify concepts from many works, thoroughly discussing the underlying assumptions and resolving technical inconsistencies. Therefore, our contribution represents a significant advancement towards a broader and deeper understanding of QKD security proofs.
\end{abstract}

\tableofcontents

\section{\label{sec:introduction}Introduction}
Quantum key distribution (QKD) is a method for remotely establishing a secure key between two parties.
While classical cryptographic methods rely on unproven computational hardness assumptions of certain mathematical problems \cite{Rivest78, Diffie76, Koblitz87, Miller85}, the protocol security of QKD is based on the laws of quantum mechanics
and does not make any assumptions about the computational power of a potential eavesdropper. In fact, no proven secure method currently exists for remotely establishing a secure key by means of classical communication alone \cite{Shannon49}.
At its core, QKD relies on carriers of quantum information
\cite{Wootters82},
which cannot be eavesdropped without causing noticeable disturbances.
With its security proofs formulated in the universal composability framework, QKD can be combined with other composable cryptographic primitives, such as encryption algorithms, to yield \textit{information-theoretically secure} protocols \cite{Shor00, Canetti2001, Mayers01, Gisin02, Scarani09, Portmann22}. Nevertheless, it is important to distinguish between \textit{protocol security} and \textit{implementation security}. The former refers to the theoretical security guarantees of the QKD protocol, whereas the latter deals with security implications of the physical implementations, including device imperfections and the resulting side-channels. In this work, we aim to present a baseline security proof and thus focus on protocol security.

The concept of QKD was proposed in 1984 by Bennett and Brassard, inspired by earlier works from Wiesner\cite{Brassard05}, leading to the formulation of what is today recognized as the \textit{BB84 protocol} \cite{Bennett84}.
While the protocol itself remains largely unchanged to this day, its security analysis has undergone significant improvements and refinements.
The original protocol assumed single photons as information carriers.
Today, however, owing to their greater practicality, weak coherent states, for example generated by lasers in gain-switched operation, are used in practical implementations\cite{boaronSecureQuantumKey2018}.
Due to the intrinsic non-zero probability for multi-photon emissions by coherent sources, the protocol was adjusted to mitigate so-called photon-number splitting attacks \cite{Bennet92, Huttner95,Lutkenhaus02}.
This was achieved by the \textit{decoy-state method}, first introduced in 2003 \cite{Hwang03, wang2005, Lo05}.
Its core idea involves monitoring the photon-number-dependent channel transmission by preparing the weak coherent states with randomly chosen intensities unknown to an adversary.
A rigorous security proof was in turn presented in 2005 by Lo et al. \cite{Lo05} for asymptotic key rates and by Wang for a finite amount of decoy states \cite{wang2005}.
While the security of QKD with coherent sources can also be maintained by other means
\cite{
    lutkenhausSecurityIndividualAttacks2000,
    Scarani04},
the decoy-state method remains the only one with a linear relation between the channel transmission and secure-key rate.
In 2005, security proofs were reformulated within the universal composability framework \cite{Canetti2001} by Renner \cite{Renner05}, taking into account the effects caused by finite post-processing block sizes, thereby laying the foundation for the formulation of modern QKD security proofs.

There have been a variety of papers on a finite-size security proof for decoy-state BB84 against coherent attacks \cite{kamin2024decoygeat,LimCurtyWalenta14,Rusca18}.
In 2014, a notable advancement towards practical applications was made by Lim et al. \cite{LimCurtyWalenta14}, who presented a proof using entropic uncertainty relations \cite{Tomamichel11}, requiring just two additional decoy-state intensities, also called the \textit{2-decoy state BB84 protocol} \cite{maPracticalDecoyState2005}. This approach was applied to the \textit{1-decoy state BB84 protocol} by Rusca et al. \cite{Rusca18} in 2018 (see also Ref.~\cite{kamin2024improveddecoystate}).
Even though many other protocols have been proposed
\cite{
    Ekert91,
    Scarani04,
    Lo12,
    lucamariniOvercomingRateDistance2018,
    zengModepairingQuantumKey2022
},
the family of decoy-state BB84 protocols remains the most widespread protocol family, both in academia and industry, and will thus be the focus of this work. In particular, we will focus on Refs.~\cite{LimCurtyWalenta14,Rusca18}.

The knowledge necessary for a complete understanding of modern QKD security proofs is scattered across numerous publications, often exhibiting technical inconsistencies.
Therefore, it has proven difficult to gain a comprehensive understanding of each step involved in the security proof process and their limitations without considerable effort and attention to detail.
This problem has been addressed in 2017 by Tomamichel and Leverrier, who presented a self-contained security proof for entanglement-based and prepare-and-measure protocols \cite{Tomamichel17}.
However, they explicitly do not cover the case of signals generated by weak coherent pulses and therefore exclude the treatment of decoy states.
Furthermore, their work requires a solid understanding of the mathematical foundations of quantum information theory, which restricts its accessibility.

\paragraph{Contributions.} 
We aim to present a rigorous security proof for the 1-decoy and 2-decoy BB84 protocols, resolving various technical flaws found in previous works. We provide a rigorous treatment for fixed-length protocols with acceptance testing, focusing on the subtle but important fact that, following the fixed-length definition of security, the length of the secure-key must be fixed prior to running the protocol. We also account for the distinction in the 1-decoy approach where acceptance testing occurs after error correction, rather than based on public announcements directly after sifting, as is common in traditional protocols. These arguments, overlooked in earlier analyses, are explicitly addressed in this work. To this end, we make it a point to explicitly clarify the aspect of conditioning on states throughout the analysis. Security proofs for variable-length protocols require additional arguments and do not trivially follow from fixed-length statements. This issue extends to works such as Refs.~\cite{LimCurtyWalenta14, Rusca18} which do not rigorously handle fixed-length protocols (via acceptance testing) or variable-length protocols. In this work, we aim to address the issues stated above and present a rigorous fixed-length security proof in the entropic uncertainty relation framework using smooth min-entropies. Moreover, we slightly improve the secure-key length by using a suitable combination of min-entropy chain rules and update the protocol description to address the issue of iterative sifting observed in earlier works \cite{Tomamichel12, LimCurtyWalenta14}. We also address several technical gaps pointed out in Ref.~\cite{Tupkary24}.

We also aim at largely increasing the accessibility of the security proof by employing a constructive approach:
starting from the general definition of security, we systematically expand and bound each term, providing a clear, step-by-step framework for the proof. We discuss underlying concepts in detail, providing a proof outlined in a coherent manner.
Underlying and often hidden assumptions are named and listed, cf. App.~\ref{ap:assumptions}.
These assumptions should be kept in mind when designing practical QKD systems as the discrepancies between the security proof and the practical implementation can lead to side-channel attacks compromising the security of QKD systems \cite{BSI24, Jain16, Sajeed21, Makarov23}.
With the inclusion of decoy states, we derive equations expressed in terms of experimentally accessible parameters, thus allowing for direct applications in practical systems.
The proof builds upon the works by Rusca et al. \cite{Rusca18} and Lim et al. \cite{LimCurtyWalenta14}, focusing on the derivation of an expression for the secure-key length in the 1-decoy protocol, but also deriving the necessary bounds for the 2-decoy variant. We take finite-size effects into account and prove security against coherent attacks, the most general form of attacks. 

The proof is performed in Renner's framework, i.e. by applying the quantum leftover hash lemma and bounding the relevant entropies using the entropic uncertainty relation (EUR) \cite{TomamichelRenner11}.
We note that other approaches to security proofs exist, such as using the post-selection technique reduction \cite{Nahar24,Christandl09} to IID attacks followed by an IID security proof, the generalized entropy accumulation theorem \cite{metgerSecurityQuantumKey2023}, and the phase error correction framework \cite{Koashi09}, each with their advantages and drawbacks. With numerous examples and remarks, we hope to largely improve the comprehensibility of the security proof.
While assuming general knowledge of QKD and quantum mechanics, no prior exposure to security proofs is required.
For an overview of quantum key distribution, we refer to the extensive reviews \cite{Gisin02, Scarani09, Pirandola20} and recent theoretical books \cite{Wolf21, Grasselli21}, and for an introduction to quantum mechanics and quantum information theory, we refer to the standard textbooks \cite{Audretsch07, ChuangNielsen10}.
We strive to present a rigorous yet accessible security proof for the 1-decoy and 2-decoy state BB84 protocols that can serve as a foundation for the discussion of their security and for the identification of potential vulnerabilities.
Additionally, this work aims at providing a robust reference for practical implementations of the protocol.

\paragraph{Outline.}
The structure of the security proof is illustrated in Fig.~\ref{fig:structure_security_proof}. Section~\ref{sec:preliminaries} first
discusses fixed-length protocols and formally describes the 1-decoy state protocol, which we slightly adjust to rigorously treat acceptance testing. Then, the theoretical background is laid out and core
concepts such as density operators, the trace distance and bipartite quantum systems are introduced. Information theoretic security is then defined and the security parameters are constructively introduced as a metric to evaluate the distance to a perfectly secure system. The assumptions of quantum key distribution and its composition with other cryptographic primitives such as the one-time pad are also described. 

The terms appearing in the definition of security are then separately addressed to prove protocol security. We first bound the correctness parameter in Sec.~\ref{sec:preliminaries:universal_2_hashing} using universal$_2$ hashing, which is introduced as part of the error verification and privacy amplification steps of the post-processing.
The quantum leftover hash lemma, first introduced by Renner in Ref.~\cite{Renner05}, is then presented in Sec.~\ref{sec:quantum_leftover_hash_lemma}.
It provides a bound on the secrecy parameter and is expanded in terms of the number of photon events and errors using the entropic uncertainty relation by bounding the relevant entropies.
For this, the min-entropy as well as its smoothed version are introduced. The result of this section is an expression linking the secure-key length to the min-entropy, which represents the information a potential eavesdropper could have gathered about the key.

Since the number of $m$-photon events and errors is not directly accessible, the aim of Sec.~\ref{sec:decoy_state_bounds}
is to estimate and bound these quantities in terms of experimentally accessible parameters using the Poissonian statistics of weak coherent sources. Putting everything together, an operational expression for the secure-key length is derived in Sec.~\ref{sec:secret_key_length}. This expression solely depends on experimental parameters (the acceptance bounds on the number of photon events and errors), the predefined security parameters, and the information leaked during error correction and verification. 

\vspace*{0.5cm}

\tikzstyle{level 1}=[level distance=2cm, sibling distance=2.5cm]
\tikzstyle{level 2}=[level distance=2.5cm, sibling distance=2.5cm]
\tikzstyle{level 3}=[level distance=2.5cm, sibling distance=2.5cm]

\tikzstyle{bag} = [text width=8em, text centered]
\tikzstyle{end} = [circle, minimum width=3pt,fill, inner sep=0pt]
\begin{figure}[h]
    \centering
    \begin{tikzpicture}[grow=down, sloped, edge from parent/.style={->, draw, thin, -Latex}]
        \node[draw, anchor = north, fill = black!8] (c1) {Assumptions of QKD\strut};

        \node[draw, fill = black!8, align = center] (c2) at ([xshift=6cm, yshift=2cm]c1) {Protocol description\strut}
        child{
            node[draw, fill = black!8, align = center, very thick] (security) {\textbf{Definition of security}\strut}
            child{
                node[draw, fill = black!8, align = center, very thick] (EV) {\textbf{Error correction}\strut\\ \textbf{and verification}\strut}
                        child{
                    node[draw, fill = black!8, align = center, very thick] (leftover) {\textbf{Quantum leftover}\strut\\ \textbf{hash lemma}\strut}
                    child{
                        node[draw, fill = black!8, align = center] (decompose) {Expanding\strut \\the min-entropy\strut}
                        child{
                            node[draw, fill = black!8, align = center] (bounds) {Bounds on the number of\strut\\ photon events and errors\strut}
                            child{
                                node[draw, fill = black!8, align = center, very thick] (skl) { \textbf{Expression for the}\strut \\ \textbf{secure-key length}\strut}
                            }
                        }
                    }
                }
            }
        };

        \node[draw, fill = black!8, align = center] (c3) at ([xshift=6cm, yshift=0cm]security) {Distance between real\strut\\ and ideal protocol\strut};

        \node[draw, anchor = north, fill = black!8] (minentropy) at ([xshift=-6cm, yshift=11pt]leftover) {Min-entropy\strut};

        \node[draw, anchor = north, fill = black!8] (hashing) at ([xshift=6cm, yshift=1.6cm]leftover) {Universal$_2$ hashing\strut};

        \node[draw, anchor = north, fill = black!8, align = center] (EUR) at ([xshift=6cm, yshift=18pt]decompose) {Entropic uncertainty\strut\\ relation \strut};

        \node[draw, anchor = north, fill = black!8, align = center] (chain) at ([xshift=-6cm, yshift=18pt]decompose) {Chain rule for\strut\\ smooth min-entropies \strut};

        \node[draw,fill = black!8, align=center] (finite) at ([xshift=-6cm, yshift=0cm]bounds) {Finite-size photon\strut \\events statistics\strut};

        \node[draw, anchor = north, fill = black!8, align = center] (information) at ([xshift=-6cm,  yshift=17pt]skl) {Information disclosed\strut\\ during post-processing\strut};

        \node[draw, anchor = north, fill = black!8] (analysis) at ([xshift=6cm,  yshift=11pt]skl) {Security analysis\strut};

        \draw[->, thin, -Latex] (c1) -- (security);
        \draw[->, thin, -Latex] (c3) -- (security);
        \draw[->, thin, -Latex] (minentropy) -- (leftover);
        \draw[->, thin, -Latex] (hashing) |- (leftover);
        \draw[->, thin, -Latex] (hashing) |- (EV);
        \draw[->, thin, -Latex] (EUR) -- (decompose);
        \draw[->, thin, -Latex] (chain) -- (decompose);
        \draw[->, thin, -Latex] (finite) -- (bounds);
        \draw[->, thin, -Latex] (information) -- (skl);
        \draw[->, thin, -Latex] (analysis) -- (skl);

        \draw[densely dashed] (-2.3, 2.25) rectangle (14.3,-1.7);
        \node[] at (12.2, -1.4) {\textbf{\small{Preliminaries (Sec.~\ref{sec:preliminaries})}}};
        \draw[densely dashed] (-2.3, -1.9) rectangle (14.3,-6.8);
        \node[] at (9.6, -6.5) {\textbf{\small{Link security and secure-key length (Secs.~\ref{sec:preliminaries:universal_2_hashing} and \ref{sec:quantum_leftover_hash_lemma})}}};
        \draw[densely dashed] (-2.3, -7) rectangle (14.3, -14.25);
        \node[] at (8.1, -13.95) {\textbf{\small{Link secure-key length and experimental parameters (Secs.~\ref{sec:decoy_state_bounds} and \ref{sec:secret_key_length})}}};
            
        \end{tikzpicture}
    \caption{Structure of the 1-decoy and 2-decoy state security proofs.}
    \label{fig:structure_security_proof}
\end{figure}

\section{\label{sec:preliminaries}Preliminaries}
The aim of this section is twofold.
First, the 1-decoy state BB84 protocol is formally described in Sec.~\ref{sec:preliminaries:protocol_description}.
Then, mathematical concepts and the notation used throughout this work are introduced in Secs.~\ref{sec:preliminaries:theoretical_background} and \ref{sec:preliminaries:def_security}.

\subsection{\label{sec:preliminaries:protocol_description}The 1-decoy state BB84 protocol}

The \textit{1-decoy state protocol} \cite{Rusca18} described in Fig.~\ref{fig:protocol_description} is based on the original \textit{BB84 protocol} proposed by Bennett and Brassard \cite{Bennett84} and constitutes a slight variation of the \textit{2-decoy state protocol} \cite{LimCurtyWalenta14} in the sense that it only requires two intensity\footnote{More precisely, this denotes the mean photon number. Nevertheless, in the following we adhere to the name conventionally used in the literature.}
levels $\mu_1, \mu_2$ for the state preparation instead of three. In this scenario, Alice and Bob are linked by a quantum channel as well as an authenticated classical channel (see Ref.~\cite[Sec.~VII.A]{Portmann22} for how to construct an authenticated channel). Their goal is to create a symmetric binary key that is unknown to a potential eavesdropper, Eve, who has access to both the quantum and classical channels. In order to generate a key, Alice sends signals encoded in weak coherent optical pulses that are in turn measured by Bob. In the so-called post-processing, the resulting bit strings are distilled into a secure key.

The authenticity of the classical channel ensures that Eve cannot perform man-in-the-middle attacks by tampering with the channel (swapping, modifying, adding or removing messages).
Although authenticated, information transmitted over the classical channel is considered insecure as it can be read by Eve without Alice and Bob noticing. As such, in the following, it will be assumed that Eve has knowledge about all classical communications and that all messages exchanged during the post-processing are authenticated, e.g. using \cite{Wegman81, Stinson94, Portmann14}.

Following the no-cloning theorem of quantum mechanics (see Refs.~\cite{Wootters82} and \cite[Box 12.1]{ChuangNielsen10} for a proof), the same principle does not apply to the quantum channel, where, even though Eve has access to the channel, her interference causes measurable disturbances. This is in turn taken into account during the privacy amplification step and allows Alice and Bob to estimate bounds for the correlation of their key with Eve's quantum system. These estimates are used in the privacy amplification, which shortens the key so as to arbitrarily reduce Eve's correlation with the final key.

\subsubsection{Fixed-length protocols}
\label{sec:fixed_length_protocols}
For the entirety of this work, we consider a fixed-length protocol, as opposed to a variable-length protocol \cite{Tupkary23, Hayashi12} \cite[Ch.~3]{Kawakami17} \cite[Supplementary note A]{Curras21}. This informs the definition of security, which we introduce in Sec.~\ref{sec:security_parameters} and only holds for keys of length known prior to running the protocol, a subtle fact easily overlooked. The more general definition of security, i.e. when keys of different lengths can be generated, exhibits an additional sum over all possible key lengths, which is formalized in the references above. For fixed-length protocols, the length of the secure-key is fixed prior to running the protocol and does not depend on the statistics observed during the protocol run. The protocol then outputs a secure-key pair, $S_A$ and $S_B$, shared by Alice and Bob where the keys are either finite bit strings of fixed size $l$ if the protocol run was successful, or $\perp$ if the protocol aborts. This means that any \textit{acceptance parameter} determining the secure-key length must be fixed by Alice and Bob before the protocol run\footnote{The parameters may however be adjusted between protocol runs.}. These parameters define a set of conditions that must be fulfilled by the observed statistics during the so-called \textit{acceptance test} in order for the protocol to not abort, i.e. produce a non-trivial key. This step is often also called \textit{parameter estimation} in the literature.

More formally, prior to running the protocol, Alice and Bob agree on an \textit{acceptance set} $Q$ such that the protocol aborts if the observed statistics $F_\mathrm{obs}$ are not in the acceptance set, i.e. $F_\mathrm{obs}\notin Q$\footnote{Unlike other works, where $F_\mathrm{obs}$ is a frequency vector, here $F_\mathrm{obs}$ is a set of statistics observed during the protocol.}. This means that if any of the conditions fails, the protocol aborts. See Example~\ref{ex:acceptance_test} for an illustrative example and Refs.~\cite{Tupkary23,George21,Renner05} for a more detailed discussion.

\begin{example}
\label{ex:acceptance_test}
    To illustrate the acceptance test, consider, as an example, a protocol for which two conditions need to be satisfied, e.g. $x < X$ and $y \geq Y$ with $x,y\in\mathbb{N}$. The acceptance set can then be parameterized as
    \begin{equation}
        Q = \left\{ (x, y)\in\mathbb{N}^2\,|\,x < X \land y \geq Y \right\}.
    \end{equation}
    A protocol run produces statistics $F_\mathrm{obs}=(x_0, y_0)$. During the acceptance test, the protocol aborts if $F_\mathrm{obs}\notin Q$, i.e. if at least one of the conditions in $Q$ is not fulfilled. If $F_\mathrm{obs} \in Q$, then the length of the key generated is given by
    \begin{equation}
        l = \min_{(x, y)\in Q} l(x, y) \,,
    \end{equation}
    i.e. by minimizing the secure-key length $l(x, y)$ over all possible statistics $(x, y)$ in the acceptance set $Q$. Note that this optimization is done analytically in this work.
\end{example}
\noindent If the protocol does not abort, the secure-key length is given by minimizing the expression for the secure-key length over all possible sets of observations contained in $Q$. One advantage of this approach is that instead of computing and optimizing the secure-key length during each run, it must only be minimized once for a given $Q$. Then, only a set of conditions must be verified during the protocol to produce a key of said length. In general, relaxing the conditions imposed by $Q$ lowers the probability for the protocol to abort but also decreases the secure-key length as it is minimized over a larger set. A good choice for $Q$ should thus depend on the practical implementation. For systems with stable channels, e.g. fiber-based systems, a tight set of conditions based on statistics observed in previous protocol runs might be chosen. In this case, if the observed statistics do not fluctuate much, the probability that the protocol aborts is low for appropriately chosen acceptance tests. 

For unstable systems, e.g. satellite-based systems, more relaxed bounds may be necessary for the protocol to not abort, which decreases the secure-key length. In this case, a variable-length protocol may be more suitable. For methods to prove variable-length security in the EUR framework, see Refs.~\cite[Apps.~B and F]{Tupkary24} and \cite[Supplementary Note~A]{Curras21}. Finally, as discussed in Sec.~\ref{sec:secret_key_length}, the approach taken in this work using smooth min-entropies enables an analytical lower bound on the secure-key length which does not require a numerical optimization since the expression on the secure key is monotonously dependent on the set of observed statistics. 

\subsubsection{Protocol description}
\label{sec:long_protocol_description}
The 1-decoy state BB84 protocol is formally described in Fig.~\ref{fig:protocol_description}. Here, we aim at describing the protocol more intuitively. As discussed in the previous section, we consider a fixed-length protocol.

Before the protocol run, Alice and Bob agree on the set of parameters that define the protocol and which can be chosen to optimize the performance. 
They choose two intensities, $\mu_1$ and $\mu_2$, with $\mu_1 > \mu_2 > 0$,
the probability $p_k$ for Alice to choose the intensity $k$ and
the probability\footnote{Other publications often only introduce one basis choice probability $p_\mathsf{Z} = p_\mathsf{Z}^A = p_\mathsf{Z}^B$, but this is not necessary.}
$p_\mathsf{Z}^A$ ($p_\mathsf{Z}^B$) for Alice (Bob) to choose the $\mathsf{Z}$-basis for state preparation (measurement) with 
$0 < p_\mathsf{Z}^A, p_\mathsf{Z}^B < 1$. They also agree on a value for the security parameters $\epsilon_\mathrm{sec}'$ and $\epsilon_\mathrm{cor}$, the number of bits $\mathrm{leak}_\mathrm{EC}$ to leak for error correction and the number of signals $N$ for Alice to send.

They then agree on an acceptance set $Q$ which defines a set of conditions that need to be fulfilled for the protocol to output a non-trivial key, as discussed in Sec.~\ref{sec:fixed_length_protocols}. For this purpose, they choose values for the acceptance conditions in $Q$, namely the number of bits used for key generation and parameter estimation, $N_\mathsf{Z}$ and $N_\mathsf{X}$ respectively, and choose values for $s_{\mathsf{Z},0}^\mathrm{l}$, $s_{\mathsf{Z}, 1}^\mathrm{l}$, $s_{\mathsf{X}, 1}^\mathrm{l}$ and $\Lambda_\mathsf{X}^\mathrm{u}$, which set the acceptance threshold bounds on the computed decoy statistics, as summarized in Table~\ref{tab:acceptance_testing}\cite[App. B]{Mannalath24}. The conditions are verified during the protocol, namely the sifting step and acceptance test, as described in Fig.~\ref{fig:protocol_description}. The acceptance parameters \added{and acceptance test are more thoroughly discussed in Remark~\ref{rem:acceptance_testing_summary}, and} in later sections. Finally, they agree on a secure-key length $l$, cf. Eq.~\eqref{eq:max_secret_key_formula}, which only depends on parameters fixed prior to running the protocol, including the acceptance parameters. The parameter agreement step (where Alice and Bob determine the protocol parameters and acceptance parameters) is usually not understood as part of the protocol and the acceptance set $Q$ can be adjusted each protocol run. \added{In this work, the parties fix the security parameters (as well as the remaining protocol parameters), which yields an upper bound on the achievable secure-key length $l$ via Eq.~\eqref{eq:max_secret_key_formula}. Alternatively, one may instead fix the key length $l$ and the length of the hash used for error verification (as well as the remaining protocol parameters), in which case the resulting security parameters are determined accordingly.}

The protocol starts with the \textit{state preparation}.
For each signal transmitted from Alice to Bob, Alice randomly chooses a bit value, the encoding basis according to $p_\mathsf{Z}^A$ and an intensity level according to $p_k$.
The signal is then transmitted to Bob who randomly chooses a \textit{measurement} basis according to $p_\mathsf{Z}^B$.
In practical scenarios, for most signals, Bob does not detect any photons due to losses in the quantum channel.
If more than one of Bob's detectors click, he randomly assigns a measurement outcome (see Footnote~\ref{foot:double_click_events}). The first two steps are repeated $N$ times.

Those instances for which Bob had no clicks or Alice and Bob chose different bases do not contain usable information and are discarded during \textit{sifting}.
Hence, Alice announces the signal indices and measurement bases.
Bob then discards his information about detected signals where Alice chose a different basis or he did not detect a signal, and informs Alice about the remaining measurement results to keep, i.e. those that Bob did not discard. Alice also informs Bob about her intensity choices, which will be required during the acceptance test. For the analysis in this work, it is important to not publish the basis choices before all $N$ signals have been received, cf. Remark~\ref{rem:pre_sifting}. In fact, various errors and subtleties in the analysis of protocols with so-called iterative sifting were pointed out in Ref.~\cite{Pfister16}. For methods to address these issues, we refer to Ref.~\cite{Tamaki18}. During the sifting step, Alice and Bob verify a set of conditions described in Fig.~\ref{fig:protocol_description}. The basis and detect/no-detect announcements are stored in a classical register $\tilde{C}^N$.

After the sifting step, Alice and Bob possess sifted keys $\tilde{Z}_A, \tilde{Z}_B\in\mathcal{Z}$ of length $N_\mathsf{Z}$, which are partially correlated with Eve and where $\mathcal{Z}= \{0, 1\}^{N_\mathsf{Z}}$ is the set containing all possible keys of length $N_\mathsf{Z}$.
Due to experimental limitations (imperfect state preparation, channel noise, imperfect photon detection) or the presence of an eavesdropper, the sifted keys are usually not identical, hence why during the \textit{error correction} step, Bob corrects his key in order to obtain a key $Z_B$ ideally identical to $Z_A$. To achieve this, Alice transmits $\mathrm{leak}_\mathrm{EC}$ bits of information about her key to Bob.
Error correction algorithms usually require an estimate of the error rate, which can for example be obtained from previously processed blocks\footnote{We note that a wrong estimate for the error rate does not affect the security in any way but merely increases the probability of the error verification test failing.}. We denote $\Omega_\mathrm{EC}$ as the event in which error correction succeeded and store the communications in a classical register $C_\mathrm{EC}$. It follows that $|C_\mathrm{EC}| = \mathrm{leak}_\mathrm{EC}$.
The measurement outcomes obtained in the $\mathsf{X}$-basis are publicly revealed and no error correction is applied.
The $\mathsf{X}$-basis measurements do not contribute to the final key.

Error correction algorithms are not guaranteed to succeed.
Hence, errors might remain unnoticed.
To ensure that the keys are equal up to a small probability without disclosing them, a hash of Alice's key is transmitted to Bob, who compares it with a hash of his key. We define $\Omega_\mathrm{EV}$ as the event in which error verification was successful and set $S_A = S_B = \perp$ if it fails. The choice of hash function and the hash of Alice's key are stored in a classical register $C_\mathrm{EV}$, which is communicated to Bob. After error verification, Alice's and Bob's keys are identical up to a probability $\epsilon_\mathrm{cor}$\footnote{This statement should be interpreted with care, see Sec.~\ref{sec:security_parameters} and Remark~\ref{rem:general_correctness}.}, which we show in Sec.~\ref{sec:error_verification}.

\begin{table}[h]
\centering
\renewcommand{\arraystretch}{1.2} 
\begin{tabular}{ |c|c|c|c| } 
 \hline
 \textbf{Observed statistics} & \textbf{Decoy bound} & \textbf{Acceptance parameter} & \textbf{Acceptance condition} \\ \hline
 $|X^\mathrm{S}|$ (Fig.~\ref{fig:protocol_description}) & - & $N_\mathsf{X}$ &  $|X^\mathrm{S}| \geq N_\mathsf{X}$ \\ 
 $|Z^\mathrm{S}|$ (Fig.~\ref{fig:protocol_description}) & - & $N_\mathsf{Z}$ &  $|Z^\mathrm{S}| \geq N_\mathsf{Z}$ \\ 
 $s_{\mathsf{Z},0}$ & $s_{\mathsf{Z},0}^-$ (Eq.~\eqref{eq:lower_bound_vacuum}) & $s_{\mathsf{Z},0}^\mathrm{l}$ & $s_{\mathsf{Z},0}^- \geq s_{\mathsf{Z},0}^\mathrm{l}$ \\
 $s_{\mathsf{Z},1}$ & $s_{\mathsf{Z},1}^-$ (Eq.~\eqref{eq:lower_bound_single}) & $s_{\mathsf{Z},1}^\mathrm{l}$ & $s_{\mathsf{Z},1}^- \geq s_{\mathsf{Z},1}^\mathrm{l}$ \\ 
 $s_{\mathsf{X},1}$ & $s_{\mathsf{X},1}^-$ (Eq.~\eqref{eq:lower_bound_single}) & $s_{\mathsf{X},1}^\mathrm{l}$ & $s_{\mathsf{X},1}^- \geq s_{\mathsf{X},1}^\mathrm{l}$ \\ 
 $\Lambda_\mathsf{X}$ & $\Lambda_\mathsf{X}^+$ (Eq.~\eqref{eq:upper_bound_qber}) & $\Lambda_\mathsf{X}^\mathrm{u}$ & $\Lambda_\mathsf{X}^+ \leq \Lambda_\mathsf{X}^\mathrm{u}$  \\
 \hline
\end{tabular}
\caption{Acceptance tests performed on the observed statistics and decoy bounds. If any of the conditions fails, the protocol aborts. The acceptance bounds have a pre-defined value that can be freely chosen e.g. to maximize the secure-key length or lower the probability of an abort. \added{See Remark~\ref{rem:acceptance_testing_summary} for more details.} }
\label{tab:acceptance_testing}
\end{table}

Additionally, Bob now has full knowledge over all errors in the $\mathsf{Z}$-basis by comparing the corrected key to his sifted key, allowing him to count the number of errors for a certain basis and intensity,
$c_{\mathsf{X},k}$, $c_{\mathsf{Z}, k}$, where $k\in\{\mu_1, \mu_2\}$\footnote{Notice that Bob only obtained the correct number of errors if error correction succeeded.}.
Bob then verifies an additional set of conditions (cf. Fig.~\ref{fig:protocol_description} \added{and Remark~\ref{rem:acceptance_testing_summary})}. We call this step \textit{acceptance test}.
We denote $\Omega_\mathrm{AT}$ as the event in which $F_\mathrm{obs}\in Q$, i.e. all conditions verified in the sifting step and acceptance test are satisfied. Acceptance testing is summarized in Table~\ref{tab:acceptance_testing} and Remark~\ref{rem:acceptance_testing_summary}. To simplify the notation, we define $\Omega_\top\coloneqq\Omega_\mathrm{AT}\land\Omega_\mathrm{EV}$ such that $\Pr[\Omega_\top]$ denotes the probability that the protocol does not abort. The acceptance decision (pass/fail) following the acceptance test and error verification is stored in a classical register $C_\mathrm{AT}$. 

In contrast to the 2-decoy state protocol, the 1-decoy state protocol requires the number of errors in the $\mathsf{Z}$-basis to compute a bound on the number of single-photon events. Consequently, Bob requires the number of $\mathsf{Z}$-basis errors for the acceptance test, hence why it is performed after error correction. This introduces subtleties in the analysis \cite[Remark~14]{Tupkary24} which are often overlooked in existing works but are explicitly addressed in this work. Bob may in theory also estimate these errors using traditional acceptance testing, which is done right after sifting, by disclosing and discarding a fraction of the key. However, the approach taken in this work allows for tighter bounds and does not disclose $\mathsf{Z}$-basis measurement outcomes, leading to overall higher secure-key rates\footnote{We note that for the 2-decoy state protocol, the number of errors in the $\mathsf{Z}$-basis are not required, cf. App.~\ref{ap:two_decoy_security_proof}. In this case, the acceptance test can be performed before error correction. The proof can however be performed analogously. }.

Finally, to reduce Eve's knowledge about the key, it is shortened during \textit{privacy amplification} using a randomly chosen hash function. The choice of hash function is stored in a classical register $C_\mathrm{PA}$ and communicated from Alice to Bob.
The secure-key length is calculated using the quantum leftover hash lemma, the entropic uncertainty relation and the decoy-state method which provides bounds on the number of photon events and errors using concentration inequalities, resulting in Eq.~\eqref{eq:max_secret_key_formula}. We denote $\Omega_\mathrm{B}$ the event where all bounds given by the decoy concentration inequalities hold. For clarity, we list all events used for the fixed-length protocol and their relation to one another in Table~\ref{tab:events_description}. We also list all classical communication registers in Table~\ref{tab:classical_registers_comm_description}. To simplify the notation, we define $C \coloneqq \tilde{C}^N C_\mathrm{EC} C_\mathrm{EV} C_\mathrm{AT} C_\mathrm{PA}$ as the classical register storing all classical communications occurring between Alice and Bob during the protocol run.

\begin{remark}
\label{rem:acceptance_testing_summary}
\added{
    The steps involved in the acceptance test for the protocol described in Fig.~\ref{fig:protocol_description} can be summarized as follows (as depicted in Table~\ref{tab:acceptance_testing} and described in Sec.~\ref{sec:fixed_length_protocols})}
    \begin{enumerate}
        \item \added{Alice and Bob agree on acceptance parameters before the protocol run (third column in Table~\ref{tab:acceptance_testing}).} 
        \item \added{During the protocol, they have access to certain statistics, e.g. the number of $\mathsf{X}$-basis and $\mathsf{Z}$-basis detections after sifting, $|X^\mathrm{S}|$ and $|Z^\mathrm{S}|$. However, they cannot determine the exact number of vacuum or single-photon events, denoted $s_{\mathsf{Z},0}, s_{\mathsf{Z},1}$ and $s_{\mathsf{X},1}$, nor the exact error rate $\Lambda_\mathsf{X}$. Therefore, they compute bounds for the inaccessible statistics using the decoy-state method (second column in Table~\ref{tab:acceptance_testing}). We denote these bounds with subscripts + and - throughout this work to differentiate them from acceptance bounds which are denoted with subscripts $\mathrm{u}$ and $\mathrm{l}$. As such, the bounds denoted with subscripts + and - are random variables observed during the protocol while the acceptance bounds have fixed values.}
        \item \added{They verify the corresponding acceptance conditions for the observed and the computed statistics by comparing them to the acceptance parameters (fourth column in Table~\ref{tab:acceptance_testing}), and abort the protocol if any of the conditions is not met. The acceptance set $Q$ is defined by the set of acceptance conditions, see Example~\ref{ex:acceptance_test}.}
    \end{enumerate}
\end{remark}

\begin{remark}
\label{rem:pre_sifting}
    Usually, Alice and Bob disclose the basis choices and whether or not Bob detected a signal only after all signals have been received. For practical reasons, e.g. to lower memory usage, it may be convenient to perform on-the-fly announcements during the signal transmission and measurement stage of the protocol. This complicates the analysis \cite{Pfister16,Tamaki18}. In this work, we perform traditional sifting, as described in Fig.~\ref{fig:protocol_description}, where all announcements take place after all signals have been measured. 
\end{remark}

\begin{remark}
    In Fig.~\ref{fig:protocol_description}, part of the acceptance conditions are verified during the sifting step and not in the acceptance test. The advantage of this approach is that the protocol may already abort before error correction and verification if insufficient signals are detected. For the analysis, we assume the scenario where all conditions are verified during the acceptance test, but note that both scenarios are equivalent for the purposes of proving security. 
\end{remark}

\begin{center}
\setlength\fboxsep{0.4cm}

\begin{minipage}{0.98\textwidth}
{\captionsetup{justification=raggedright, singlelinecheck=false}
\captionof{figure}{\underline{\textbf{Protocol description}}}
\label{fig:protocol_description}
}

\added{\textbf{\underline{Inputs:}}} \\
\vspace*{-1mm}

\begin{tabularx}{\textwidth}{lX}

    \added{\textbf{Security parameters:}} &
    $\epsilon_{\mathrm{sec}'} \in (0,1),\;
     \epsilon_{\mathrm{cor}} \in (0,1),\;
     \epsilon_{\mathrm{sec}'} + \epsilon_{\mathrm{cor}} \le \epsilon$ \\[1mm]
    
    \added{\textbf{Setup parameters:}} &
    $\mu_1 > \mu_2 > 0,\;
     p_k\in (0,1),\; p_\mathsf{Z}^A\in (0,1),\; p_\mathsf{Z}^B \in (0,1),\;
     N \ge 0$ \\[1mm]
    
    \added{\textbf{Acceptance parameters:}} &
    $N_\mathsf{X},\;
     N_\mathsf{Z},\;
     s_{\mathsf{Z},0}^{\mathrm{l}},\;
     s_{\mathsf{Z},1}^{\mathrm{l}},\;
     s_{\mathsf{X},1}^{\mathrm{l}},\;
     \Lambda_\mathsf{X}^{\mathrm{u}} \ge 0$ \\[1mm]
    
    \added{\textbf{Post-processing parameters:}} &
    $\mathrm{leak}_{\mathrm{EC}} \ge 0,\;
     l>0\ \text{computed from Eq.~\eqref{eq:max_secret_key_formula}}$ \\

\end{tabularx}

\end{minipage}
\end{center}

\begin{enumerate}    
    \item \textbf{State preparation:} For the $i$-th signal, Alice chooses a basis $b_i^A\in\{\mathsf{X},\mathsf{Z}\}$ 
    with probability $p_\mathsf{Z}^A$ of choosing the $\mathsf{Z}$-basis, which is used to generate the key, and probability
    $p_\mathsf{X}^A$ of choosing the $\mathsf{X}$-basis, which is used for monitoring the phase error rate.
    She then chooses an intensity $k_i\in\{\mu_1,\mu_2\}$ according to $p_k$ and encodes a randomly chosen bit value $y_i^A\in\{0,1\}$ in a weak coherent and phase-randomized optical pulse she sends to Bob through the quantum channel.
    
    \item \textbf{Measurement:} For each signal, Bob measures in the basis $b_i^B\in\{\mathsf{X}, \mathsf{Z}\}$, randomly chosen with probabilities $p_\mathsf{X}^B$ and $p_\mathsf{Z}^B$, yielding $y_i^B\in \{0, 1, \varnothing\}$, with $\varnothing$ denoting the case where he does not detect a signal. In the case of multiple detections, Bob randomly assigns one of the measurement outcomes to the signal\footnote{For the canonical model of ideal threshold detectors, this is mandatory to satisfy the basis-efficiency mismatch condition from the phase error rate estimation discussed in Sec.~\ref{sec:decomposing_min_entropy}. Discarding these detections also makes the systems vulnerable to an attack exploiting this \cite[Table~4.3]{BSI24}.\label{foot:double_click_events}}. 
    
    \item \textbf{Sifting:} The two first steps are repeated $N$ times. Alice and Bob then communicate their choice of basis and intensity in the classical channel and only keep the signals for which $b_i^A =b_i^B$. As such, we define the sets $X^\mathrm{S} \coloneqq \{i: b_i^A = b_i^B = \mathsf{X} \land y_i^B \neq \varnothing\}$ and $Z^\mathrm{S}\coloneqq\{i: b_i^A = b_i^B = \mathsf{Z} \land y_i^B \neq \varnothing\}$. Here, they also sift out all signals that were not detected, i.e. resulting in $y_i^\mathrm{B}\neq \varnothing$. \added{They verify the following conditions:
    \begin{align}
        |X^\mathrm{S}| \geq  N_\mathsf{X} \qquad\text{and}\qquad |Z^\mathrm{S}| \geq  N_\mathsf{Z}\,.
    \end{align}
    If any condition does not hold, they} set $S_A = S_B = \perp$ and abort the protocol. Otherwise, they choose random subsets $\Gamma_\mathsf{X} \subset X^\mathrm{S}$  and $\Gamma_\mathsf{Z} \subset Z^\mathrm{S}$ of sizes $N_\mathsf{X}$ and $N_\mathsf{Z}$, respectively\footnote{\added{This step involves classical communications between Alice and Bob as they agree on the same subsets $\Gamma_\mathsf{X}$ and $\Gamma_\mathsf{Z}$. Nevertheless, since the subsets are chosen at random, they are not correlated with the raw key and it can be shown that they do not provide any information to Eve, analogously to the argument used for the error verification hash function, cf. Eq.~\eqref{eq:remove_EC_tildetilde}.}}. Alice's and Bob's sifted keys in the $\mathsf{Z}$-basis are defined as the bit sequences $\tilde{Z}_A\coloneqq (y_i^A)_{i\in \Gamma_\mathsf{Z}}$ and $\tilde{Z}_B\coloneqq (y_i^B)_{i\in \Gamma_\mathsf{Z}}$. For the $\mathsf{X}$-basis, we define $\tilde{X}_A\coloneqq (y_i^A)_{i\in \Gamma_\mathsf{X}}$ and $\tilde{X}_B\coloneqq (y^B_i)_{i\in \Gamma_\mathsf{X}}$. Additionally, we define $\Gamma_{\mathsf{Z},k}\coloneqq\{i \in \Gamma_\mathsf{Z} : k_i = k\}$ and $\tilde{Z}_{B,k} \coloneqq (y_i^B)_{i\in\Gamma_{\mathsf{Z},k}}$, which corresponds to the part of Bob's sifted key that originated from detections with intensity choice $k$. We analogously define $\tilde{X}_{A,k}$ and $\tilde{X}_{B,k}$.
    
    \item \textbf{Error correction:} An error correction scheme that publishes $\mathrm{leak}_\mathrm{EC}$ bits of information is applied in order for Bob to compute an estimate $Z_B$ of $\tilde{Z}_A$, also called corrected key. \added{Let $f$ be a helper function that maps an index $j\in\{1, ...,N_\mathsf{Z}\}$ to the corresponding index in terms of the signal rounds $\{1, ..., N\}$. Then, we define $\Gamma_{\mathsf{Z},k}^{\mathrm{cor}} \coloneqq \{j: k_{f(j)}=k\}$ and therefore $Z_{B,k} \coloneq (Z_B^j)_{j\in\Gamma_{\mathsf{Z},k}^{\mathrm{cor}}}$ is the error-corrected bit string corresponding to detections with intensity choice $k$, in analogy to $\tilde{Z}_{B,k}$, where $Z_B^j$ is the $j$-th element of $Z_B$.} Alice's key remains unchanged and we define $Z_A=\tilde{Z}_A$ as Alice's key after error correction. The $\mathsf{X}$-basis detections $\tilde{X}_A$ and $\tilde{X}_B$ also remain unchanged.

    \item \textbf{Error verification:} Alice then computes the hash of length $\lceil \log_2{1/\epsilon_\mathrm{cor}} \rceil$ corresponding to $Z_A$ given a randomly chosen universal$_2$ hash function. She sends the hash as well as her choice of hash function to Bob who in turn computes the hash of $Z_B$ using the same hash function as Alice. If both hashes disagree, the protocol aborts and they set $S_A = S_B = \perp$. Otherwise, the protocol continues and the corrected keys are called verified keys.
    
    \item \textbf{Acceptance test:}
    Alice discloses her $\mathsf{X}$-basis bits. For each intensity $k$, Bob compares $\tilde{Z}_{B,k}$ with $Z_{B,k}$ to determine the number of errors, $c_{\mathsf{Z},k} = |\tilde{Z}_{B,k}\oplus Z_{B,k}|$, and $\tilde{X}_{A,k}$ with $\tilde{X}_{B,k}$ to determine $c_{\mathsf{X},k} = |\tilde{X}_{A,k}\oplus \tilde{X}_{B,k}|$. The Hamming distance $|\cdot\oplus \cdot|$ corresponds to the number of bits that do not coincide in both strings. Bob computes $c_\mathsf{X} = \sum_k c_{\mathsf{X},k}$ and $c_\mathsf{Z} = \sum_k c_{\mathsf{Z},k}$. He then computes the bounds $s_{\mathsf{Z},0}^-, s_{\mathsf{Z}, 1}^-$, $s_{\mathsf{X}, 1}^-$ and $\Lambda_\mathsf{X}^+$ from the observations using the decoy-state method, cf. Sec.~\ref{sec:decoy_state_bounds}. \added{They perform the acceptance test by verifying the following conditions:
    \begin{align}
        s_{\mathsf{Z},0}^- \geq s_{\mathsf{Z},0}^\mathrm{l} \qquad\text{and}\qquad s_{\mathsf{Z}, 1}^- \geq s_{\mathsf{Z}, 1}^\mathrm{l} \qquad\text{and}\qquad s_{\mathsf{X}, 1}^- \geq s_{\mathsf{X}, 1}^\mathrm{l} \qquad\text{and}\qquad \Lambda_\mathsf{X}^+ \leq \Lambda_\mathsf{X}^\mathrm{u}\,.
    \end{align}
    If any condition does not hold, they set $S_A = S_B = \perp$ and abort the protocol.}

    \item \textbf{Privacy amplification:} Alice extracts a secure key $S_A$ of fixed length $l$, cf. Eq.~\eqref{eq:max_secret_key_formula}, from $Z_A$ using a randomly chosen universal$_2$ hash function. The choice of function is communicated to Bob who uses it to compute a secure key $S_B$ from $Z_B$. 
\end{enumerate}

\begin{table}[h]
\centering
\renewcommand{\arraystretch}{1.2} 
\begin{tabular}{ |c|c|c| } 
 \hline
 \textbf{Event} & \textbf{Description} & \textbf{Relation} \\ \hline
 $\Omega_\mathrm{EC}$ & Error correction succeeded & - \\ 
 $\Omega_\mathrm{EV}$ & Error verification passed & - \\
 $\Omega_\mathrm{AT}$ & Acceptance test passed & - \\ 
 $\Omega_\top$ & Protocol did not abort & $\Omega_\top = \Omega_\mathrm{EV} \land \Omega_\mathrm{AT}$ \\ 
 $\tilde{\Omega}$ & Protocol did not abort and error correction succeeded & $\tilde{\Omega} = \Omega_\top \land \Omega_\mathrm{EC}$  \\ 
 $\Omega_\mathrm{B}$ & Decoy bounds hold & - \\ 
 $\Omega_\circ$ & True values of the statistics are in the acceptance set & $\Omega_\mathrm{AT}\land\Omega_\mathrm{EC}\land \Omega_\mathrm{B}\Rightarrow \Omega_\circ$ \\ 
 \hline
\end{tabular}
\caption{Overview of the various events used in the protocol described in Fig.~\ref{fig:protocol_description}.}
\label{tab:events_description}
\end{table}

\begin{table}[h]
\centering
\renewcommand{\arraystretch}{1.2} 
\begin{tabular}{ |c|c| } 
 \hline
 \textbf{Register} & \textbf{Description} \\ \hline
 $\tilde{C}^N$ & Basis and detect/no-detect announcements  \\ 
 $C_\mathrm{EC}$ & Bits leaked for error correction, $|C_\mathrm{EC}| = \mathrm{leak}_\mathrm{EC}$ \\
 $C_\mathrm{EV}'$ & Hash of Alice's corrected key, $|C_\mathrm{EV}'| = \log_2 2 / \epsilon_\mathrm{cor}$ \\ 
 $C_\mathrm{EV}''$ &Choice of hash function for error verification  \\
 $C_\mathrm{EV}$ & $C_\mathrm{EV} = C_\mathrm{EV}' C_\mathrm{EV}''$    \\
 $C_\mathrm{AT}$ & Acceptance decision (pass/fail) \\ 
 $C_\mathrm{PA}$ & Choice of hash function for privacy amplification \\ 
 $C$ & $C = \tilde{C}^N C_\mathrm{EC} C_\mathrm{EV} C_\mathrm{AT} C_\mathrm{PA}$ \\ 
 $C'$ & Generic classical register   \\
 \hline
\end{tabular}
\caption{Overview of the classical communication registers used in the protocol described in Fig.~\ref{fig:protocol_description}.}
\label{tab:classical_registers_comm_description}
\end{table}

\subsection{\label{sec:preliminaries:theoretical_background}Theoretical background}
The aim of this section is to provide a theoretical background and thus form a foothold for the concepts discussed in this work but does not aim at deriving each concept introduced.
We refer to the textbooks \cite{ChuangNielsen10, Audretsch07} or more recently \cite{Wolf21, Grasselli21} for a more detailed discussion. These textbooks also serve as a good introduction to the basics of the quantum mechanical formalism and quantum information theory.
For a more mathematical description, Tomamichel's book \cite{Tomamichel16} as well as his thesis \cite{TomamichelThesis12} serve as good references and discuss many of the concepts used in this work.

\subsubsection{Representing quantum systems}
\label{sec:theoretical_background_density_distance}

We assume that the reader is familiar with the Dirac notation of quantum mechanics and the density operator formalism.
In the following, the density operator
\begin{equation}
\label{eq:def_density_operator}
    \rho_A = \sum_{a}P_A(a)\ketbra{a}{a}\in S(\mathcal{H}_A)\, ,
\end{equation}
acting on a Hilbert space $\mathcal{H_A}$ describes the quantum state of a quantum system $A$, where $P_A(a)$ can be interpreted as the probability of preparing the system in the state $\ket{a}$ and $S(\mathcal{H}_A)$ is the set of positive semi-definite matrices of trace one acting on $\mathcal{H}_A$.

Consider a \textit{classical random variable} $Z$ following a distribution $P_Z$ on some set $\mathcal{Z}$ where the probability of each outcome $z\in\mathcal{Z}$ is given by $P_Z(z)$. The classical values $z$ can be represented as orthogonal states $\ket{z}\in\mathcal{H}_Z$ reflecting their distinguishability, which is a property common to classical variables \cite[Sec.~2.1]{Renner05}. Indeed, distinguishability describes the ability to differentiate two states given the outcome of a measurement. In the quantum realm, states are reliably distinguishable if the outcome of a measurement can be uniquely assigned to a quantum state. In theory, this can always be done for orthogonal states $\{\ket{z}\}_z$, as we can define a measurement operator $M = \sum_z M_z$ with $M_z = z\ketbra{z}{z}$ for which the measurement results are unique for each $\ket{z}$ and given by the eigenvalue problem $M\ket{z} = z\ket{z}$. A proof that non-orthogonal states cannot be reliably distinguished can be found in Ref.~\cite[Sec.~2.2.4]{ChuangNielsen10}. Given these properties, the density operator representation of a classical distribution $P_Z$ can be written as
\begin{equation}
    \rho_Z = \sum_{z}P_Z(z)\ketbra{z}{z} \, ,
    \label{eq:density_representation_classical_variable}
\end{equation}
in analogy to Eq.~\eqref{eq:def_density_operator}.

\subsubsection{Distance between quantum states}

In order to quantify the distinguishability of two density operators, one introduces a distance measure which is directly related to the probability of distinguishing them. For two density operators $\rho$ and $\sigma$, the \textit{trace distance} is defined as
\begin{equation}
    D(\rho, \sigma)\coloneqq\frac{1}{2}\norm{\rho-\sigma}_1 \, ,
    \label{eq:trace_distance_def}
\end{equation}
where $\norm{\tau}_1 = \Tr{\sqrt{\tau^\dagger\tau}}$ denotes the \textit{trace norm} (also called Schatten 1-norm) of a density operator $\tau$.
The trace distance generalizes the \textit{total variation distance} from the field of statistics, i.e. classical probability distributions to density operators.

Now, the density operators are said to be \textit{$\zeta$-close} if $D(\rho, \sigma) \le \zeta$.
An intuitive operational interpretation of the trace distance is given by the following scenario.
We assume a source capable of preparing two known quantum states $\rho$ and $\sigma$ with equal probability, for which $D(\rho, \sigma) \le \zeta$. The task is to determine which state was prepared, only by measuring the resulting quantum state. The probability with which the measurement result can be assigned to the correct quantum state is directly related to the trace distance and upper-bounded by
\begin{equation}
    P_\mathrm{distinguish}(\rho, \sigma) \leq \frac{1}{2} + \frac{\zeta}{2}\, ,
    \label{eq:distinguishing_advantage}
\end{equation}
regardless of the measurement strategy adopted, cf. Ref.~\cite[Sec.~9.1.4]{Wilde16} for a proof, and Refs.~\cite[Sec.~9.1]{ChuangNielsen10} \cite[Sec.~3.2.1]{TomamichelThesis12} \cite[Sec.~2.4.2]{Tsybakov2009} for discussions.
In other words, two $\zeta$-close operators cannot be distinguished with a probability more than $\zeta$ \cite[Sec.~2.1.4]{Renner05}, meaning that the trace distance can be interpreted as a \textit{distinguishing advantage}. This statement will have profound implications for the intuition behind the security of a protocol, as will be discussed in Sec.~\ref{sec:preliminaries:def_security}.

Intuitively, in the case of pure orthonormal states, $D(\rho, \sigma) = 1$ as they can be reliably distinguished. Identical quantum states, on the other hand, cannot be distinguished and $D(\rho, \sigma) = 0$. For a more detailed discussion about the intuition behind the trace distance, we refer to App.~\ref{ap:intuition_trace_norm}.

As any density operator $\rho$ is Hermitian, $\rho^\dagger=\rho$ by definition and $\rho$ is diagonalizable, so that $\norm{\rho}_1 = \sum_i |p_i|$ where $\{p_i\}_i$ is the set of eigenvalues of $\rho$. This representation of $\norm{\rho}_1$ as a function of the probability mass functions $\{p_i\}_i$ directly connects the density operator to the classical distributions discussed in App.~\ref{ap:intuition_trace_norm} and will be used when discussing the entropy of a quantum system in Sec.~\ref{sec:quantum_leftover_hash_lemma}. As such, classical distributions and their quantum analogues exhibit the same properties regarding the interpretation of the trace distance as distinguishing advantage.
For a more detailed discussion, we refer to Refs.~\cite[App.~A]{Portmann22} \cite[Sec.~3.2]{Tomamichel16} \cite[ Sec.~9.2]{ChuangNielsen10} \cite[Sec.~3.2]{TomamichelThesis12}.

\subsubsection{Bipartite quantum systems}
\label{sec:bipartite_qm_systems}
In the scenario where Alice and Bob exchange information through a quantum channel, the density operators $\rho_A$ and $\rho_B$ describe their respective quantum information.
The bipartite system of two quantum systems $A$ and $B$ describing the joint quantum state of the systems is represented by $\rho_{AB}\in S(\mathcal{H}_{AB})$, where $\mathcal{H}_{AB} = \mathcal{H}_A \otimes \mathcal{H}_B$, $S(\mathcal{H}_{AB})$ is the set of positive semi-definite Hermitian matrices of trace one acting on $\mathcal{H}_{AB}$ and $\otimes$ denotes the tensor product.

If $\rho_{A|Z}$ represents a quantum system $A$ that depends on a classical variable $Z$, we can write the \textit{classical-quantum bipartite system} as
\begin{equation}
    \rho_{ZA}= \sum_z P_Z(z)\ketbra{z}{z}\otimes \rho_{A|Z=z}\, ,
    \label{eq:bipartite_system}
\end{equation}
where $\rho_{A|Z=z}$ describes quantum system $A$ conditioned on $Z=z$ \cite[Sec.~2.1.3]{Renner05}. Intuitively, the bipartite system describes two correlated quantum systems, where $A$ depends on the outcome of $Z$. In turn, this means that any measurement on $Z$ also influenced the outcome of $A$. For an event $\Omega$ on the classical register $Z$, (where an event is defined to be any set of classical outcomes), the state $\rho_{ZA | \Omega }$ \textit{conditioned on the event $\Omega$}, is written as
\begin{equation}
 \rho_{ZA | \Omega} = \frac{\rho_{ZA \wedge \Omega}}{\Tr{\rho_{ZA \wedge \Omega}}}\, , \qquad \mathrm{with} \qquad
    \rho_{ZA \wedge \Omega}= \sum_{z \in \Omega} P_Z(z)\ketbra{z}{z}\otimes \rho_{A|Z=z}\, .   
\label{eq:bipartite_system_conditionoing}
\end{equation}
Note that $\rho_{ZA | \Omega} $ is a normalized density matrix, while $\rho_{ZA \wedge \Omega}$ (sometimes referred to as \textit{partial} state) is sub-normalized. In the following, the operator $\land$ has higher precedence than the operator $|$, i.e. it acts before the operator $|$. \added{An expression such as $\rho_{ZA|\Omega_1\land\Omega_2}$ is to be interpreted as $\rho_{ZA|(\Omega_1\land\Omega_2)}$}.

\begin{remark}
The steps involved in a QKD protocol run depend on the specific events that occur during that run, following the protocol description in Fig.~\ref{fig:protocol_description}. Therefore, for rigorous security proofs, it is essential to compute relevant entropies and statistics on states that are properly conditioned on these events. This is crucial because conditioning can affect entropies in significant ways. In this work, we almost exclusively use the normalized variant of conditioning on events, i.e. $\rho_{ZA | \Omega} $. Furthermore, we make it a point to explicitly clarify this aspect of conditioning in all our calculations, addressing a detail that is often overlooked.
\end{remark}

\begin{remark}
    When conditioning on events, it is also important to ensure that the event is well-defined. In particular, there must exist (in theory) a classical register that determines whether or not the event occurred. Note that we are free to trace out this classical register after conditioning on it. For example, $\rho_{A \wedge \Omega} = \Tr_Z\{\rho_{ZA \land \Omega}\}$ is a well-defined state conditioned on $\Omega$, which does not include the register $Z$ determining the event.
\end{remark}
\subsection{\label{sec:preliminaries:def_security}Definition of security}

The goal of a quantum key distribution protocol is to establish a shared random key known only to Alice and Bob. This is achieved by making use of the laws of quantum mechanics combined with classical information theory and cryptography.

This section aims at defining the security of the protocol and is structured as follows.
First, we introduce the concept of information-theoretic security in Sec.~\ref{sec:information-theoretic_security} \added{and state the assumptions of quantum key distribution, underlying all further analyses, in Sec.~\ref{sec:assumptions_of_qkd}.}
Using the trace distance defined in Sec.~\ref{sec:theoretical_background_density_distance}, we are then able to define the secrecy, correctness and security of the protocol in Sec.~\ref{sec:security_parameters} in terms of so-called $\epsilon$-parameters. We will introduce these parameter constructively starting out with the general definition of security. The secrecy parameter is then expanded in terms of the decoy concentration inequalities holding and failing in Sec.~\ref{sec:expansion_secrecy_ci}.
Finally, the concept of composable security is introduced in Sec.~\ref{sec:composable_security} and the one-time pad discussed in Sec~\ref{sec:one-time-pad} as an information-theoretically secure cryptographic primitive for message encryption.

\subsubsection{\label{sec:information-theoretic_security}Information-theoretic security}
A key generation protocol is said to be \textit{perfectly secure} if an adversary with infinite computing power cannot gather any information about the key \cite{Smart04}. This condition is guaranteed if the key output meets following criteria \cite{Vernam26, Shannon49}:

\begin{enumerate}
    \item The key is secret.
    \item It has been randomly generated, following a uniform distribution.
\end{enumerate}
The second condition can be fulfilled by using an ideal quantum random number generator \cite{Herrero17}. In contrast, the first condition is more difficult to fulfill when the generated key is shared by multiple parties. In fact, the only known method to classically share a perfectly secure key is by transporting it physically, e.g. from Alice to Bob\footnote{This is also known as the \textit{courier problem}.}.
There is no known and proven way to accomplish this task using only a classical channel.

However, QKD provides a solution based on the laws of quantum mechanics.
Even though perfect security can also not be guaranteed by QKD, it allows for an arbitrarily close approximation of the ideal case. The remaining deviation from perfect security is described in terms of $\epsilon$-parameters, which are introduced in \added{Sec.~\ref{sec:security_parameters}, therefore enabling \textit{information-theoretic security}. In classical cryptography, there is no proven method to share an information-theoretically secure key remotely. Remotely shared symmetric keys can only be generated algorithmically, when replacing the requirements for information-theoretical security by the stronger assumption of a computational bound on Eve.}

We note that, in the case assumed in the following, where the key is shared between multiple parties, i.e. Alice and Bob, an additional condition stating that Alice's and Bob's keys must be identical is appended to the list above.

\subsubsection{\label{sec:assumptions_of_qkd}Assumptions of quantum key distribution}
\added{As mentioned in the previous section, information-theoretical security is made possible by quantum key distribution, which does not make any assumptions about Eve's computing power. This relaxation enables the sharing of keys that fulfill the conditions for information-theoretic security as stated in Sec.~\ref{sec:information-theoretic_security}. However, QKD is not deprived of assumptions, but rather based on a set of three assumptions, namely \cite[Sec.~IV.A]{Portmann22}}
\begin{enumerate}
    \item Quantum physics is a correct and complete theory\cite{colbeckNoExtensionQuantum2011}.
    \item The classical channel used by Alice and Bob is authenticated.
    \item The devices used during the protocol perform exactly as instructed\footnote{In our case, this implies that we assume no device imperfections are present.}.
\end{enumerate}
\added{The first condition is rather fundamental and widely accepted.}
\added{In the QKD security-proof literature, the term \emph{authenticated channel} is typically used to refer to an idealized classical communication channel in which all messages sent by one party are eventually received correctly by the other party. This is the model assumed throughout this work. We emphasize that such a guarantee cannot be achieved in practice. Instead, practical authentication schemes ensure that \emph{either} a transmitted message is received correctly, \emph{or} the receiving party obtains a special symbol indicating authentication failure (see, e.g., Refs.~\cite{Portmann14,ferradini2025definingsecurityquantumkey}). Importantly, the security of QKD protocols using such  authentication mechanisms can be reduced to the security analysis in the idealized authenticated channel model assumed here, using Ref.~\cite{inprep_authentication}.}
\added{For subsequent rounds, a part of the secure key from previous rounds can be used for authentication \cite{Portmann14, Portmann22}}\footnote{Quantum key distribution can thus be described as a key expansion method since it requires a seed key for authentication, cf. Refs.~\cite{Mosca12, Portmann22} for a more thorough discussion.}.

\added{The last condition is perhaps the most difficult to ensure as devices, such as single-photon detectors and photon sources, are necessarily imperfect, making room for a wide variety of \textit{side-channel attacks} that specifically aim to exploit these loopholes in order to obtain more information about the key than initially modeled in the security proof.
In recent years, \textit{quantum hacking} has become a prominent field of research as QKD systems become more viable and are commercialized, strengthening the importance of characterizing them and ensuring no loophole can be exploited by potential eavesdroppers \cite{Curras24, Zapatero23, Makarov23,Sajeed21}.
In general, these loopholes can be addressed by adjusting the device modeling in the security proof or by redesigning practical implementations}\footnote{\textit{Device-independent} protocols \cite{Pironio09, Lo12} reduce the assumptions on the devices by partly removing the need to trust them, which can also be used to address the issue of device imperfections and loopholes.}.
\added{Our work aims to establish a baseline security proof and therefore does not consider these complications.
For a more thorough discussion, we refer to Refs.~\cite{BSI24, Jain16}.}

\subsubsection{\label{sec:security_parameters}Security parameters}

Due to the probabilistic nature of quantum key distribution, information theory and the cryptographic post-processing, perfect security of a QKD protocol cannot be guaranteed.
Therefore, we define so-called \textit{security parameters}, or \textit{$\epsilon$-parameters}, that dictate how close the protocol should be from a perfectly secure protocol which generates keys that fulfill the criteria stated in the previous section.
Choosing sufficiently small $\epsilon$-parameters, the protocol can be chosen to be arbitrarily close to a perfectly secure protocol.
The goal of this section is to introduce these parameters and discuss their direct operational meaning. We start with the general definition of security and constructively introduce the correctness and secrecy parameters by conditioning the trace distance on various events.

Let $S_A$ and $S_B$ be Alice's and Bob's secure keys, output by the protocol after having performed all the steps described in Fig.~\ref{sec:preliminaries:protocol_description}. Recall that we use the convention that the key registers store $\perp$ whenever the protocol aborts. In the following, $A$, $B$ and $E$ represent Alice's, Bob's and Eve's quantum systems, respectively, with $C$ denoting the classical register storing all classical communications, cf. Table~\ref{tab:classical_registers_comm_description}.

\begin{definition}[Security]
    A protocol is called $\epsilon$-secure if 
    \begin{equation}
        D(\rho_{S_A S_B EC}, \rho^\mathrm{ideal}_{S_A S_BEC})\leq \epsilon\, , 
        \label{eq:general_security_def}
    \end{equation}
    where $\rho_{S_A S_B EC}$ describes the quantum system of Alice's key, Bob's key, Eve and the classical communications at the output of the protocol. The state $\rho^\mathrm{ideal}_{S_A S_B EC}$ denotes the output state obtained for the actual protocol, but where the key registers are replaced with perfect keys $U_{S_AS_B}$ if the protocol accepts, and $\perp$ if the protocol aborts, and where
    \begin{equation}
        U_{S_A S_B}=\frac{1}{|\mathcal{S}|}\sum_{s\in\mathcal{S}}\ketbra{s}{s}_A \otimes \ketbra{s}{s}_B
    \end{equation}
    is the uniform distribution of all possible keys $s$ represented by $\ket{s}$ \cite[Sec.~2.2.2]{Renner05}.
    \label{sec:security_criterion}
\end{definition}
The definition of security stated above holds for fixed-length protocols where the size of the key registers is fixed and known prior to running the protocol, a subtle fact often overlooked. A more general definition of security, i.e. for variable-length protocols, requires an additional sum over all possible key lengths in Eq.~\eqref{eq:general_security_def}, as mentioned in Sec.~\ref{sec:fixed_length_protocols}. We discuss the operational interpretation of this definition later in this section. The main goal of the security proof is precisely to determine $\epsilon$ and relate it to the length of the key produced by the protocol whenever it does not abort.

Throughout the protocol, Alice and Bob perform the steps described in Fig.~\ref{fig:protocol_description} on their system. As such, in the following, the description of the system $\rho_{S_A S_B EC}$ involves careful conditioning on various events, cf. Sec.~\ref{sec:bipartite_qm_systems}, describing the different outcomes of their actions and decisions. This structure can be visualized in a probability tree depicted in Fig.~\ref{fig:prob_tree_protocol}, which serves as a reference point that we will refer to throughout this work. To avoid cluttering the notation, we define the distance to an ideal protocol, conditioned on an event $\Omega$, as
\begin{equation}
    d_\mathrm{sec}(S_A S_B EC)_{\rho|\Omega} \coloneqq D(\rho_{S_A S_B EC | \Omega}, \rho^\mathrm{ideal}_{S_A S_B EC | \Omega})\, ,
    \label{sec:def_d_sec}
\end{equation}
where we use the normalized conditioning introduced in Sec.~\ref{sec:bipartite_qm_systems}. For simplicity, it is common to reformulate Eq.~\eqref{eq:general_security_def} in terms of the \textit{correctness} and \textit{secrecy} parameters\cite{Portmann22}. Therefore, we expand the definition of security to constructively introduce these parameters in the following. We use the triangle inequality of the trace distance to write
\begin{equation}
    d_\mathrm{sec}(S_A S_B EC)_\rho \leq \Pr[\Omega_\mathrm{EC}] d_\mathrm{sec}(S_A S_B EC)_{\rho|\Omega_\mathrm{EC}} + \Pr[\neg\Omega_\mathrm{EC}] d_\mathrm{sec}(S_A S_B EC)_{\rho|\neg\Omega_\mathrm{EC}}\, ,
    \label{eq:def_sec_split_EC}
\end{equation}
where we have split the trace distance in components conditioned on error correction succeeding and failing. The probabilities $\Pr[\Omega_\mathrm{EC}]$ and $\Pr[\neg\Omega_\mathrm{EC}]$ are not known in practice as they depend on the sifted key (which depends on Eve's attack) and the error correction algorithm used. The above inequality directly corresponds to the first node in Fig.~\ref{fig:prob_tree_protocol}. 
Expanding the second term in Eq.~\eqref{eq:def_sec_split_EC} using the triangle inequality again yields
\begin{equation}
    \Pr[\neg\Omega_\mathrm{EC}] d_\mathrm{sec}(S_A S_B EC)_{\rho|\neg\Omega_\mathrm{EC}} \leq \Pr[\neg\Omega_\mathrm{EC}\land \Omega_\mathrm{EV}] \cdot 1 + \Pr[\neg\Omega_\mathrm{EC}\land \neg\Omega_\mathrm{EV}] \cdot 0\,,
    \label{eq:branch_a}
\end{equation}
where we used the fact that when the protocol aborts, i.e. $\neg \Omega_\mathrm{EV}$ occurs, the distance to an ideal protocol is zero as Alice's and Bob's key are trivially $S_A = S_B = \perp$ \cite[Sec.~III.B]{Portmann22}. We also substituted $d_\mathrm{sec}(S_A S_B EC)_{\rho | \neg\Omega_\mathrm{EC}\land \Omega_\mathrm{EV}} \leq 1$ as no better bound is known when error correction fails. The term $ \Pr[\neg\Omega_\mathrm{EC}\land \Omega_\mathrm{EV}]$ corresponds to branch (a) in Fig.~\ref{fig:prob_tree_protocol} and will explicitly be addressed in Sec.~\ref{sec:error_verification} when introducing universal$_2$ hashing for error verification. We may now rewrite the first term in Eq.~\eqref{eq:def_sec_split_EC}. Since it is conditioned on error correction succeeding, we can write 
\begin{equation}
    D(\rho_{S_A S_B EC|\Omega_\mathrm{EC}}, \rho^\mathrm{ideal}_{S_A S_B EC | \Omega_\mathrm{EC}}) = D(\rho_{S_A EC|\Omega_\mathrm{EC}}, \rho^\mathrm{ideal}_{S_A EC | \Omega_\mathrm{EC}})
    \label{eq:omit_bob_system}
\end{equation}
and omit Bob's system for this term in the following. Analogously to above, applying the triangle inequality yields
\begin{equation}
    \Pr[\Omega_\mathrm{EC}] d_\mathrm{sec}(S_A EC)_{\rho|\Omega_\mathrm{EC}} \leq \Pr[\Omega_\mathrm{EC}\land \Omega_\mathrm{EV}] d_\mathrm{sec}(S_A EC)_{\rho|\Omega_\mathrm{EC}\land \Omega_\mathrm{EV}}\,,
    \label{eq:branch_b}
\end{equation}
where we used $d_\mathrm{sec}(S_A EC)_{\rho|\Omega_\mathrm{EC}\land \neg \Omega_\mathrm{EV}} = 0$ as the protocol aborts\footnote{We also note that $\Pr[\Omega_\mathrm{EC}\land \neg\Omega_\mathrm{EV}] = 0$ as error verification always succeeds if the keys are identical, which follows from the definition of universal$_2$ hashing discussed in Sec.~\ref{sec:error_verification}.\label{foot:remark_error_verification}}. This inequality can be rewritten by applying the triangle inequality one last time, yielding
\begin{equation}
     \Pr[\Omega_\mathrm{EC}\land \Omega_\mathrm{EV}] d_\mathrm{sec}(S_A EC)_{\rho|\Omega_\mathrm{EC}\land \Omega_\mathrm{EV}} \leq\, \Pr[\Omega_\top\land \Omega_\mathrm{EC}] d_\mathrm{sec}(S_A EC)_{\rho|\Omega_\top\land \Omega_\mathrm{EC}}
\end{equation}
where we substituted $d_\mathrm{sec}(S_A EC)_{\rho|\Omega_\mathrm{EC}\land \Omega_\mathrm{EV} \neg\Omega_\mathrm{AT}} = 0$ when $\neg\Omega_\mathrm{AT}$ occurs, i.e. the protocol aborts during the acceptance test. We recall that $\Omega_\top = \Omega_\mathrm{EV}\land \Omega_\mathrm{AT}$ describes the event that the protocol does not abort, cf. Table~\ref{tab:acceptance_testing}. We can explicitly write the ideal state conditioned on the protocol not aborting and error correction succeeding as
\begin{equation}
    \rho^\mathrm{ideal}_{S_A EC | \Omega_\top\land \Omega_\mathrm{EC}} = U_{S_A}\otimes \rho_{EC | \Omega_\top\land \Omega_\mathrm{EC}}\, ,
\end{equation}
following Def.~\ref{sec:security_criterion}. Finally, combining the above expressions, we can rewrite the definition of security as
\begin{empheq}[box=\fcolorbox{black}{equation_box}]{align}
    d_\mathrm{sec}(S_A S_B EC)_{\rho} \leq  \underbrace{\Pr[\neg\Omega_\mathrm{EC}\land \Omega_\mathrm{EV}]}_{\leq\epsilon_\mathrm{cor}} + \underbrace{\Pr[\Omega_\top\land \Omega_\mathrm{EC}]  d_\mathrm{sec}(S_A EC)_{\rho|\Omega_\top\land \Omega_\mathrm{EC}}}_{\leq\epsilon_\mathrm{sec}'} \leq \epsilon\, ,
    \label{eq:expanded_security_def}
\end{empheq}
which is a more convenient expression to prove security in the rest of this work. We now define the correctness and secrecy parameters using the convention commonly used, which slightly differs from the expression derived above, cf. Remark~\ref{rem:secrec_correctness_not_seperated}.

\begin{definition}[Correctness]
    A protocol is called $\epsilon_\mathrm{cor}$\textit{-correct} if $\Pr[\neg\Omega_\mathrm{EC}\land \Omega_\top] = \Pr[S_A \neq S_B \land \Omega_\top]\leq\epsilon_\mathrm{cor}$, i.e. the probability that the protocol does not abort and the keys are not identical is upper-bounded\footnote{Setting $S_A = S_B = \perp$ if the protocol aborts ensures that the more general definition $\Pr[S_A\neq S_B]\leq\epsilon_\mathrm{cor}$ follows from the one stated above, cf. Remark~\ref{rem:general_correctness}.}.
    \label{def:correctness_criterion}
\end{definition}

\begin{definition}[Secrecy]
    A protocol is called $\epsilon_\mathrm{sec}$\textit{-secret} if
    \begin{equation}
        \Pr[\Omega_\top] D(\rho_{S_AEC|\Omega_\top}, U_{S_A}\otimes\rho_{EC|\Omega_\top})\leq \epsilon_\mathrm{sec} \, ,
        \label{eq:def_secrecy}
    \end{equation}
    where $\Pr[\Omega_\top] = 1 - \Pr[S_A = \perp]$ is the probability that the protocol does not abort, $\rho_{S_AEC|\Omega_\top}$ describes the bipartite quantum system of Alice's key and Eve as well as the classical communications, conditioned on the protocol not aborting, and
    \begin{equation}
        U_{S_A}=\frac{1}{|\mathcal{S}|}\sum_{s\in\mathcal{S}}\ketbra{s}{s}
    \end{equation}
    is the uniform distribution of all possible secure keys $s$ represented by $\ket{s}$ \cite[Sec.~2.2.2]{Renner05}.
    \label{def:secrecy}
\end{definition}
\begin{remark}
\label{rem:secrec_correctness_not_seperated}
When considering traditional acceptance testing, which is performed directly after sifting, Eq.~\eqref{eq:expanded_security_def} further simplifies to $\epsilon_\mathrm{cor} + \epsilon_\mathrm{sec} \leq \epsilon$. 
Typically, a security proof for QKD protocols is then obtained by showing that the protocol satisfies the $\epsilon_\mathrm{cor}$-correctness and $\epsilon_\mathrm{sec}$-secrecy criteria separately, and argue that this implies the $(\epsilon_\mathrm{cor}+\epsilon_\mathrm{sec})$-security of the QKD protocol. In this work, we do not prove secrecy and correctness as defined above separately, but still include a discussion of these concepts in this section for pedagogical reasons. This is because the 1-decoy state protocol requires error correction to succeed for Bob to obtain the correct number of errors in the $\mathsf{Z}$-basis and therefore make the correct acceptance decision. Thus, the second term in Eq.~\eqref{eq:expanded_security_def} is additionally conditioned on $\Omega_\mathrm{EC}$ compared to Def.~\ref{def:secrecy}, hence why we define it as $\epsilon_\mathrm{sec}'$. This is different from typical QKD protocols, where the accept decision is only based on public announcements. Nevertheless, the first term in Eq.~\eqref{eq:expanded_security_def} directly corresponds to the correctness parameter as $\Pr[\neg\Omega_\mathrm{EC}\land\Omega_\mathrm{EV}]\geq \Pr[\neg \Omega_\mathrm{EC}\land \Omega_\top]$.
We emphasize that the discrepancy in the definition of the secrecy parameter does not alter the security of the protocol in any way, as the resulting bounds satisfy the definition of security, cf. Def.~\ref{sec:security_criterion}. Defining correctness and secrecy is merely a common intermediate step performed for convenience.
\end{remark}
\tikzstyle{level 1}=[level distance=3cm, sibling distance=8cm]
\tikzstyle{level 2}=[level distance=3cm, sibling distance=4cm]
\tikzstyle{level 3}=[level distance=3cm, sibling distance=4cm]
\tikzstyle{level 4}=[level distance=3cm, sibling distance=4cm]

\tikzstyle{bag} = [text width=8em, text centered]
\tikzstyle{end} = [circle, minimum width=3pt,fill, inner sep=0pt]
\begin{figure}[h]
    \centering
    \begin{tikzpicture}[grow=down, sloped]
        \node[draw, fill=black!8, very thick] {\textbf{Error correction}}
            child {
                node[draw, fill=black!8, very thick] {\textbf{Error verification}}
                child {
                    node[draw, fill=black!8] {Abort}
                    edge from parent
                    node[above] {Failed}
                    node[below] {$\neg\Omega_\mathrm{EV}$}
                }
                child {
                    node[draw, fill=black!8] (EVfail) {Key is not correct} 
                    edge from parent
                    node[above] {Passed}
                    node[below] {$\Omega_\mathrm{EV}$}
                }
                edge from parent
                node[above] {Failed}
                node[below] {$\neg\Omega_\mathrm{EC}$}
            }
            child {
                node[draw, fill=black!8, very thick] {\textbf{Error verification}}
                child {
                    node[draw, fill=black!8, very thick] {\textbf{Acceptance test}}
                    child {
                    node[draw, fill=black!8] {Abort}
                    edge from parent
                    node[above] {Failed}
                    node[below] {$\neg\Omega_\mathrm{AT}$}
                     }
                child {
                    node[draw, fill=black!8, very thick] {\textbf{Decoy CI hold?}}
                    child {
                            node[draw, fill=black!8, very thick] (QLHL) {\textbf{Quantum LHL}} 
                            edge from parent
                            node[above] {Yes}
                            node[below] {$\Omega_\mathrm{B}$}
                        }
                        child {
                            node[draw, fill=black!8] (CInothold) {Key is not secret}
                            edge from parent
                            node[above] {No}
                            node[below] {$\neg\Omega_\mathrm{B}$}
                        }
                    edge from parent
                    node[above] {Passed}
                    node[below] {$\Omega_\mathrm{AT}$}
                     }
                    edge from parent
                    node[above] {Passed}
                    node[below] {$\Omega_\mathrm{EV}$}
                }
                child {
                    node[draw, fill=black!8] (EVfailECsuccess) {Abort}
                    edge from parent
                    node[above] {Failed}
                    node[below] {$\neg\Omega_\mathrm{EV}$}
                }
                edge from parent
                node[above] {Succeeded}
                node[below] {$\Omega_\mathrm{EC}$}
            };

            \node[shift=({0,-0.8})] at (EVfail) {(a) Sec.~\ref{sec:error_verification}};
            \node[shift=({0,-0.8})] at (QLHL) {(b) Secs.~\ref{sec:quantum_leftover_for_smooth_min_entropies} and \ref{sec:decomposing_min_entropy}};
            \node[shift=({0,-0.8})] at (CInothold) {(c) Secs.~\ref{sec:decoy_state_bounds} and \ref{sec:operational_expression_min_entropy}};
            
        \end{tikzpicture}
    \caption{Probability tree diagram representing the different steps of the protocol and conditional bounds on the various events. The branches denoted by (a), (b) and (c) are addressed in separate sections. LHL: leftover hash lemma, CI: concentration inequalities.}
    \label{fig:prob_tree_protocol}
\end{figure}

\noindent Note that, following Eq.~\eqref{eq:omit_bob_system}, we don't need to define a secrecy criterion for Bob. Indeed, if the correctness and secrecy criteria are met, then Bob's key is also necessarily secret. As such, to avoid cluttering the notation, in the following, we use $S$ to describe the secure key when $S_A$ and $S_B$ can be used interchangeably. In the following, we assume an authenticated channel and focus on the correctness and secrecy parameters as they are usually at the heart of QKD security proofs.

The secrecy criterion states that the product of the probability of not aborting the protocol and the distance to an ideal key is smaller than $\epsilon_\mathrm{sec}$, where $\epsilon_\mathrm{sec}$ is usually chosen to be small. As an example (inspired by the manner in which security is proven), the secrecy criterion can be interpreted as either the key $\rho_{SEC|\Omega_\top}$ being close to an ideal system $U_S\otimes\rho_{EC|\Omega_\top}$ or the protocol aborting with high probability. For this ideal system, the key is uniformly distributed, $\rho_S = U_S$, and independent of Eve, as we can write the combined quantum system as a tensor product $U_S\otimes\rho_{EC|\Omega_\top }$, meaning that Eve's quantum system does not depend on the secure key. It describes an information-theoretically secure key, where $S$ is uniformly distributed and independent of $E$, as discussed in Sec.~\ref{sec:information-theoretic_security}. Stating that a key is close to ideal is not to confuse with an $\epsilon_\mathrm{sec}$-secret protocol,
as it is conditioned on the probability $\Pr[\Omega_\top]$ of the protocol not aborting. The secrecy of a QKD system must be given according to $\epsilon_\mathrm{sec}$ as defined in Def.~\ref{def:secrecy}, meaning that the probability for the protocol to pass and for Eve to have significant information about the key is small \cite{Biham05}. This definition takes into accounts attacks where the protocol aborts most of the time but when it passes, Eve is certain to have information about the key. One example is the so-called \textit{swap attack} where Eve intercepts all of Alice's signals, sends random ones to Bob and measures her signals only when the basis choices are published. Most of the time, Bob's error rate will be too high and the protocol will abort, but when Eve gets lucky and the protocol does not abort, she has full information about the key. 

Similarly, we note that the correctness parameter describes the correctness of the protocol but not the key itself. As a simple example, consider an error correction algorithm that always outputs two different keys. In this case, the protocol will abort most of the time during error verification, meaning the protocol is $\epsilon_\mathrm{cor}$-correct, but whenever it does not abort, which inherently occurs with non-zero probability, the keys are not identical. Following this discussion, it is important to emphasize that no claims about the secrecy or correctness of the key can be made when the protocol does not abort. Instead, the security parameters describe the security of \textit{the protocol as a whole}, not the specific instances where it succeeds in producing a non-trivial key. Indeed, intuitively, the security parameters also include the cases where the protocol aborts and the keys are trivially correct and secret, cf. Def.~\ref{sec:security_criterion}. Given that the probability of aborting is unknown, we cannot extend the security claims about the protocol to the non-trivial keys. 

\added{The value of $\epsilon_\mathrm{sec}'$ can be freely chosen depending on the practical use case. To give this parameter a more tangible interpretation, assume there exists the following bound on the second term in Eq.~\eqref{eq:expanded_security_def},
\begin{equation}
    d_\mathrm{sec}(S_A EC)_{\rho|\Omega_\top\land \Omega_\mathrm{EC}} \leq \Delta\,.
    \label{eq:def_delta}
\end{equation}
Then the probability for Eve to guess the entirety of the key when the protocol does not abort is upper-bounded by \cite[Lemma~11]{Portmann22}
\begin{equation}
    P_\mathrm{guess}(S|EC)_{\rho} \leq \frac{1}{|\mathcal{S}|} + \epsilon_\mathrm{sec}' \, ,
    \label{eq:probability_eve_guess_key}
\end{equation}
where $\Delta \leq \epsilon_\mathrm{sec}'$. However, any non-trivial bound $\Delta$ depends on the probability to abort the protocol, which is under Eve's control. Therefore, we cannot obtain a practical bound on Eq.~\eqref{eq:def_delta} directly. What we can bound is the second term in Eq.~\eqref{eq:expanded_security_def}, which includes the abort probability.} Nevertheless, we can see that the parameter $\epsilon_\mathrm{sec}'$ loosely has an operational meaning in the sense that it describes the deviation from an ideal key for which $P_\mathrm{guess}(S|EC)_{U_{S}\otimes\rho_{EC}} = \frac{1}{|\mathcal{S}|}$ and $|\mathcal{S}| = 2^l$ is the number of possible keys of length $l$.
This intuitive expression can serve as a general guideline to estimate the value of the security parameter $\epsilon$ in practical scenarios.

All in all, based on the triangle inequality type arguments discussed above, the security parameter $\epsilon$ upper-bounds the probability for at least one of the following events occurring: the protocol did not abort and the key is distinguishable from an ideal key $U_S\otimes \rho_{EC}$, or the protocol did not abort and the keys $S_A$ and $S_B$ are not identical. The main goal of the security proof is to bound all the terms appearing in Fig.~\ref{fig:prob_tree_protocol}, i.e. Eq.~\eqref{eq:expanded_security_def}, and derive an expression for the secure-key length $l$ such that the protocol is $\epsilon$-secure by either aborting or generating a key of length $l$. Deriving an operational expression for the first term in Eq.~\eqref{eq:expanded_security_def} is the goal of Sec.~\ref{sec:error_verification}. The second term requires more work, involving the quantum leftover hash lemma and bounding min-entropies, as discussed in Secs.~\ref{sec:preliminaries:distance_from_ideal_key}, \ref{sec:quantum_leftover_hash_lemma}, \ref{sec:decoy_state_bounds} and \ref{sec:secret_key_length}. 

\begin{remark}
    An additional parameter, called robustness parameter, is sometimes introduced to quantify the resilience of the protocol or the expected key rate. Indeed, a protocol that consistently aborts is inherently secure but fails to produce a useful output. In fact, a protocol is called $\epsilon_\mathrm{rob}$\textit{-robust} if, in the absence of an adversary, the probability that the protocol aborts is $1 - \Pr[\Omega_\top]\leq\epsilon_\mathrm{rob}$, meaning that the protocol outputs a non-trivial key with a probability of at least $1-\epsilon_\mathrm{rob}$\cite{Mueller-Quade09}. 
\end{remark}

\subsubsection{Conditional bounds expansion} 
\label{sec:expansion_secrecy_ci}
The main goal of the security proof is to ensure that the security criterion, i.e. Def.~\ref{sec:security_criterion}, is fulfilled by the protocol described in Sec.~\ref{sec:preliminaries:protocol_description}. To achieve this, concentration inequalities are used to derive bounds on the detection statistics which are used to perform the acceptance test and determine the secure-key length, as will be discussed in Sec.~\ref{sec:decoy_state_bounds}. These bounds come with an associated probability of failure, which is not explicitly reflected in the second term in Eq.~\eqref{eq:expanded_security_def}. \added{Recall that we denote $\Omega_\mathrm{B}$ the event where all bounds given by the decoy concentration inequalities hold.} To simplify the notation, we define $\tilde{\Omega}\coloneqq \Omega_\top \land \Omega_\mathrm{EC}$ in the following. We thus reformulate this term in preparation for Sec.~\ref{sec:quantum_leftover_hash_lemma}, by applying the triangle inequality of the trace distance
\begin{equation}
    \Pr[\tilde{\Omega}] d_\mathrm{sec}(S_AEC)_{\rho|\tilde{\Omega}} \leq\, \Pr[\tilde{\Omega} \land \Omega_\mathrm{B}] d_\mathrm{sec}(S_AEC)_{\rho|\tilde{\Omega} \land \Omega_\mathrm{B}} + \Pr[\tilde{\Omega} \land \neg\Omega_\mathrm{B}]\, , \label{eq:reformulation_security_def}
\end{equation}
where we have split the terms conditioned on the decoy concentration inequalities holding and failing, represented by the branches (b) and (c), respectively, in Fig.~\ref{fig:prob_tree_protocol}. We have used $d_\mathrm{sec}(S_AEC)_{\rho|\tilde{\Omega}\land \neg\Omega_\mathrm{B}}\leq 1$ as no better bound is known when the concentration inequalities do not hold. This reformulation is better suited for the analysis as we can separately bound each term appearing in the above inequality. A bound on the first term in Eq.~\eqref{eq:reformulation_security_def} is provided by the quantum leftover hash lemma and will be tackled in Secs.~\ref{sec:quantum_leftover_for_smooth_min_entropies} and \ref{sec:decomposing_min_entropy}. An expression for the probability $\Pr[\tilde{\Omega} \land \neg\Omega_\mathrm{B}]$ will be derived in Sec.~\ref{sec:operational_expression_min_entropy} in terms of the probability that the concentration inequalities do not hold.

\subsubsection{\label{sec:composable_security}Composable security}
QKD protocols are almost always combined with other cryptographic schemes as the mere generation of a secure key is of little interest.
Hence, in this scenario, we are interested in describing how the security of the combined system behaves. The security definition in terms of the trace distance inherits an intuitive property from the \textit{universal composability framework} that enables the description of combined systems\cite{Canetti2001, Portmann22}. 

Assume, as an example, that the key produced by the QKD protocol is $\epsilon$-secure and used to encrypt a message with an $\epsilon_\mathrm{enc}$-secure encryption scheme (in the universal composability framework)\footnote{For one-time pad encryption, as described in Sec.~\ref{sec:one-time-pad}, $\epsilon_\mathrm{enc}=0$.}.
This means that the QKD protocol is indistinguishable from an ideal QKD protocol up to a probability $\epsilon$, as discussed in Sec.~\ref{sec:security_parameters}, and the message encrypted is indistinguishable from a perfect encryption procedure up to a probability $\epsilon_\mathrm{enc}$.
Now, following the properties of universal composability, the combined system is $(\epsilon + \epsilon_\mathrm{enc})$-secure and $\epsilon + \epsilon_\mathrm{enc}$ upper-bounds the probability of either the key generation or encryption failing. 
The same applies when concatenating $n$ different $\epsilon$-secure keys, where the resulting key is $n\epsilon$-secure \cite{Mueller-Quade09}.

\subsubsection{The one-time pad}
\label{sec:one-time-pad}
As discussed in the previous section, the key produced by a QKD protocol is almost always used as part of other cryptographic schemes. In fact, the key is often used to establish secure communication between two parties. In this case, the message, also known as \textit{plaintext}, is encrypted and transformed into a \textit{ciphertext}, which can in turn be publicly communicated to a receiver who is also in possession of the key. Making use of its key, the receiver can decrypt the ciphertext and retrieve the plaintext. A straightforward method to achieve this is the so-called \textit{one-time pad}, where the encryption and decryption processes involve performing an XOR-operation $\oplus$ between the key $S$, the plaintext $M$ and the ciphertext $M'$. As such, the ciphertext is computed using
\begin{equation}
    M' = S \oplus M
\end{equation}
and can be decrypted by the receiver using the key to retrieve the plaintext,
\begin{equation}
    M = S \oplus M' \, .
\end{equation}
This encryption procedure is perfectly secure, i.e. $\epsilon_\mathrm{enc} = 0$, if the key used is secure (cf. Sec~\ref{sec:information-theoretic_security}) and the following conditions are met \cite{Vernam26, Shannon49}:

\begin{enumerate}
    \item The key has the same length as the plaintext to be encrypted.
    \item It is only used once (hence the name one-time pad).
\end{enumerate}
If these conditions are met and the key is generated by an $\epsilon$-secure QKD protocol, as described in Sec.~\ref{sec:preliminaries:protocol_description}, then the resulting communications are $\epsilon$-secure. Although mathematically sound, we note that, in practice, the one-time pad is not the preferred encryption scheme as it is expensive in terms of key rates \cite{Schneier96}.

\section{\label{sec:preliminaries:universal_2_hashing}\texorpdfstring{Universal$_2$ hashing}{Two-universal hashing}}
This section serves to introduce \textit{universal$_2$ hashing}, also called \textit{two-universal hashing}, originally presented in Ref.~\cite{Carter79}, and describe its use during the error verification and privacy amplification steps, cf. Sec.~\ref{sec:preliminaries:protocol_description}. In fact, using universal$_2$ hashing, we can compare Alice's and Bob's corrected keys $Z_A$ and $Z_B$ and ensure that they are identical up to a small probability without revealing them. The properties of universal$_2$ hashing directly provide a bound on branch (a) from Fig.~\ref{fig:prob_tree_protocol} and enable the derivation of an operational expression for $\epsilon_\mathrm{cor}$ in Sec.~\ref{sec:error_verification}. \added{Additionally, using a universal$_2$ family of hash functions, the keys are mapped to shorter hashes and, as formalized by the quantum leftover hash lemma, their secrecy regarding an eavesdropper is increased, which is specifically the goal of the privacy amplification step, as discussed in Sec.~\ref{sec:preliminaries:distance_from_ideal_key} and further developed in Sec.~\ref{sec:quantum_leftover_hash_lemma}.} 

\subsection{Error correction and verification}
\label{sec:error_verification}
During the error correction step, $\mathrm{leak}_{\mathrm{EC}}$ bits are disclosed to Eve over the classical communication channel and stored in $C_\mathrm{EC}$ in order for Alice to correct her sifted key $\tilde{Z}_A$, cf. Fig.~\ref{fig:protocol_description}. For numerical analyses, this quantity can be approximated by \added{\cite{Tomamichel17b}}
\begin{equation}
    \mathrm{leak}_\mathrm{EC} \approx N_\mathsf{Z} f_\mathrm{EC} h\left(\frac{1}{N_\mathsf{Z}}|\tilde{Z}_A\oplus Z_A|\right) \, ,
    \label{eq:approx_leak_EC}
\end{equation}
where $f_\mathrm{EC} > 1$ is the error correction inefficiency, which depends on the error rate and the correction scheme used\footnote{Usually, $f_\mathrm{EC}\in[1.05, 1.2]$\cite{Tomamichel17b}.}. Alice and Bob then possess a corrected key pair $Z_A$ and $Z_B$.

The aim of the error verification step is to ensure that Alice's and Bob's corrected keys are identical up to a small probability without disclosing them. In fact, if $Z_A$ and $Z_B$ are not identical, the protocol aborts with high probability and $S_A = S_B = \perp$. This corresponds to the first two branches in Fig.~\ref{fig:prob_tree_protocol}.
The idea is that instead of comparing $Z_A$ and $Z_B$ directly, we compare $f_y(Z_A)$ and $f_y(Z_B)$ where $f_y$ is a randomly chosen hash function\footnote{An example of widely used hash functions is Toeplitz matrices \cite{tyagiUniversalHashingInformationTheoretic2015}.}, resulting in two possibilities:
\begin{align}
    Z_A = Z_B & \Rightarrow f_y(Z_A) = f_y(Z_B) \, , \\
    Z_A \neq Z_B &\Rightarrow \Pr[f_y(Z_A) = f_y(Z_B)] \leq \delta \, ,
\end{align}
\added{where $\delta$ is formally defined in Def.~\ref{def:universal_2}.} In other words, if $Z_A = Z_B$ we get with certainty $f_y(Z_A) = f_y(Z_B)$ and if $Z_A \neq Z_B$, we observe $f_y(Z_A) \neq f_y(Z_B)$ up to a small probability $\delta$, which correspond to the correctness parameter introduced in Sec.~\ref{sec:security_parameters}. Additionally, $Z_A$ and $Z_B$ cannot reliably be reconstructed from $f_y(Z_A)$ and $f_y(Z_B)$ as hash functions are chosen to be deterministically random and non-invertible, and leak at maximum the length of the hash in bits. For universal$_2$ hashing, the index $2$ denotes the fact that the comparison occurs between two elements and the properties need only to hold for pairs of elements.  This is more formally described in the following definition.

\begin{definition}[Universal$_2$]
\label{def:universal_2}
    Let $\mathcal{F} =\{f_y\}_y$ be a family of hash functions mapping a set $\mathcal{Z}$ onto a set $\mathcal{S}$, where $|\mathcal{Z}|>|\mathcal{S}|$. $\mathcal{F}$ is called universal$_2$ if for all $z, \tilde{z}\in \mathcal{Z}$ with $z \neq \tilde{z}$, we have $f_y(z) = f_y(\tilde{z})$ for at most a fraction $\delta \coloneqq 1 / |\mathcal{S}|$ of the functions $f_y$ (see Refs.~\cite{Carter79} and \cite[Sec.~2.1]{RennerKoenig05}). 
    In this context, $\mathcal{Z}$ is called the set of possible \textit{keys} and $\mathcal{S}$ the set of \textit{hashes} (also called \textit{indices} in the original work).
\end{definition}
\noindent In the following, the term hash of $z$ will be used to describe $f_y(z)$.
Using the definition above, we can bound the first term in Eq.\eqref{eq:expanded_security_def}, i.e. branch (a) in Fig.~\ref{fig:prob_tree_protocol}. We recall that $\Omega_\mathrm{EC}$ denotes the event where error correction succeeds, i.e. $Z_A = Z_B$, and $\Omega_\mathrm{EV}$ the error verification passing, i.e. $f_y(Z_A) = f_y(Z_B)$. Following Def.~\ref{def:universal_2}, we can state that $\Pr[\Omega_\mathrm{EV} | \neg\Omega_\mathrm{EC}] \leq \delta$ is bounded. Using Bayes' theorem, this directly bounds branch (a) as $\Pr[\neg\Omega_\mathrm{EC}\land \Omega_\mathrm{EV}] \leq \Pr[\Omega_\mathrm{EV} | \neg\Omega_\mathrm{EC}]$, cf. Eq.~\eqref{eq:branch_a}. We can thus choose $\delta \leq \epsilon_\mathrm{cor}$, cf. Def.~\ref{def:correctness_criterion}, as $\Pr[\neg\Omega_\mathrm{EC}\land \Omega_\top] \leq \Pr[\neg\Omega_\mathrm{EC}\land \Omega_\mathrm{EV}]\leq \delta$. 

Recalling that $\delta = 1 / |\mathcal{S}|$ is defined in terms of the number of possible hash function $f_y$, we can determine the length $c$ of the hash required for the correctness criterion to hold. Indeed, we note that for a bit string of length $c$, $|\mathcal{S}| = 2^c$. Thus, in order to fulfill the correctness criterion, we can see that $c \geq \lceil \log_2{1/\epsilon_\mathrm{cor}} \rceil$ as
\begin{empheq}[box=\fcolorbox{black}{equation_box}]{align}
    \Pr[\Omega_\mathrm{EV} \land \neg\Omega_\mathrm{EC}] \leq \Pr[\Omega_\mathrm{EV}\,|\,\neg\Omega_\mathrm{EC}]
    \leq
    \delta
    \leq
    2^{-\lceil \log_2{1/\epsilon_\mathrm{cor}}\rceil} 
    \leq
    \epsilon_\mathrm{cor}\, ,
    \label{eq:bound_epsilon_cor}
\end{empheq}
which bounds branch (a) in Fig.~\ref{fig:prob_tree_protocol}, i.e. the first term in Eq.~\eqref{eq:expanded_security_def}, and thus fulfills the correctness criterion, cf. Def.~\ref{def:correctness_criterion}. 
Or, by adding one bit to the hash function length to avoid rounding up, $c= \log_2{2/\epsilon_\mathrm{cor}}$,
\begin{equation}
    \Pr[\Omega_\mathrm{EV}\,|\,\neg\Omega_\mathrm{EC}] \leq 2^{-\log_2{2/\epsilon_{\mathrm{cor}}}} = 2^{-1}\epsilon_\mathrm{cor} \leq \epsilon_\mathrm{cor}\, .
\end{equation}
It follows that $\log_2{2/\epsilon_\mathrm{cor}}$ bits are used for the hash function and disclosed to Eve through the classical channel for error verification. We store the choice of hash function as well as the hash of Alice's corrected key in a classical register $C_\mathrm{EV}$.
Error verification using universal$_2$ hashing thus directly provides a bound on branch (a), i.e. the protocol is $\epsilon_\mathrm{cor}$-correct, as summarized by Theorem~\ref{th:epsilon_cor}. We may now focus on branches (c) and (d), i.e. Eq.~\eqref{eq:reformulation_security_def}, which require more work and will be the aim of the following sections.
\added{
\begin{theorem}
    Consider the protocol described in Fig.~\ref{fig:protocol_description}, where a universal$_2$ family of hash functions $\mathcal{F}=\{f_y\}_y$ is used during the error verification step, with $f_y:\mathcal{Z}\rightarrow\mathcal{S}$. Then, the protocol is $\epsilon_\mathrm{cor}$-correct if the length $\log_2|\mathcal{S}|$ of the hashes is at least $\lceil \log_2{1/\epsilon_\mathrm{cor}} \rceil$.
    \label{th:epsilon_cor}
\end{theorem}
\begin{proof}
    The proof is given above.
\end{proof}}

\begin{remark}
\label{rem:general_correctness}
    Setting $S_A = S_B = \perp$ when the protocol aborts leads to a more general definition for the correctness parameter, namely $\Pr[S_A \neq S_B] \leq \epsilon_\mathrm{cor}$ as \cite{Wolf21}
    \begin{equation}
        \Pr[S_A \neq S_B] = \Pr[S_A \neq S_B \land \Omega_\top] + \Pr[S_A \neq S_B \land \neg \Omega_\top] \leq \epsilon_\mathrm{cor}
    \end{equation}
    where $\Pr[S_A \neq S_B \land \Omega_\top] \leq \epsilon_\mathrm{cor}$ and $\Pr[S_A \neq S_B \land \neg \Omega_\top] =0$ as upon aborting the protocol, we trivially have $\Pr[S_A=S_B]=1$ since $S_A = S_B = \perp$. As discussed in Sec.~\ref{sec:security_parameters}, we may however not make any claims about the correctness of the key when conditioning on the protocol not aborting.
\end{remark}

\subsection{\label{sec:preliminaries:distance_from_ideal_key}Privacy amplification}
Another application of universal$_2$ hashing is privacy amplification. As perfect security can never be ensured, we \added{introduced the definition of security in Sec.~\ref{sec:preliminaries:def_security}, which allows us to} quantified how close the key output is from a perfectly secure key. By performing privacy amplification, a key arbitrarily close to a perfectly secure key can be produced.
In fact, the privacy amplification step is performed by taking the verified key $Z$ and a hash function $f_y$ chosen at random following a uniform distribution $U_Y$ from a universal$_2$ family of hash functions\footnote{A more general definition of universal$_2$ hashing exists that does not require keys to be chosen uniformly at random. However, in this work we assume uniformity, as it is the most common approach in practical implementations.}, in order to produce a secure key $S$ of length $l$ which is shorter than $Z$ and (ideally) independent of $E$. The systems $Z$ and $S$ represent classical distributions and can hence be represented using Eq.~\eqref{eq:density_representation_classical_variable}.

Let $\mathcal{F} = \{f_y\}_y$ be a universal$_2$ family of hash functions from $\mathcal{Z}$ to $\mathcal{S}$, as defined in Def.~\ref{def:universal_2}, describing the privacy amplification step.
The output string is given by \cite{Renner05}
\begin{equation}
    S = f_Y(Z) = \sum_{f_y\in\mathcal{F}} U_Y(f_y) f_y(Z)\,, 
\end{equation}
where $l = |S| < |Z| = N_\mathsf{Z}$ and the hash function $f_y$ is chosen uniformly at random from $\mathcal{F}$. Alice (or Bob) publicly communicates her choice of hash function to Bob (Alice) and both compute the hash of their verified key using the same hash function. The choice of hash function should not be disclosed before Bob has received all signals from Alice.
It can be shown that for the resulting key $S$, $\epsilon_\mathrm{sec}'$ exponentially decreases with the length difference $N_\mathsf{Z} - l$, meaning that virtually any $\epsilon_\mathrm{sec}'$ can be chosen if
the length $N_\mathsf{Z}$ of the verified key is long enough and the length $l$ of the secure key is short enough\footnote{Note, however, that $l$ needs to be finite and positive for key generation. Thus, one cannot increase $N_\mathsf{Z} - l$ arbitrarily.} \cite[Sec.~4.1]{RennerKoenig05}.
This relation is given by the \textit{quantum leftover hash lemma}, linking the extractable secure-key length to $\epsilon_\mathrm{sec}'$, which we introduce and discuss in the next section in order to bound branch (b) in Fig.~\ref{fig:prob_tree_protocol}.

\section{The quantum leftover hash lemma}
\label{sec:quantum_leftover_hash_lemma}
In Sec.~\ref{sec:preliminaries:def_security}, we have defined the security of the protocol and expanded it by conditioning on various events and applying the triangle inequality to obtain Eq.~\eqref{eq:expanded_security_def}, which we recall to be
\begin{equation} 
    d_\mathrm{sec}(S_A S_B EC)_{\rho} \leq  \Pr[\neg\Omega_\mathrm{EC}\land \Omega_\mathrm{EV}] + \Pr[\Omega_\top\land \Omega_\mathrm{EC}]  d_\mathrm{sec}(S_A EC)_{\rho|\Omega_\top\land \Omega_\mathrm{EC}} \leq \epsilon\, ,
\end{equation}
where we have bound the first term with $\epsilon_\mathrm{cor}$ in Sec.~\ref{sec:error_verification}.
We now focus on the second term. Therefore, in the following sections, we assume that the keys have already passed all steps in the protocol, as described in Fig.~\ref{fig:protocol_description}, up to but not including privacy amplification. 

We consider the event $\Omega_{\top} \land \Omega_\mathrm{EC}$. Alice and Bob thus possess a verified key pair $Z_A$ and $Z_B$ of length $N_Z$ which is correct but still potentially strongly correlated with Eve. This means that Eve potentially holds substantial information about their keys. The goal of the privacy amplification step is to produce a key that is ideally independent of Eve, i.e. fulfilling the secrecy criterion, as discussed in Sec.~\ref{sec:preliminaries:distance_from_ideal_key}. In this section, we use $Z$ to describe $Z_A$ and $Z_B$ interchangeably as we condition on the error correction succeeding, i.e. $Z_A$ and $Z_B$ are identical. We have further expanded the above expression in Sec.~\ref{sec:expansion_secrecy_ci} in terms of the decoy concentration inequalities holding and failing (Eq.~\eqref{eq:reformulation_security_def}) to obtain
\begin{equation}
    \Pr[\tilde{\Omega}] d_\mathrm{sec}(SEC)_{\rho|\tilde{\Omega}} \leq\, \Pr[\tilde{\Omega} \land \Omega_\mathrm{B}] d_\mathrm{sec}(SEC)_{\rho|\tilde{\Omega} \land \Omega_\mathrm{B}} + \Pr[\tilde{\Omega} \land \neg\Omega_\mathrm{B}]\, , 
    \label{eq:recall_expanded_decoy_ci}
\end{equation}
where we recall that $\tilde{\Omega} = \Omega_\top \land \Omega_\mathrm{EC} = \Omega_\mathrm{AT} \land \Omega_\mathrm{EV} \land \Omega_\mathrm{EC}$, cf. Table~\ref{tab:events_description}. The quantum leftover hash lemma in fact provides a bound on $d_\mathrm{sec}(SEC)_{\rho|\tilde{\Omega}\land\Omega_\mathrm{B}}$, conditioned on the decoy concentration inequalities holding, in terms of the secure-key length and will thus be the focus of this section. This term corresponds to branch (b) in Fig.~\ref{fig:prob_tree_protocol}. The term corresponding to the concentration inequalities not holding, i.e. the second term in Eq.~\eqref{eq:recall_expanded_decoy_ci} (branch (c)), will be tackled in Sec.~\ref{sec:operational_expression_min_entropy}.

First, the min-entropy will be introduced in Secs.~\ref{sec:guessing_prob_min_entropy} and \ref{sec:smooth_min_entropy} as a conservative way to describe Eve's uncertainty about the sifted key.
The quantum leftover hash lemma will then formally be introduced in Sec.~\ref{sec:quantum_leftover_for_smooth_min_entropies}, relating $\epsilon_\mathrm{sec}'$ to the min-entropy and the secure-key length. A bound on the min-entropy is finally derived in terms of the numbers of photon events and errors in Sec.~\ref{sec:decomposing_min_entropy} by applying the chain rule for smooth min-entropies and using the entropic uncertainty relation.

\subsection{Eve's guessing probability and the min-entropy}
\label{sec:guessing_prob_min_entropy}
The \textit{von Neumann entropy} $S(\rho)$ of a quantum system $\rho$ is a measure for the average uncertainty of an observer about the system \cite{Audretsch07}. It constitutes a generalization of the Shannon entropy from classical information theory and is defined as
\begin{equation}
    S(\rho)=-\Tr{\rho \log_2 \rho}\, .
\end{equation}
Using the spectral decomposition of $\rho$, the von Neumann entropy can be rewritten equivalently to the Shannon entropy as 
\begin{equation}
    S(\rho)=-\sum_{z}p_z\log_2 p_z\, ,
\end{equation}
where $p_z$ are the eigenvalues of $\rho$. The von Neumann entropy is bounded by (for a proof, see Ref.~\cite[Sec.~5.1]{Audretsch07})
\begin{equation}
    0\leq S(\rho) \leq \log_2 \dim \mathcal{H}
    \label{eq:bounds_neumann_entropy}
\end{equation}
and is maximal when considering a uniform distribution $\{p_x\}_x$ as it constitutes the case of maximum uncertainty. In fact, in this case, $\rho = \frac{1}{\dim\mathcal{H}}I$ and $S(\rho) = \log_2\dim\mathcal{H}$, where $I$ is the identity matrix acting on $\mathcal{H}$. For example, in the case of $l$ uniformly distributed qubits, $\dim\mathcal{H} = 2^l$ and $S(\rho) = l$.

When working in the realm of cryptography, the worst case scenario is typically considered and thus a more conservative entropy measure is introduced.
In contrast to the von Neumann entropy, the \textit{min-entropy}
does not consider the average uncertainty of an adversary Eve, but her best possible attack strategy and quantifies the entropy in the case where her uncertainty is minimal and she is the most likely to guess the secure key \cite[Sec.~I.B]{Tomamichel11}. \added{Formally, the min-entropy of A conditioned on B of the state $\sigma_{AB}$ is defined as \cite[Def.~6.2]{Tomamichel16}
\begin{equation}
    H_\mathrm{min}(A|B)_{\sigma} = \sup_{\tau_B\in S_\leq (\mathcal{H}_B)} \sup\{\lambda \in \mathbb{R} : \sigma_{AB} \leq \exp(-\lambda) I_A \otimes \tau_B\}\, ,
\end{equation}
where $I_A$ is the identity matrix acting on $A$ ($\mathcal{H}_A$) and $S_\leq(\mathcal{H}_B)$ is the set of positive semi-definite matrices of trace less or equal to one acting on $\mathcal{H}_B$. When $A$ is a classical register, the min-entropy has an operational interpretation in terms of the maximum guessing probability as follows.} 

\added{Let $C'$ denote a generic classical register storing classical communications. In the case where $\rho_{ZEC'}$ describes potentially correlated systems $Z$ (describing the verified key) and $EC'$, the min-entropy given side information $E$ and $C'$ is defined as
\begin{equation}
    H_\mathrm{min}(Z|EC')_{\rho}=-\log_2p_\mathrm{guess}(Z|EC')_{\rho} \, ,
    \label{eq:min_entropy}
\end{equation}
where $p_\mathrm{guess}(Z|EC')_{\rho}$ denotes Eve's maximum probability of guessing the key $Z$ given her side information $EC'$ ($p_\mathrm{guess}$ is formally defined in Ref. \cite[Theorem 1]{Konig09}).} A concrete comparison between the von Neumann and min-entropy can be found in App.~\ref{ap:comparison_neumann_min_entropy}.  In the following, we use the notation $H_\mathrm{min}(Z|EC')_{\rho} = H_\mathrm{min}(\rho_{Z|EC'})$. The notation $f(A|B)$ describes $f(A)$ given side information $B$. In the case of $H_\mathrm{min}(Z|EC')_{\rho}$, it is assumed that Eve has all information about the classical communications $C'$ and she possesses the information $E$ she gathered during the key distribution. As such, $H_\mathrm{min}(Z|EC')_{\rho}$ describes her uncertainty about $Z$ with access to side information $EC'$.

\begin{remark}     
    An illustrative example for Eve's guessing probability is given in the following. Assume that the verified keys $Z$ are uniformly distributed but correlated with Eve. Then, for each measurement on her system $E$, the outcome influences her perspective of the distribution of the verified keys as she acquires information about $Z$. As such, from her point of view, some keys become more or less likely given that she measured a particular outcome. For a visual representation of how the verified key distribution changes from her perspective after performing a measurement on her system, see Fig.~\ref{fig:measurement_influence_probability_keys}. Assuming perfect state preparation, before measuring $E$, $Z$ is a uniform distribution of all possible keys. After measuring E, the probability distribution of the keys is altered as she is able to make certain predictions about the verified key given the measurement outcome. If Eve's quantum system is not correlated with $Z$, then any measurement on her system yields an outcome that does not give her any information about $Z$ and the verified key distribution remains uniform from her perspective.
    The min-entropy assumes the worst case scenario where Eve picks the key that, in her perspective, has the highest probability of being generated, as marked by a red cross in Fig.~\ref{fig:measurement_influence_probability_keys}. The maximum probability of guessing the key, cf. Eq.~\eqref{eq:min_entropy}, is then given by the probability of generating the most probable key (from Eve's perspective).

    \begin{figure}
        \centering
        \includegraphics{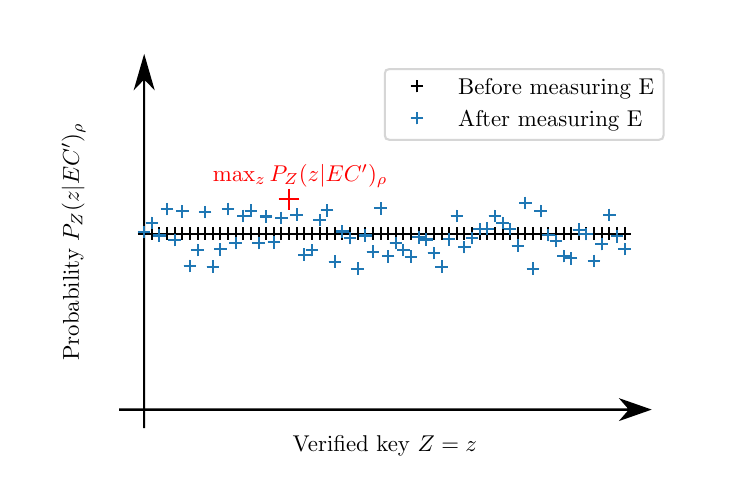}
        \caption{Illustration of the probability distribution of the verified key $Z$ before and after Eve measures her quantum system, where the keys are originally uniformly distributed but correlated with Eve. The correlation translates into the probability distribution being affected by the measurement outcome, leading to some keys being more or less likely given Eve's perspective. Here, the abscissa represents all possible $6$-bit verified keys in an arbitrary order.}
        \label{fig:measurement_influence_probability_keys}
    \end{figure}
\end{remark}

\noindent Hence, the min-entropy is a conservative measure of entropy as it considers the worst case scenario where Eve performs the best possible attack and has the highest probability of guessing $Z$ while the von Neumann entropy describes the average uncertainty about the system. Using the bounds from Eq.~\eqref{eq:bounds_neumann_entropy} we find
\begin{equation}
    0\leq H_\mathrm{min}(\rho)\leq S(\rho)\leq \log_2{\dim{\mathcal{H}}} \, ,
\end{equation}
where the maximum is attained for both entropy measures if $\rho$ is a uniform distribution and is not correlated with another system.

\begin{example}
    If $Z$ and $EC'$ are uncorrelated, then $P_\mathrm{guess}(Z|EC')_{\rho} = \max_z P_{Z}(z)$ does not depend on Eve's measurement outcome or strategy.
    Hence, $H_\mathrm{min}(Z|EC')_{\rho}= -\log_2 \max_z P_{Z}(z)$, following Eq.~\eqref{eq:min_entropy}. In the case of $l$ uniformly distributed qubits, $H_\mathrm{min}(Z|EC')_{\rho}=-\log_2{2^{-l}}=l$, which is the length of the verified key and corresponds to the maximum uncertainty. In this situation, the entirety of the verified key is already secret.
    Note, that this is never the case when considering finite-size effects, cf. Sec.~\ref{sec:decoy_state_bounds}.
\end{example}

\subsection{\label{sec:smooth_min_entropy}The smooth min-entropy}

The \textit{smooth min-entropy} is a more optimal measure of entropy than the min-entropy, leading to greater secure-key lengths as it maximizes the min-entropy $H_\mathrm{min}(\rho)$ over all states $\bar{\rho}$ that are $\zeta$-close to $\rho$ in terms of a distance measure called the purified distance \cite[Secs.~3.1 and 5.2]{TomamichelThesis12}\footnote{Indeed, the min-entropy, as defined in Eq.~\eqref{eq:min_entropy}, can significantly change from small modifications of the system's state. The smoothed version of the min-entropy resolves this issue.}. More formally, it is evaluated at $\rho$ by maximizing $H_\mathrm{min}(\bar{\rho})$ over all $\bar{\rho}\in\mathcal{B}^\zeta (\rho)$ where $\mathcal{B}^\zeta (\rho)$ is a set of density operators $\zeta$-close to $\rho$, also called $\zeta$-ball, such that 
\begin{equation}
    \mathcal{B}^\zeta (\rho) \coloneqq \{\sigma\in S_\leq (\mathcal{H}) : P(\sigma,\rho)\leq \zeta\}\,,
\end{equation}
where \cite[Def.~3.15]{Tomamichel16}
\begin{equation}
    P(\sigma, \rho) \coloneqq \sqrt{1-\left( \Tr{|\sqrt{\sigma}\sqrt{\rho} |}+ \sqrt{(1-\Tr{\rho})(1-\Tr{\sigma})}  \right)^2_1}
\end{equation}
is the purified distance.
Here, a new distance measure, the \textit{purified distance}, is introduced as a more optimal metric to define $\zeta$-balls. In fact, $P(\sigma,\rho)$ represents the minimum trace distance between purifications of $\sigma$ and $\rho$ (see Refs.~\cite[Sec.~3.4]{Tomamichel16} and \cite[Def.~4]{Tomamichel10}). The definition of the smooth entropies in terms of the purified distance is applied to derive the duality relation between smooth min- and max-entropies \cite{Tomamichel10}, which is in turn used for the uncertainty relation for smooth entropies \cite{TomamichelRenner11}, cf. Sec.~\ref{sec:entropic_uncertainty_relation}.
For a more detailed discussion about the purified distance, we refer to Ref.~\cite{TomamichelThesis12}. The purified distance inherits the same interpretation as the trace distance as distinguishing advantage \cite[Sec.~3.2.3]{TomamichelThesis12}.

As such, for some $0\leq \zeta < 1$, also called \textit{smoothing parameter}\footnote{\added{Technically, the smoothing parameter of the smooth min-entropy evaluated for a state $\rho$ lies in $[0, \sqrt{\Tr{\rho}})$. Throughout this work we typically have $\Tr{\rho} = 1$, but this becomes relevant when considering sub-normalized conditional states, such as in Eq.~\eqref{eq:subnormalized_state_in_entropy}. \label{foot:smoothing_trace}}}, the smooth min-entropy of $Z$, given Eve's side information $E$ and $C'$, is defined as
\begin{equation}
    H_\mathrm{min}^\zeta(Z|EC')_{\rho}
    \coloneqq \max_{\bar{\rho}_{ZEC'}\in\mathcal{B}^\zeta (\rho_{ZEC'})} H_\mathrm{min}(Z|EC')_{\bar{\rho}}\, .
    \label{eq:definition_smooth_min_entropy}
\end{equation}
Intuitively, a small error probability $\zeta$ that $\bar{\rho}_{SEC'}$ can be distinguished from $\rho_{SEC'}$ is tolerated in return for optimizing the min-entropy over an $\zeta$-ball around $\rho_{SEC'}$. Intuitively, this can be done as two $\zeta$-close states cannot be distinguished with a probability greater than $\zeta$, meaning that for small $\zeta$, both states are effectively indistinguishable (see Sec.~\ref{sec:preliminaries:theoretical_background}). This optimization can lead to a much greater secure-key length than the non-smoothed min-entropy would allow. It should be noted that this does not weaken the protocol security in any way because the smoothing parameter is taken into account in $\epsilon_\mathrm{sec}'$. In a sense, a part of $\epsilon_\mathrm{sec}'$ is used to optimize the smooth min-entropy over an $\zeta$-ball around $\rho_{ZEC'}$. Indeed, in Sec.~\ref{sec:quantum_leftover_for_smooth_min_entropies} it will be seen that the term $2\zeta$ will appear in $\epsilon_\mathrm{sec}'$ due to this optimization. Note that if no error is tolerated, then $\zeta = 0$ and the smooth and non-smoothed versions of the min-entropy coincide:
\begin{equation}
    H_\mathrm{min}^{\zeta=0}(Z|EC')_{\rho} =  H_\mathrm{min}(Z|EC')_{\rho}\, .
\end{equation}
Now that we have formally introduced the smooth min-entropy, we can state and use the quantum leftover hash lemma in the next section to bound branch (b), i.e. the first term in Eq.~\eqref{eq:reformulation_security_def}.

\begin{remark}
    Often, a subindex $\rho$ is introduced, such that $H_\mathrm{min}(Z|EC')_\rho$ describes Eve's uncertainty given access to side information $E$ and $C'$. In this context, the subindex represents the state the entropy is evaluated at.
\end{remark}

\subsection{The quantum leftover hash lemma}
\label{sec:quantum_leftover_for_smooth_min_entropies}
We recall that in order to fulfill the security criterion, we require a bound on Eq.~\eqref{eq:general_security_def}, which we expanded in terms of various events, as illustrated in Fig.~\ref{fig:prob_tree_protocol}. We have bounded branch (a) in Sec.~\ref{sec:error_verification} using the properties of universal$_2$ hash functions. In Sec.~\ref{sec:expansion_secrecy_ci}, we have further expanded in terms of the decoy concentration inequalities holding and failing, which corresponds to branches (b) and (c), respectively. Branch (c) will be tackled in Sec.~\ref{sec:operational_expression_min_entropy}. The quantum leftover hash lemma provides a bound on branch (b), i.e. the first term in Eq.~\eqref{eq:reformulation_security_def}, in terms of the secure-key length and the smooth min-entropy introduced in the last section.

We recall that the quantum state in branch (b) is conditioned on $\tilde{\Omega}\land \Omega_\mathrm{B}$. However, in Sec.~\ref{sec:entropic_uncertainty_relation}, we will see that we cannot directly bound the min-entropy of such state when applying the entropic uncertainty relation\footnote{As discussed in Sec.~\ref{sec:entropic_uncertainty_relation}, this has to do with the subtle point that we cannot condition on $\Omega_\mathrm{AT}, \Omega_\mathrm{EC}$ nor $\Omega_\mathrm{EV}$ while using the EUR because those events are not well-defined for certain states that show up in the EUR statement.}. To avoid this issue, we define $\Omega_\circ$ as the event where the true values of the observed statistics fulfill the acceptance condition, i.e. they are in the acceptance set, cf. Table~\ref{tab:acceptance_testing}. This event is not something we observe in the protocol, but is nevertheless required in the theoretical security proof. By definition, if the event $\Omega_\mathrm{AT}\land\Omega_\mathrm{EC}\land\Omega_\mathrm{B}$ is true, then $\Omega_\circ$ is also true, i.e. we can write
\begin{equation}
    \Omega_\mathrm{AT}\land\Omega_\mathrm{EC}\land\Omega_\mathrm{B} \Rightarrow \Omega_\circ\,,
    \label{eq:implication_omega_circ}
\end{equation}
cf. Table~\ref{tab:events_description}. Using this, we show below that the bounds derived for states conditioned on $\Omega_\circ$ also hold for states conditioned on $\tilde{\Omega}\land \Omega_\mathrm{B}$ and we can thus bound branch (b). For now, however, we assume a conditioning on $\Omega_\circ$.

Assuming that a universal$_2$ family of hash functions is used in the privacy amplification step to extract a key $S$ from $Z$, cf. Sec.~\ref{sec:preliminaries:distance_from_ideal_key}, the quantum leftover hash lemma \cite[App.~B]{Tomamichel17} \cite[ Corollary~5.6.1]{Renner05} provides an upper bound on the distance of the real key to an ideal key in terms of the secure-key length $l$ and Eve's uncertainty about the key,
\begin{empheq}[box=\fcolorbox{black}{equation_box}]{align}
    d_\mathrm{sec}(SEC)_{\rho|\Omega_\circ}\leq 2\zeta + \frac{1}{2}\sqrt{2^{l-H_\mathrm{min}^\zeta \left(Z\,|\,E\tilde{C}^N C_\mathrm{EC} C_\mathrm{EV} C_\mathrm{AT}\right)_{\rho|\Omega_\circ}}}\,,
    \label{eq:quantum_leftover_hash_smooth}
\end{empheq}
where $d_\mathrm{sec}(SEC)_{\rho|\Omega_\circ}$ is defined in Eq.~\eqref{sec:def_d_sec}, with $0 \leq \zeta < 1$ and we recall that $C = \tilde{C}^N C_\mathrm{EC} C_\mathrm{EV} C_\mathrm{AT} C_\mathrm{PA}$, cf. Table~\ref{tab:classical_registers_comm_description}. Notice that the choice of hash function for privacy amplification, stored in $C_\mathrm{PA}$, does not appear on the right-hand side of the equation above. This can be written as
\begin{equation}
    d_\mathrm{sec}(SEC)_{\rho|\Omega_\circ}\leq 2\zeta + \Delta^\zeta_\mathrm{pa} \eqqcolon \Delta_\mathrm{pa} \,,
    \label{eq:delta_dsec_bound}
\end{equation}
where we define
\begin{equation}
    \Delta^\zeta_\mathrm{pa} \coloneqq \frac{1}{2}\sqrt{2^{l-H_\mathrm{min}^\zeta \left(Z\,|\,E\tilde{C}^N C_\mathrm{EC} C_\mathrm{EV} C_\mathrm{AT}\right)_{\rho|\Omega_\circ}}}\,.
    \label{eq:def_delta_pa_zeta}
\end{equation}
The term $2\zeta$ results from the fact that the smooth min-entropy is optimized on a $\zeta$-ball around $\rho_{ZE\tilde{C}^N C_\mathrm{EC} C_\mathrm{EV} C_\mathrm{AT}|\Omega_\circ}$ where an error $2\zeta$ is tolerated. Indeed, in addition to $\Delta^\zeta_\mathrm{pa}$ which directly results from the privacy amplification, we have a contribution $2\zeta$ due to the fact that $\rho_{ZE\tilde{C}^N C_\mathrm{EC} C_\mathrm{EV} C_\mathrm{AT}|\Omega_\circ}$ and $\bar{\rho}_{ZE\tilde{C}^N C_\mathrm{EC} C_\mathrm{EV} C_\mathrm{AT}|\Omega_\circ} \in \mathcal{B}^\zeta(\rho_{ZE\tilde{C}^N C_\mathrm{EC} C_\mathrm{EV} C_\mathrm{AT}|\Omega_\circ})$ are not perfectly indistinguishable, as discussed in Sec.~\ref{sec:smooth_min_entropy}. If we set $\zeta = 0$, the min-entropy without smoothing is used.

\added{We consider two cases in the following. First, assume that $\zeta < \Tr{(\rho_{ZEC|\Omega_\circ})_{\land \tilde{\Omega} \land \Omega_\mathrm{B}}}$, where the normalized ($\Omega_\circ$) and sub-normalized ($\tilde{\Omega} \land \Omega_\mathrm{B}$) conditioning are used, as introduced in Sec.~\ref{sec:bipartite_qm_systems}.} We now show that applying the quantum leftover hash lemma for states conditioned on $\Omega_\circ$ also provides a bound for states conditioned on $\tilde{\Omega}\land \Omega_\mathrm{B}$, which is required to bound branch (b), i.e. the first term in Eq.~\eqref{eq:reformulation_security_def}. To see this, we can write
\begin{align}
    \Pr[\tilde{\Omega}\land\Omega_\mathrm{B}] d_\mathrm{sec}(SEC)_{\rho|\tilde{\Omega}\land \Omega_\mathrm{B}} &= \Pr[\tilde{\Omega}\land\Omega_\mathrm{B} \land \Omega_\circ] d_\mathrm{sec}(SEC)_{\rho|\tilde{\Omega}\land \Omega_\mathrm{B} \land \Omega_\circ}\label{eq:argument_trace_qlhl}\\
    &= \Pr[\Omega_\circ] d_\mathrm{sec}(SEC)_{(\rho| \Omega_\circ)\land\tilde{\Omega}\land \Omega_\mathrm{B}} \label{eq:alpha_3_directly_bound} \\
    &\leq \Pr[\Omega_\circ] \left(2\zeta + \frac{1}{2}\sqrt{2^{l-H_\mathrm{min}^\zeta \left(Z\,|\,E\tilde{C}^N C_\mathrm{EC} C_\mathrm{EV}\right)_{(\rho| \Omega_\circ)\land\tilde{\Omega}\land \Omega_\mathrm{B}} }}\right)\,,
    \label{eq:subnormalized_state_in_entropy}
\end{align}
where we used $\Pr[\tilde{\Omega}\land\Omega_\mathrm{B} \land \neg\Omega_\circ] = 0$ for the first equality, as, by definition, $\Omega_\circ$ follows from $\tilde{\Omega}\land\Omega_\mathrm{B}$, and we substituted the quantum leftover hash lemma, i.e. Eq.~\eqref{eq:quantum_leftover_hash_smooth}, in the last inequality. We also removed the register $C_\mathrm{AT}$ in the last step as we condition on the protocol not aborting, i.e. $\Omega_\top$, such that $C_\mathrm{AT}$ takes a fixed value. We may now use $H_\mathrm{min}^\zeta(A)_{\rho\land \Omega}\geq H_\mathrm{min}^\zeta(A)_{\rho}$ \cite[Lemma 10]{Tomamichel17} to find an upper bound on the above inequality without the event $\tilde{\Omega}\land \Omega_\mathrm{B}$, \added{yielding Eq.~\eqref{eq:lower_bound_branch_b_circ} in Lemma~\ref{lem:omega_circ_bound}.} 

\added{On the other hand, if $\zeta\geq \Tr{(\rho_{ZEC|\Omega_\circ})_{\land \tilde{\Omega} \land \Omega_\mathrm{B}}}$, then $\zeta$ cannot be used as smoothing parameter, cf. Footnote~\ref{foot:smoothing_trace}. Fortunately, in this case we can bound Eq.~\eqref{eq:alpha_3_directly_bound} without the use of the quantum leftover hash lemma. First, we note that the ideal state can be written as the result of applying a CPTP map $\mathcal{E}$ on the real output such that 
\begin{equation}
    \mathcal{E}\left((\rho_{SEC|\Omega_\circ})_{\land \tilde{\Omega} \land \Omega_\mathrm{B}}\right) =  U_S\otimes (\rho_{EC|\Omega_\circ})_{\land \tilde{\Omega} \land \Omega_\mathrm{B}} \,,
\end{equation}
by replacing the key register with a perfect key. Additionally, the trace distance between two states $\rho$ and $\sigma$ is bounded by $(\Tr{\rho} + \Tr{\sigma}) / 2$ via the triangle inequality \cite[Sec.~9.2]{ChuangNielsen10}. Finally, we can write
\begin{equation}
    \Tr{(\rho_{ZEC|\Omega_\circ})_{\land \tilde{\Omega} \land \Omega_\mathrm{B}}} = \Tr{(\rho_{SEC|\Omega_\circ})_{\land \tilde{\Omega} \land \Omega_\mathrm{B}}} = \Tr{\mathcal{E}\left((\rho_{SEC|\Omega_\circ})_{\land \tilde{\Omega} \land \Omega_\mathrm{B}}\right)} \leq \zeta\,,
\end{equation}
for any map from $Z$ to $S$. Putting the arguments together yields an upper bound on Eq.~\eqref{eq:alpha_3_directly_bound} without using the quantum leftover hash lemma,  
\begin{align}
    d_\mathrm{sec}(SEC)_{(\rho| \Omega_\circ)\land\tilde{\Omega}\land \Omega_\mathrm{B}} = \frac{1}{2}\norm{(\rho_{SEC|\Omega_\circ})_{\land\tilde{\Omega}\land \Omega_\mathrm{B}} - \mathcal{E}\left((\rho_{SEC|\Omega_\circ})_{\land\tilde{\Omega}\land \Omega_\mathrm{B}}\right)}_1 
    \leq \zeta\,.
\end{align}
We emphasize that this bound holds for any map from $Z$ and $S$. In particular, if a privacy amplification step is performed to map $Z$ to a key $S$ of length $l$ using a universal$_2$ family of hash functions, as done in the protocol described in Fig.~\ref{fig:protocol_description}, the bound still holds. We also observe that $\Pr[\Omega_\circ]\zeta$ is a lower bound on Eq.~\eqref{eq:subnormalized_state_in_entropy}, which results from using the quantum leftover hash lemma in the first case considered. Therefore, we may restrict our attention to the first case as the bounds on the security parameters subsequently derived apply to both cases.}

In the following sections, we derive an operational expression for the min-entropy appearing in Eq.~\eqref{eq:lower_bound_branch_b_circ} by using the chain rule for smooth min-entropies and the entropic uncertainty relation.
\added{
\begin{lemma}
\label{lem:omega_circ_bound}
    Let $\rho_{SEC}$ be the state describing Eve and Alice's key $S$ after privacy amplification, i.e. resulting from applying a randomly chosen hash function on her verified key $Z$, for the protocol described in Fig.~\ref{fig:protocol_description}. Let $\tilde\Omega$ denote the event that the protocol does not abort and error correction succeeds, and $\Omega_B$ denote the event that the decoy bounds hold, cf. Table~\ref{tab:events_description}. Finally, let $0 \leq \zeta < 1$. If $\zeta < \Tr{(\rho_{ZEC|\Omega_\circ})_{\land \tilde{\Omega} \land \Omega_\mathrm{B}}}$, then
    \begin{empheq}[box=\fcolorbox{black}{equation_box}]{align}
    \Pr[\tilde{\Omega}\land\Omega_\mathrm{B}] d_\mathrm{sec}(SEC)_{\rho|\tilde{\Omega}\land \Omega_\mathrm{B}} \leq \Pr[\Omega_\circ] \left(2\zeta + \frac{1}{2}\sqrt{2^{l-H_\mathrm{min}^\zeta \left(Z\,|\,E\tilde{C}^N C_\mathrm{EC} C_\mathrm{EV}\right)_{\rho| \Omega_\circ}}}\right),
    \label{eq:lower_bound_branch_b_circ}
    \end{empheq}
    where $\Omega_\circ$ is defined as the event where the true values of the observed statistics fulfill the acceptance condition, cf. Eq.~\eqref{eq:implication_omega_circ}, and the classical registers are listed in Table~\ref{tab:classical_registers_comm_description}. On the other hand, if $\zeta \geq \Tr{(\rho_{ZEC|\Omega_\circ})_{\land \tilde{\Omega} \land \Omega_\mathrm{B}}}$, then 
    \begin{equation}
        \Pr[\tilde{\Omega}\land\Omega_\mathrm{B}] d_\mathrm{sec}(SEC)_{\rho|\tilde{\Omega}\land \Omega_\mathrm{B}} \leq \Pr[\Omega_\circ]\zeta\,.
    \end{equation}
\end{lemma}
\begin{proof}
    The proof is given above.
\end{proof}
}

\begin{remark}
    It is also possible to minimize $\Delta_\mathrm{pa}$ for a given secure-key length $l$ by optimizing $\zeta$, resulting in \cite[Eq.~(S2)]{Tomamichel12}
    \begin{equation}
        \Delta_\mathrm{pa} = \min_\zeta \left\{2\zeta + \Delta^\zeta_\mathrm{pa}\right\}.
        \label{eq:maximizing_delta}
    \end{equation}
    However, in the following, we consider the case where $\Delta_\mathrm{pa}$ has a predefined upper bound and we are interested in determining the extractable secure-key length $l$ as it is usually the case considered.
\end{remark}

\subsection{Expanding the smooth min-entropy}
\label{sec:decomposing_min_entropy}
The quantum leftover hash lemma, cf. Eq.~\eqref{eq:lower_bound_branch_b_circ}, relates the secure-key length and the smooth min-entropy to $\epsilon_\mathrm{sec}'$ through Eq.~\eqref{eq:reformulation_security_def}.
The aim of this section is to expand the smooth min-entropy in terms of operational parameters, namely the acceptance bounds on the number of photon events and errors, cf. Sec.~\ref{sec:long_protocol_description} and Table~\ref{tab:acceptance_testing}.

In fact, in the decoy-state protocol, a weak coherent photon source, which either emits multiple photons, one photon or no photon, following the Poisson distribution, is usually used (as will be discussed in Sec.~\ref{sec:photon_event_statistics_and_bounds}).
The number of $m$-photon events detected by Bob in the $\mathsf{Z}$-basis is represented by $s_{\mathsf{Z},m}$ and the number of errors, due to experimental limitations or the presence of an eavesdropper, represented by $c_{\mathsf{Z},m}$. \added{These quantities are formally defined in Sec.~\ref{sec:photon_event_statistics_and_bounds}.}  Note that Bob obviously cannot determine these quantities experimentally as, following a detector click, the photon number remains unknown.
However, using the decoy-state method, i.e. suitable concentration inequalities, we can estimate the amount of occurrences and derive bounds on the number of photon events and errors, namely $s_{\mathsf{Z},0}^-, s_{\mathsf{Z},1}^-$, $s_{\mathsf{X},1}^-$ and $\Lambda_\mathsf{X}^+$. This will be the aim of Sec.~\ref{sec:decoy_state_bounds}. In this section, we assume that $\Omega_\circ$ is true, which, by definition, implies that $s_{\mathsf{Z},0} \geq s_{\mathsf{Z},0}^\mathrm{l}$, $s_{\mathsf{Z},1} \geq s_{\mathsf{Z},1}^\mathrm{l}$, $s_{\mathsf{X},1} \geq s_{\mathsf{X},1}^\mathrm{l}$ and $\Lambda_\mathsf{X} \leq \Lambda^\mathrm{u}_\mathsf{X}$ hold. Throughout this work, we use superscripts $+$ and $-$ to denote decoy bounds and superscripts $\mathrm{u}$ and $\mathrm{l}$ to denote the acceptance bounds defining the conditions in the acceptance set $Q$, \added{cf. Table~\ref{tab:acceptance_testing} and Remark~\ref{rem:acceptance_testing_summary}}.

\subsubsection{Chain rule for smooth min-entropies}
\label{sec:chain_rule_smooth_min_entropies}

First, we recall that $C_\mathrm{EV}$ stores the hash of Alice's corrected key as well as her choice of hash function for error verification, which we denote $C_\mathrm{EV}'$ and $C_\mathrm{EV}''$ respectively, such that $C_\mathrm{EV} = C_\mathrm{EV}' C_\mathrm{EV}''$. We can now use the fact that the choice of hash function is not correlated to the corrected key $Z$, as it is chosen randomly and independently of $Z$, in order to remove the register $C_\mathrm{EV}''$ from the min-entropy appearing in  Eq.~\eqref{eq:lower_bound_branch_b_circ}, yielding \cite[Theorem 6.2]{Tomamichel16}
\begin{equation}
    H_\mathrm{min}^\zeta (Z|E\tilde{C}^N C_\mathrm{EC} C_\mathrm{EV}' C_\mathrm{EV}'')_{\rho|\Omega_\circ} = H_\mathrm{min}^\zeta (Z|E\tilde{C}^N C_\mathrm{EC} C_\mathrm{EV}')_{\rho|\Omega_\circ}\, .
    \label{eq:remove_EC_tildetilde}
\end{equation}
Then, we can rewrite the min-entropy from the equation above in terms of the bits leaked during error correction and error verification, stored in $C_\mathrm{EC}$ and $C_\mathrm{EV}'$ respectively, by using the chain rule for smooth min-entropies (see Refs.~\cite[Theorem~3.2.12]{Renner05} and \cite{Tomamichel12})
\begin{equation}
    H_\mathrm{min}^\zeta (Z|E\tilde{C}^N C_\mathrm{EC} C_\mathrm{EV}')_{\rho|\Omega_\circ} \geq H_\mathrm{min}^\zeta (Z|E\tilde{C}^N)_{\rho|\Omega_\circ} - \log_2 |C_\mathrm{EC} C_\mathrm{EV}'|\,,
    \label{eq:decompose_classical}
\end{equation}
where $\log_2|C_\mathrm{EC} C_\mathrm{EV}'| = \mathrm{leak}_\mathrm{EC} + \log_2{\frac{2}{\epsilon_\mathrm{cor}}}$ describes the classical communications and corresponds to the number of bits disclosed during error correction and error verification, cf. Sec.~\ref{sec:error_verification}. 
\begin{remark}
    Intuitively, Eve's uncertainty about the sifted key $Z$ decreases during the protocol as she intercepts the bits disclosed through the classical channel during the error correction and error verification steps. Inequality~\eqref{eq:decompose_classical} states that her uncertainty cannot decrease by more than the number of bits disclosed during these steps, i.e. $\log_2|C_\mathrm{EC} C_\mathrm{EV}'|$.
\end{remark}
\noindent Now, $Z=Z_\mathrm{v}Z_\mathrm{s}Z_\mathrm{m}$ can be split into three subsystems, i.e. the vacuum, single-photon and multi-photon events $Z_\mathrm{v}$, $Z_\mathrm{s}$ and $Z_\mathrm{m}$ respectively. \added{Here, an $m$-photon event is defined as Alice sending $m$ photons and Bob registering at least one click. Therefore, for example, a vacuum event may occur if Alice prepared a vacuum state but Bob nevertheless registered a click due to dark counts.} Note that the number of rounds in $Z_\mathrm{v}$ and $Z_\mathrm{s}$ is given by $s_{\mathsf{Z},0} = |Z_\mathrm{v}|$ and $s_{\mathsf{Z},1} = |Z_\mathrm{s}|$ respectively, and is a random variable. However, since we condition on the event $\Omega_\circ$, we are guaranteed that $s_{\mathsf{Z},0} \geq s_{\mathsf{Z},0}^\mathrm{l}$ and $s_{\mathsf{Z},1} \geq s_{\mathsf{Z},1}^\mathrm{l}$ where $s_{\mathsf{Z},0}^\mathrm{l}$ and $s_{\mathsf{Z},1}^\mathrm{l}$ have fixed values that determine the accept conditions. We use this below.

Multi-photon events are entirely unsafe due to being prone to photon number splitting attacks, i.e. Eve has no uncertainty about multi-photon events. As such, we can remove the classical register $Z_\mathrm{m}$ using \cite[Lemma 6.17]{Tomamichel16}, yielding
\begin{equation}
    H_\mathrm{min}^\zeta (Z_\mathrm{v}Z_\mathrm{s}Z_\mathrm{m}|E\tilde{C}^N)_{\rho|\Omega_\circ} \geq H_\mathrm{min}^\zeta (Z_\mathrm{v}Z_\mathrm{s}|E\tilde{C}^N)_{\rho|\Omega_\circ} \geq H_\mathrm{min}^\zeta (Z^{s_{\mathsf{Z},0}^\mathrm{l}}Z^{s_{\mathsf{Z},1}^\mathrm{l}}|E\tilde{C}^N)_{\rho|\Omega_\circ} \,, 
    \label{eq:removed_multi_photon_events}
\end{equation}
where, in the second inequality, we apply the Lemma again on the registers $Z_\mathrm{v}$ and $Z_\mathrm{s}$, which we shrink according to the acceptance bounds on the number of vacuum and single-photon numbers. This step is necessary to remove any random variable and have registers of well-defined size $|Z^{s_{\mathsf{Z},0}^\mathrm{l}}| = s_{\mathsf{Z},0}^\mathrm{l}$ and $|Z^{s_{\mathsf{Z},1}^\mathrm{l}}| = s_{\mathsf{Z},1}^\mathrm{l}$, i.e. given by the acceptance set, as we consider a fixed-length protocol, cf. Sec.~\ref{sec:fixed_length_protocols}. To simplify the notation, we define $Z_{\mathrm{v}}^\mathrm{l} \coloneqq Z^{s_{\mathsf{Z},0}^\mathrm{l}}$ and $Z_{s}^\mathrm{l} \coloneqq Z^{\mathrm{s}_{\mathsf{Z},1}^\mathrm{l}}$. Additionally, the chain rule for smooth min-entropies can be used to decompose the min-entropy\footnote{A tighter expression for the chain rule for smooth min-entropies can be found in Ref.~\cite[Theorem~13]{Vitanov13}.}
\begin{equation}
    H_\mathrm{min}^{2\alpha_1 + \alpha_2 + \alpha_3}(A'B'|C')_\rho \geq H_\mathrm{min}^{\alpha_1}(A'|B'C')_\rho + H_\mathrm{min}^{\alpha_3}(B'|C')_\rho - \log_2\frac{2}{\alpha_2^2}\, ,
\end{equation}
where $A', B', C'$ are generic classical and quantum systems, $\alpha_1, \alpha_3 \geq0$ and $\alpha_2 > 0$, \added{cf. Ref.~\cite[Theorem~13]{Vitanov13}, and we used the fact that $\log_2\frac{2}{\alpha^2} \geq \log_2\frac{1}{1-\sqrt{1-\alpha^2}}$ for all $\alpha\in[0, 1]$.}
Applying the chain rule to Eq.~\eqref{eq:removed_multi_photon_events} with conveniently defined $\zeta = 2\alpha_1 + \alpha_2 + \alpha_3$ yields\footnote{It is also possible to completely remove the vacuum contributions, analogously to the multi-photon contributions, which may slightly increase the secure-key length if their contribution is small (or zero) as the term $\alpha_2$ then vanishes.}
\begin{equation}
    H_\mathrm{min}^\zeta (Z_{\mathrm{v}}^\mathrm{l}Z_{\mathrm{s}}^\mathrm{l}|E\tilde{C}^N)_{\rho|\Omega_\circ} \geq H_\mathrm{min}^{\alpha_1} (Z_{\mathrm{v}}^\mathrm{l}|Z_{\mathrm{s}}^\mathrm{l}E\tilde{C}^N)_{\rho|\Omega_\circ} + H_\mathrm{min}^{\alpha_3} (Z_{\mathrm{s}}^\mathrm{l}|E\tilde{C}^N)_{\rho|\Omega_\circ}-\log_2{\frac{2}{\alpha_2^2}}\, .
\end{equation}
Written in this form, we can determine an operational bound for the expression above in terms of the acceptance bounds. Due to the fact that vacuum events contain no information about $Z$ and the min-entropy is given according to Alice's key as Bob corrects his key to match Alice, we can substitute $H_\mathrm{min}^{\alpha_1} (Z_{\mathrm{v}}^\mathrm{l}|Z_{\mathrm{s}}^\mathrm{l}E\tilde{C}^N)_{\rho|\Omega_\circ} = \log_2{2^{s_{\mathsf{\mathsf{Z}},0}^\mathrm{l}}} = s_{\mathsf{\mathsf{Z}},0}^\mathrm{l}$, where we use the fact that in the case of absolute uncertainty, as is the case for vacuum events, the min-entropy is maximal (as discussed in Sec.~\ref{sec:guessing_prob_min_entropy}). \added{The smoothing parameter $\alpha_1$ is a free parameter, and as the bound on the smooth min-entropy for vacuum events is independent of $\alpha_1$, we set $\alpha_1 = 0$.} Thus, interestingly, vacuum events that passed the sifting step contribute to the extractable secure key. Putting everything together yields \added{Lemma~\ref{lem:entropy_classical_registers}.}

\added{
\begin{lemma}
    Consider the protocol described in Fig.~\ref{fig:protocol_description} and the following classical registers: the basis choice announcements $\tilde{C}^N$, error correction communications $C_\mathrm{EC}$, hash of Alice's corrected key $C_\mathrm{EV}'$, and choice of hash function for error verification $C_\mathrm{EV}''$, cf. Table~\ref{tab:classical_registers_comm_description}. Then, we can bound the min-entropy of the state $\rho_{SE\tilde{C}^N C_\mathrm{EC} C_\mathrm{EV}' C_\mathrm{EV}''}$ describing Eve and Alice's key $S$ after privacy amplification,
    \begin{align}
    H_\mathrm{min}^\zeta (Z|E\tilde{C}^N C_\mathrm{EC} C_\mathrm{EV}' C_\mathrm{EV}'')_{\rho|\Omega_\circ} &\geq s_{\mathsf{\mathsf{Z}},0}^\mathrm{l} +  H_\mathrm{min}^{\alpha_3} (Z_{\mathrm{s}}^\mathrm{l}|E\tilde{C}^N)_{\rho|\Omega_\circ} - \log_2{\frac{2}{\alpha_2^2}} - \mathrm{leak}_\mathrm{EC} - \log_2{\frac{2}{\epsilon_\mathrm{cor}}}\, ,
    \label{eq:entropy_bound_partial}
    \end{align} 
    where $\mathrm{leak}_\mathrm{EC}$ is the number of bits leaked during error correction, $s_{\mathsf{\mathsf{Z}},0}^\mathrm{l}$ is the acceptance bound on the number of vacuum events and $\Omega_\circ$ is defined as the event where the true values of the observed statistics fulfill the acceptance condition, cf. Eq.~\eqref{eq:implication_omega_circ}.
    \label{lem:entropy_classical_registers}
\end{lemma}
\begin{proof}
    The proof is given above.
\end{proof}
}

\noindent We note that the bound we derived is slightly tighter than Refs.~\cite{LimCurtyWalenta14,Rusca18} as we remove the multi-photon events using Ref.~\cite[Lemma 6.17]{Tomamichel16} instead of the chain rule, resulting in fewer smoothing parameters. 
In the following sections, we address the single-photon min-entropy in Eq.~\eqref{eq:entropy_bound_partial} using the entropic uncertainty relation. This term requires more technical arguments than the ones previously discussed. 

\subsubsection{Source-replacement scheme}
\label{sec:source_replacement_scheme}
In this section, we discuss a technical argument required to apply the entropic uncertainty relation used to bound the single-photon min-entropy from the previous section, cf. Eq.~\eqref{eq:entropy_bound_partial}. In fact, the single-photon rounds of the \textit{prepare-and-measure scheme} discussed until now, where Alice locally prepares the signal states and sends them to Bob over the quantum channel, can be replaced by an equivalent entanglement-based scenario, following the \textit{source-replacement scheme}\cite{Curty04, Ferenczi12, Tupkary24}. Essentially, for the rounds where Alice sends single photons, we consider a virtual setup where Alice prepares entangled states
\begin{equation}
    \ket{\Psi_+}=\frac{1}{\sqrt{2}}\left(\ket{00}_{AA'} + \ket{11}_{AA'}\right)\,,
\end{equation}
keeps system $A$ and sends system $A'$ to Bob. Eve may then intercept system $A'$ and send a system $B$ to Bob. Alice and Bob then share a potentially entangled system $\rho_{AB}$, where Eve holds a purification of this system. Alice and Bob can estimate how much information Eve holds about $A$ by performing a \textit{von Neumann measurement}, i.e. locally measuring their systems in incompatible bases, the $\mathsf{X}$-basis and $\mathsf{Z}$-basis. This will be formalized in Sec.~\ref{sec:entropic_uncertainty_relation} when introducing the entropic uncertainty relation, which requires the source replacement. In the following, we may delay Alice's measurements until after Eve has performed her attack. 

In the scenario where Alice prepares perfect signal states, her measurement POVMs are given by  $\mathbb{X} = \{M_x\}_x$ and $\mathbb{Z} = \{N_z\}_z$ for the $\mathsf{X}$-basis and $\mathsf{Z}$-basis, respectively. Note that one can only view Alice's measurements as having active basis choice  if signal preparation is perfect, which is important when using the EUR statement (see Ref.~\cite[Remark 1]{Tupkary24}), as it assumes active basis choice on Alice's measurements. For the BB84 protocol with perfectly diagonal bases, considering the single-photon rounds, we have $\mathbb{X} = \{\ketbra{+}{+}, \ketbra{-}{-}\}$ and $\mathbb{Z} = \{\ketbra{0}{0}, \ketbra{1}{1}\}$ where $\ket{\pm} = (\ket{0} \pm \ket{1}) / \sqrt{2}$. In the next section, we derive the bounds for the virtual entanglement-based scheme described in this section but note that the bounds equivalently hold for the prepare-and-measure scheme considered throughout this work and described in Sec.~\ref{sec:long_protocol_description}.

\subsubsection{Entropic uncertainty relation}
\label{sec:entropic_uncertainty_relation}
A lower bound on the single-photon min-entropy $H_\mathrm{min}^{\alpha_3} (Z_{\mathrm{s}}^\mathrm{l}|E\tilde{C}^N)_{\rho|\Omega_\circ}$ can be found by using the fact that the secrecy of the $\mathsf{Z}$-basis single-photon events is given by how well Bob is able to estimate Alice's key if he would have measured in the $\mathsf{X}$-basis instead\footnote{This is rather intuitive in the sense that if Eve gathers information about the $\mathsf{Z}$-basis (e.g. through an intercept-and-resend attack), she necessarily induce some error in Bob's detections in the $\mathsf{X}$-basis.}. This is expressed by the \textit{entropic uncertainty relation} (EUR) for smooth entropies \cite{TomamichelRenner11}. The EUR statement considers an arbitrary state  $\rho_{ABE'}$, and relates the min-entropy of the state obtained via $\mathsf{Z}$-basis measurements ($\rho_{ZBE'}$) to the max-entropy of the state obtained by $\mathsf{X}$-basis measurements ($\rho_{XBE}$), on Alice's system. This is given by
\begin{equation}
    H_\mathrm{min}^{\alpha}(Z|E')_{\rho_{ZBE'}} +  H_\mathrm{max}^{\alpha}(X|B)_{\rho_{XBE'}^\mathrm{virt}}
    \geq
    q \, ,
    \label{eq:entorpic_uncertainty_relation}
\end{equation}
where $q = -\log_2 c$ and $c = \max_{x,z}\norm{\sqrt{M_x}\sqrt{N_z}}_\infty^2$ is a measure for the incompatibility of the bases. Additionally, $E'$ denotes a generic quantum state of an adversary who also has access to classical communications $C'$. Here, $\rho_{ZBE'}$ describes the post-measurement state for the actual protocol, given that Alice measures in the $\mathsf{Z}$-basis, and $\rho_{XBE'}^\mathrm{virt}$ describes the virtual protocol where Alice measures in the $\mathsf{X}$-basis. Considering perfectly diagonal bases, as discussed in the previous section, we have $q = -\log_2 1/2 = 1$\cite{Coles16, Tomamichel12}\footnote{In practical implementations, the bases might not be perfectly diagonal, leading to a quality factor $q<1$.}. See also Refs.~\cite{Konig09, Ekert14} for additional discussions. In the following, we assume that $q=1$\footnote{This is required to apply the random sampling argument discussed below.}.

One advantage of the EUR framework is that the smooth entropies appearing in the entropic uncertainty relation quantify measures beyond the i.i.d. scenario\cite{TomamichelRenner11}. This approach thus inherently proves security against coherent attacks, the most general form of attack in which Eve can apply any strategy permitted by quantum mechanics on the channel. Thus, it yields tighter finite-size key rates than proof techniques that involve pessimistic lifts from collective to coherent attacks, such as the post-selection technique \cite{Nahar24,Christandl09}.

We apply the EUR statement on the set of rounds where Alice chose the $\mathsf{Z}$-basis and sent a single photon, and Bob measured in the $\mathsf{Z}$-basis and got a detection. We know that there are exactly 
$s_{\mathrm{Z},1}$ of such rounds, although the exact value of $s_{\mathrm{Z},1}$ is not known.  Since we condition on the event $\Omega_\circ$, we have $s_{\mathrm{Z},1} \geq s^\mathrm{l}_{\mathrm{Z},1}$, and we only apply the EUR statement on $s^\mathrm{l}_{\mathrm{Z},1}$ rounds. We recall that the key registers were shrunk to match the acceptance bounds in Sec.~\ref{sec:chain_rule_smooth_min_entropies}. 
Moreover, the state on which the EUR statement is applied can be obtained by reformulating Bob's measurements to consist of several steps, where he first measures to see whether he obtains a detection or not, determines his basis, and only later completes the measurement procedure (see Ref.~\cite[Section III]{Tupkary24}). \added{We consider the $s^\mathrm{l}_{\mathrm{Z},1}$ rounds where Alice measured in the $\mathsf{Z}$-basis and denote Alice's POVM element for the $i$-th round as $N_{z_i} \in\mathbb{Z}$ with $z_i\in\{0, 1\}$, cf. Sec.~\ref{sec:source_replacement_scheme}. Then, the POVM describing the full $s^\mathrm{l}_{\mathrm{Z},1}$-round measurement is given by $\mathbb{N}^{\mathrm{full}}=\{N_\textbf{z}^{\mathrm{full}}\}_\textbf{z}$, where $\textbf{z}=(z_1, ..., z_{s^\mathrm{l}_{\mathrm{Z},1}})$ and $N_\textbf{z}^{\mathrm{full}} =\bigotimes_i N_{z_i}$. Analogously, we define the $s^\mathrm{l}_{\mathrm{Z},1}$-round POVM for the case where Alice measures in the $\mathsf{X}$-basis as $\mathbb{M}^{\mathrm{full}} =\{M_\textbf{x}^{\mathrm{full}}\}_\textbf{x}$. Then we can apply the EUR statement, i.e. Eq.~\eqref{eq:entorpic_uncertainty_relation}, for the multi-round POVMs $\mathbb{N}^{\mathrm{full}}$ and $\mathbb{M}^{\mathrm{full}}$ and obtain
\begin{equation}
    H_\mathrm{min}^{\alpha_3} (Z_{\mathrm{s}}^\mathrm{l}|E)_{\rho_{Z_{\mathrm{s}}^\mathrm{l} B E\tilde{C}^N|\Omega_\circ}} +  H_\mathrm{max}^{\alpha_3}(X_{\mathrm{s}}^\mathrm{l}|B)_{\rho_{X_{\mathrm{s}}^\mathrm{l} B E\tilde{C}^N|\Omega_\circ}^\mathrm{virt}}
    \geq s_{\mathsf{Z},1}^\mathrm{l} \,,
\end{equation}
where we defined $X_{\mathrm{s}}^\mathrm{l} \coloneqq X^{s_{\mathsf{Z}, 1}^\mathrm{l}}$ and used the fact that
\begin{equation}
    -\log_2 \max_{\textbf{x},\textbf{z}}\norm{\sqrt{M_\textbf{x}^{\mathrm{full}}}\sqrt{N_\textbf{z}^{\mathrm{full}}}}_\infty^2 = -\log_2 \left(\Pi_{i=1}^{s_{\mathsf{Z},1}^\mathrm{l}} \max_{x,z}\norm{\sqrt{M_x}\sqrt{N_z}}_\infty^2\right) = s_{\mathsf{Z},1}^\mathrm{l}\,,
\end{equation}
with $\max_{x,z}\norm{\sqrt{M_x}\sqrt{N_z}}_\infty^2 =1/2$.} Now, we can use data processing inequalities \cite[Theorem 6.19]{Tomamichel16} to make Bob measure his register $B$ in the $\mathsf{X}$-basis and obtain the classical register $\tilde{X}^\mathrm{l}_\mathrm{s}$. This yields
\begin{equation}
    H_\mathrm{min}^{\alpha_3} (Z_{\mathrm{s}}^\mathrm{l}|E)_{\rho_{Z_{\mathrm{s}}^\mathrm{l} B E\tilde{C}^N|\Omega_\circ}} +  H_\mathrm{max}^{\alpha_3}(X_{\mathrm{s}}^\mathrm{l}| \tilde{X}^\mathrm{l}_\mathrm{s})_{\rho_{X_{\mathrm{s}}^\mathrm{l} \tilde{X}^\mathrm{l}_\mathrm{s} |\Omega_\circ}^\mathrm{virt}}
    \geq s_{\mathsf{Z},1}^\mathrm{l} \,.
      \label{eq:EUR_multiple_rounds}
\end{equation}
The max-entropy term can now be related to the error rate in the virtual $\mathsf{X}$-basis measurements, which we formalize below. This is known as the \textit{phase error rate}, and we denote it $\Lambda_\mathsf{Z}$. In this way, Eve's uncertainty about the $\mathsf{Z}$-basis detections (quantified by the min-entropy) is related to the correlation between Alice's and Bob's $\mathsf{X}$-basis detections, if they had measured in the $\mathsf{X}$-basis instead (quantified by the max-entropy). Obviously, when measuring in the $\mathsf{Z}$-basis, the correlation in the $\mathsf{X}$-basis is not directly accessible in the protocol. However, if we assume that Eve's attack strategy does not depend on Alice's and Bob's basis choice, we can use the statistics from the events where Alice and Bob choose the $\mathsf{X}$-basis in order to estimate the correlation in the case where they measured in the $\mathsf{Z}$-basis. This is effectively a sampling without replacement problem where one considers a virtual scenario in which Alice and Bob only measure in the $\mathsf{X}$-basis, cf. Refs. \cite[Sec.~2.2]{Lim14} and \cite[Section 3]{Tupkary24} for a more thorough discussion. Note that this assumption can be rigorously argued in the scenario where Alice prepares perfect signal states, and Bob's probability of detection is independent of his basis choice (i.e absence of basis-efficiency mismatch).   

We note that the EUR is applied to quantum states and is not applicable to a state that was already measured. Intuitively, the virtual scenario described above cannot be assumed if the measurements have already been performed in any basis. Hence, we cannot condition Eq.~\eqref{eq:EUR_multiple_rounds} on $\Omega_\mathrm{EC}$, $\Omega_\mathrm{EV}$, $\Omega_\mathrm{AT}$ nor $\Omega_\mathrm{B}$ as these events require the state $\rho_{AB}$ to already be measured. This motivates why we introduced the event $\Omega_\circ$. In Sec.~\ref{sec:quantum_leftover_for_smooth_min_entropies}, we have shown that the bounds derived for Eq.~\eqref{eq:quantum_leftover_hash_smooth} conditioned on $\Omega_\circ$ also hold when conditioning on the other events listed above, resulting in Eq.~\eqref{eq:lower_bound_branch_b_circ}. We defined the \textit{single-photon quantum bit error rate} (QBER) in the $\mathsf{X}$-basis as
\begin{equation}
    \Lambda_\mathsf{X} \coloneqq \frac{c_{\mathsf{X},1}}{s_{\mathsf{X},1}} \,,
\end{equation}
where $s_{\mathsf{X},1}$ is the number of single-photon events and $c_{\mathsf{X},1}$ the corresponding number of single-photon errors. We note again that these variables are not directly accessible to Alice and Bob, and it will be the aim of Sec.~\ref{sec:decoy_state_bounds} to bound them and verify that the bounds satisfy the acceptance conditions, cf. Table~\ref{tab:acceptance_testing}. We can now use Serfling's inequality to bound the virtual single-photon error rate $\Lambda_\mathsf{Z}$, or phase error rate, using the statistics from the $\mathsf{X}$-basis. In fact, following Refs.~\cite[Corollary 1.1]{Serfling74} and \cite[Supplementary Notes 2]{Tomamichel12},
\begin{equation}
    \Pr\left[\Lambda_\mathsf{Z} \geq \Lambda_\mathsf{X} + \gamma(\nu, s_{\mathsf{Z},1}, s_{\mathsf{X},1})\right] \leq \nu^2 \, ,
    \label{eq:general_serfling}
\end{equation}
where 
\begin{equation}
    \gamma(a, b, c) = \sqrt{\frac{b + c}{bc}\frac{c + 1}{c}\ln\frac{1}{a}} \, ,
    \label{eq:def_gamma}
\end{equation}
i.e. the probability that the phase error rate deviates by more than $\gamma(\nu, s_{\mathsf{Z},1}, s_{\mathsf{X},1})$ from the single-photon QBER is bounded by $\nu^2$. The above equation implies
\begin{equation}
         \Pr[\Lambda_\mathsf{Z} \geq \Lambda_\mathsf{X} + \gamma(\nu, s_{\mathsf{Z},1}, s_{\mathsf{X},1})  \land \Omega_\circ]  \leq \nu^2\,.
         \label{eq:probability_upper_bound_phase_error}
\end{equation}
Using the acceptance bounds from $Q$ and the properties of $\gamma(a,b,c)$, which is decreasing in $b$ and $c$, we obtain
\begin{equation}
         \Pr[\Lambda_\mathsf{Z} \geq \Lambda_\mathsf{X}^\mathrm{u} + \gamma(\nu, s^\mathrm{l}_{\mathsf{Z},1}, s^\mathrm{l}_{\mathsf{X},1})\,|\,\Omega_\circ]  \leq \frac{\nu^2}{\Pr[\Omega_\circ]}\,,
         \label{eq:serfling_with_bounds}
\end{equation}
where we used that conditioning on $\Omega_\circ$ implies $s_{\mathsf{Z},1} \geq s^\mathrm{l}_{\mathsf{Z},1}$, $s_{\mathsf{X},1} \geq s^\mathrm{l}_{\mathsf{X},1}$ and $\Lambda_\mathsf{X} \leq \Lambda_\mathsf{X}^\mathrm{u}$.
To simplify the notation, we define $\gamma \coloneqq \gamma(\nu, s^\mathrm{l}_{\mathsf{Z},1}, s^\mathrm{l}_{\mathsf{X},1})$ in the following. Using the results from Ref.~\cite[Lemma 3]{Tomamichel12}, the max-entropy appearing in Eq.~\eqref{eq:EUR_multiple_rounds} can be rewritten in terms of the binary entropy function $h$\footnote{More precisely, $h$ denotes the truncated binary entropy function defined as $h: x \mapsto -x \log_2 x - (1-x) \log_2 (1-x)$ if $x \leq 0.5$ and $h(x)=1$ if $x > 0.5$ \cite{Tomamichel12}.\label{foot:binary_entropy}} and the bound on the phase error rate, namely
\begin{equation}
    H_\mathrm{min}^{\alpha_3} (Z_{\mathrm{s}}^\mathrm{l}|E\tilde{C}^N)_{\rho_{Z_{\mathrm{s}}^\mathrm{l} B E\tilde{C}^N|\Omega_\circ}}
    \geq
    s_{\mathsf{Z}, 1}^\mathrm{l} - H_\mathrm{max}^{\alpha_3}(X_{\mathrm{s}}^\mathrm{l}|\tilde{X}^\mathrm{l}_\mathrm{s})_{\rho_{X_{\mathrm{s}}^\mathrm{l} \tilde{X}^\mathrm{l}_\mathrm{s} E|\Omega_\circ}^\mathrm{virt}}
    \geq
    s_{\mathsf{Z}, 1}^\mathrm{l} - s_{\mathsf{Z}, 1}^\mathrm{l}h(\Lambda_\mathsf{X}^\mathrm{u}+\gamma) \, ,
    \label{eq:single_photon_entropy}
\end{equation}
where
\begin{equation}
    \alpha_3 \coloneqq \frac{\nu}{\sqrt{\Pr[\Omega_\circ]}}\,.
    \label{eq:def_alpha_3}
\end{equation}
Intuitively, if $\Lambda_\mathsf{X}^\mathrm{u}+\gamma$ is small, we can deduce that the correlation between Alice's and Bob's single-photon bit values is high and thus the max-entropy is small. We must distinguish two cases in the following. If $\alpha_3 < 1$, we continue the analysis. However, if $\alpha_3 \geq 1$, we cannot smooth the entropies using $\alpha_3$, cf. Sec.~\ref{sec:smooth_min_entropy}. We discuss the latter case below Eq.~\eqref{eq:branch_c_bound_distance_real_ideal} and assume $\alpha_3 < 1$ for now. Using Lemma~\ref{lem:entropy_classical_registers} and substituting in Eq.~\eqref{eq:single_photon_entropy} yields an operational bound on the smooth min-entropy,
\begin{empheq}[box=\fcolorbox{black}{equation_box}]{align}
    H_\mathrm{min}^{\zeta}(Z|E\tilde{C}^N C_\mathrm{EC} C_\mathrm{EV})_{\rho|\Omega_\circ} \geq s_{\mathsf{Z},0}^\mathrm{l} + s_{\mathsf{Z}, 1}^\mathrm{l}\bigl(1 - h(\Lambda_\mathsf{X}^\mathrm{u}+\gamma)\bigr) - \log_2{\frac{2}{\alpha_2^2}} - \mathrm{leak}_\mathrm{EC} - \log_2{\frac{2}{\epsilon_\mathrm{cor}}} \,,
    \label{eq:H_min_operational_hypothetical}
\end{empheq}
as appearing in Eq.~\eqref{eq:lower_bound_branch_b_circ}. We derived the bound by applying the chain rule for smooth min-entropies to separate the different $m$-photon event contributions in $Z$ and used the EUR to bound the single-photon contributions with the phase error rate. As discussed above, the conditioning on $\Omega_\circ$ is required to apply the EUR. 
We have thus derived an upper bound on branch (b), i.e. the first term in Eq.~\eqref{eq:reformulation_security_def}, following Eq.~\eqref{eq:lower_bound_branch_b_circ}, in terms of the min-entropy, for which we have derived an operational expression above, cf. Eq.~\eqref{eq:H_min_operational_hypothetical}. We can now rewrite Eq.~\eqref{eq:reformulation_security_def} by plugging in Eq.~\eqref{eq:lower_bound_branch_b_circ}, yielding
\begin{align}
    \Pr[\tilde{\Omega}] d_\mathrm{sec}(SEC)_{\rho|\tilde{\Omega}} &\leq\, \Pr[\tilde{\Omega} \land \Omega_\mathrm{B}] d_\mathrm{sec}(SEC)_{\rho|\tilde{\Omega}\land \Omega_\mathrm{B}} + \Pr[\tilde{\Omega} \land \neg\Omega_\mathrm{B}]\, , \\
    &\leq\,2(\alpha_2 + \nu) + \Delta_\mathrm{pa}^\zeta + \Pr[\tilde{\Omega} \land \neg\Omega_\mathrm{B}] \, ,
    \label{eq:branch_c_bound_distance_real_ideal}
\end{align}
where we use $\Pr[\Omega_\circ]\leq 1$, and $\alpha_3 \Pr[\Omega_\circ] \leq \nu$, cf. Eq.~\eqref{eq:def_alpha_3}. We recall that $\zeta = \alpha_2 + \alpha_3$ and we set $\alpha_1 = 0$ as discussed above. We note that if $\alpha_3 \geq 1$, cf. Eq.~\eqref{eq:def_alpha_3}, then we directly have a bound on Eq.~\eqref{eq:alpha_3_directly_bound} without requiring smooth entropies as this implies $\Pr[\Omega_\circ] \leq \nu^2 \leq \nu$ and using the fact that the trace distance never exceeds one, we have $\Pr[\tilde{\Omega} \land \Omega_\mathrm{B}] d_\mathrm{sec}(SEC)_{\rho|\tilde{\Omega}\land \Omega_\mathrm{B}} \leq \nu$, such that all further analyses on Eq.~\eqref{eq:branch_c_bound_distance_real_ideal} also hold if $\alpha_3 \geq 1$. Finally, we recall that the secure-key length $l$ appears in $\Delta_\mathrm{pa}^\zeta$. We choose an appropriate $l$ to fulfill the security criterion in Sec.~\ref{sec:operational_expression_min_entropy}.

\added{
\begin{lemma}
    Consider the protocol described in Fig.~\ref{fig:protocol_description}, the events listed in Table~\ref{tab:events_description} and the classical announcements listed in Table~\ref{tab:classical_registers_comm_description}. Let $\rho_{SEC}$ be the state describing Eve and Alice's key $S$ after privacy amplification. Then the following bound holds
    \begin{equation}
        \Pr[\tilde{\Omega}] d_\mathrm{sec}(SEC)_{\rho|\tilde{\Omega}} \leq 2(\alpha_2 + \nu) + \Delta_\mathrm{pa}^\zeta + \Pr[\tilde{\Omega} \land \neg\Omega_\mathrm{B}]\,,
    \end{equation}
    where $\alpha_2, \nu >0$,
    \begin{equation}
        \Delta^\zeta_\mathrm{pa} = \frac{1}{2}\sqrt{2^{l-H_\mathrm{min}^\zeta \left(Z\,|\,E\tilde{C}^N C_\mathrm{EC} C_\mathrm{EV} C_\mathrm{AT}\right)_{\rho|\Omega_\circ}}}\,,
    \end{equation}
    and the smooth min-entropy is lower bounded by
    \begin{equation}
        H_\mathrm{min}^{\zeta}(Z|E\tilde{C}^N C_\mathrm{EC} C_\mathrm{EV}C_\mathrm{AT})_{\rho|\Omega_\circ} \geq s_{\mathsf{Z},0}^\mathrm{l} + s_{\mathsf{Z}, 1}^\mathrm{l}\bigl(1 - h(\Lambda_\mathsf{X}^\mathrm{u}+\gamma)\bigr) - \log_2{\frac{2}{\alpha_2^2}} - \mathrm{leak}_\mathrm{EC} - \log_2{\frac{2}{\epsilon_\mathrm{cor}}}\,,
    \end{equation}
    where $s_{\mathsf{Z},0}^\mathrm{l}$, $s_{\mathsf{Z}, 1}^\mathrm{l}$ and $\Lambda_\mathsf{X}^\mathrm{u}$ are acceptance parameters fixed before the protocol run, cf. Fig.~\ref{fig:protocol_description}, and $\gamma = \gamma(\nu, s^\mathrm{l}_{\mathsf{Z},1}, s^\mathrm{l}_{\mathsf{X},1})$ is given by Eq.~\eqref{eq:def_gamma}.
\end{lemma}
\begin{proof}
    The proof is given above.
\end{proof}
}

\noindent Until now, we have assumed that bounds on the number of vacuum events $s_{\mathsf{Z},0}^-$, single photon events $s_{\mathsf{Z}, 1}^-$, $s_{\mathsf{X}, 1}^-$ and the single-photon QBER $\Lambda_\mathsf{X}^+$ exist and can be used by Alice and Bob to perform the acceptance test described in Table~\ref{tab:acceptance_testing} and Sec.~\ref{sec:long_protocol_description}. The aim of the following section is precisely to derive these bounds in terms of experimentally available parameters. This will also provide us with a bound on $\Pr[\tilde{\Omega} \land \neg\Omega_\mathrm{B}]$ from Eq.~\eqref{eq:branch_c_bound_distance_real_ideal}, i.e. branch (c), which is the last term we need to bound in Fig.~\ref{fig:prob_tree_protocol}.

\begin{remark}
    The random sampling argument for estimating the phase error rate, formulated in terms of the detection statistics, cf. Eq.~\eqref{eq:probability_upper_bound_phase_error}, assumes no detection-efficiency mismatch, which is experimentally unfeasible, as will be discussed in Sec.~\ref{sec:discussion}. A recent work resolves this constraint by adjusting the random sampling arguments for scenarios that do not perfectly satisfy the no-mismatch assumption. In particular, Ref.~\cite{Tupkary24} can produce key rates for scenarios with  an imperfectly characterized, bounded amount of mismatch in detector efficiencies and dark count rates.
\end{remark}

\section{Decoy bounds}
\label{sec:decoy_state_bounds}
In the last section, the smooth min-entropy has been decomposed so as to yield an expression in terms of the acceptance bounds $s_{\mathsf{Z},0}^\mathrm{l}$, $s_{\mathsf{Z},1}^\mathrm{l}$, $s_{\mathsf{X},1}^\mathrm{l}$ and $\Lambda^\mathrm{u}_\mathsf{X}$ defined by the acceptance set, cf. Eq~\eqref{eq:H_min_operational_hypothetical}. The acceptance test is performed by verifying that $s_{\mathsf{Z},0}^- \geq s_{\mathsf{Z},0}^\mathrm{l}$, $s_{\mathsf{Z},1}^- \geq s_{\mathsf{Z},1}^\mathrm{l}$, $s_{\mathsf{X},1}^- \geq s_{\mathsf{X},1}^\mathrm{l}$ and $\Lambda_\mathsf{X}^+ \leq \Lambda^\mathrm{u}_\mathsf{X}$, cf. Table~\ref{tab:acceptance_testing} and Fig.~\ref{fig:protocol_description}. We recall that the number of vacuum events, single-photon events and the phase error rate are not directly accessible to Alice and Bob. Thus, our goal in this section is to determine the decoy bounds $s_{\mathsf{Z},0}^-, s_{\mathsf{Z},1}^-$, $s_{\mathsf{X},1}^-$ and $\Lambda_\mathsf{X}^+$ solely in terms of experimentally available parameters. The acceptance test is then performed using these bounds. If the decoy bounds hold, the acceptance test passes and error correction was successful, then $s_{\mathsf{Z},0} \geq s_{\mathsf{Z},0}^- \geq s_{\mathsf{Z},0}^\mathrm{l}$ and analogously for  $s_{\mathsf{Z},1}^\mathrm{l}$, $s_{\mathsf{X},1}^\mathrm{l}$ and $\Lambda^\mathrm{u}_\mathsf{X}$. Verifying these relations is precisely the aim of the acceptance test on the decoy bounds. We recall that we used $s_{\mathsf{Z},0} \geq s_{\mathsf{Z},0}^\mathrm{l}$ and the other relations when deriving an operational expression for the min-entropy in previous section. The methods used in this section are similar to Refs.~\cite{Lim14, LimCurtyWalenta14, Rusca18}. The bounds for the 2-decoy state protocol are derived in App.~\ref{ap:two_decoy_security_proof}.

\subsection{Finite-size photon event statistics}
\label{sec:photon_event_statistics_and_bounds}
In the 1-decoy state BB84 protocol, similarly to the original BB84 prepare-and-measure protocol, two bases $\mathsf{X}$ and $\mathsf{Z}$ are chosen to transmit signals over a quantum channel. However, in order to counter so-called photon-number-splitting (PNS) attacks by an eavesdropper Eve, an additionally degree of freedom is introduced in the level of intensity $k$ used to transmit photons with a phase-randomized coherent laser source. A phase-randomized coherent state with intensity $k$ sent by Alice can be represented by \cite{Lo07}
\begin{equation}
    \rho_k = \int_0^{2\pi} d\theta f(\theta) \ketbra{\sqrt{k}e^{i\theta}}{\sqrt{k}e^{i\theta}} \, , 
    \label{eq:density_matrix_coherent_state}
\end{equation}
where $f(\theta)$ is the probability density function representing the probability of generating a state with phase $\theta$. Here, $\ket{\sqrt{k}e^{i\theta}}$ represents a coherent state with mean photon number $k$ and phase $\theta$, defined as a coherent superposition of Fock states $\ket{m}$, \added{where $\ket{m}$ denotes a photon-number (Fock) state containing exactly $m$ photons in the optical mode \cite{Scully97}}, 
\begin{equation}
    \ket{\alpha} = \sum_{m=0}^\infty e^{-|\alpha|/ 2} \frac{\alpha^m}{\sqrt{m!}}\ket{m}\,.
    \label{eq:def_coherent_state}
\end{equation}
In the case of uniform phase randomization, i.e. $f(\theta)=\frac{1}{2\pi}$, Eq.~\eqref{eq:density_matrix_coherent_state} simplifies to
\begin{equation}
    \rho_k = \sum_{m=0}^\infty e^{-k}\frac{k^m}{m!}\ketbra{m}{m}\, .
    \label{eq:simplified_coherent_state}
\end{equation}
Essentially, with uniform phase randomization, Alice's density operator is diagonal in the Fock-basis, which reduces the analysis to a discussion of the statistics of photon events as Fock states do not carry information about the intensity choice. In contrast, if the phases are not uniformly randomized, then Alice's density operator is not diagonal in the Fock-basis and her signals carry information about the intensity choice. This conflicts with the assumption that Eve has no a priori knowledge about Alice's intensity choices. Note that it is still possible to prove security in this case, but the secure-key length is significantly impacted in current proofs \cite{Curras24b, Sixto23, Nahar23}. In the following, for simplicity, we assume uniform phase randomization. Following Eq.~\eqref{eq:simplified_coherent_state}, the photon number of a uniformly phase-randomized coherent photon source follows the Poisson distribution
\begin{equation}
    \Pr[m]=e^{-k}\frac{k^m}{m!}\, ,
\end{equation}
where $m$ is the number of photons sent per signal. In the case of the 1-decoy state protocol, two levels of intensity $\mu_1$ and $\mu_2$ are used at random following pre-configured probabilities $p_{\mu_1}$ and $p_{\mu_2}$, with $\mu_1 > \mu_2$. Often, the states with intensity $\mu_1$ are called \textit{signal states}, and the states with intensity $\mu_2$ are called \textit{decoy states}\footnote{We note that in our protocol both intensities are used for testing and key generation.}.

It is useful to assume an equivalent virtual scenario in which Alice chooses the mean photon-number after Bob measured an $m$-photon state. This means that instead of considering the probability of an $m$-photon event occurring given an intensity $k\in\{\mu_1, \mu_2\}$, we consider the probability that an intensity $k$ was chosen given the detection of an $m$-photon event. Both scenarios are equivalent as Eve's attack strategy cannot depend on the chosen intensity, meaning that Bob's measurement results are also independent of Alice's intensity choice (for a more thorough discussion, see App.~\ref{sec:scenario_alice_intensity_before_detection} or Ref.~\cite[Sec.~3.1]{Lim14}). This follows from the fact that Eve has no a priori knowledge about Alice's intensity choice and that the coherent states are uniformly phase-randomized such that they don't carry any information about the intensity choice, as discussed above.

In the following, solely the $\mathsf{Z}$-basis is considered, but the equations analogously hold for the $\mathsf{X}$-basis. The total number of detections observed by Bob in the $\mathsf{Z}$-basis is given by
\begin{equation}
    N_\mathsf{Z} = \sum_{m=0}^{\infty}s_{\mathsf{Z},m}\,,
\end{equation}
where $s_{\mathsf{Z},m}$ is the number of detections of $m$-photon events in the $\mathsf{Z}$-basis. Alice chooses one of the intensities $\mu_1$ and $\mu_2$ randomly with probabilities $p_{\mu_1|m}$ and $p_{\mu_2|m}$ respectively, given that Bob measures an $m$-photon state. As such, for a given intensity $k\in \{\mu_1, \mu_2\}$, the expected number of detections is given by
\begin{equation}
    n_{\mathsf{Z},k}^*=\sum_{m=0}^{\infty} \Pr[k|m]\,s_{\mathsf{Z},m}\, .
    \label{eq:def_n_star}
\end{equation}
Analogously, the total number of errors observed in the $\mathsf{Z}$-basis is given by
\begin{equation}
    c_\mathsf{Z} = \sum_{m=0}^\infty v_{\mathsf{Z},m}\, ,
    \label{eq:total_error_detected}
\end{equation}
where $v_{\mathsf{Z},m}$ is the number of errors associated with $s_{\mathsf{Z},m}$. The expected number of bit errors observed, given a certain intensity $k$, is given by
\begin{equation}
    c_{\mathsf{Z},k}^*=\sum_{m=0}^{\infty} p_{k|m}v_{\mathsf{Z},m}\, .
\end{equation}
Let $n_{\mathsf{Z},k}$ and $c_{\mathsf{Z},k}$ be the number of detections and errors observed, respectively, given an intensity $k$. Here, $N_\mathsf{Z}$, $c_\mathsf{Z}$, $n_{\mathsf{Z},k}$ and $c_{\mathsf{Z},k}$ are known experimental values while the sets $\{s_{\mathsf{Z},m}\}$ and $\{v_{\mathsf{Z},m}\}$ are not. The goal is to derive bounds on $s_{\mathsf{Z},0}$, $s_{\mathsf{Z},1}$, $s_{\mathsf{X},1}$ and $\Lambda_\mathsf{X}$ (necessary to perform the acceptance test leading to Eq.~\eqref{eq:H_min_operational_hypothetical}) solely in terms of experimentally accessible parameters by taking finite-size effects into account.

Note that each signal is independently mapped to a specific intensity, based on the photon number of the pulse. Thus, using suitable concentration inequalities, such as Hoeffding's inequality \cite{Hoeffding63} for independent random variables, we can bound the deviation of the observed number of detections and errors, $n_{\mathsf{Z},k}$ and $c_{\mathsf{Z},k}$, from the expected number of detections and errors, $n_{\mathsf{Z},k}^*$ and $c_{\mathsf{Z},k}^*$, as follows:\footnote{Here, other concentration inequalities such as Azuma's inequality can be used.}
\begin{empheq}[box=\fcolorbox{black}{equation_box}]{align}
    \Pr\left[n_{\mathsf{Z},k}^* > n_{\mathsf{Z},k}^+ \right] &\leq \epsilon_{\mathsf{Z}, k}^{n,+}\, , \label{eq:def_n_plus} \\
    \Pr\left[n_{\mathsf{Z},k}^* < n_{\mathsf{Z},k}^- \right] &\leq \epsilon_{\mathsf{Z}, k}^{n,-}\, ,
    \label{eq:def_n_minus}
\end{empheq}
where
\begin{equation}
    n_{\mathsf{Z},k}^\pm \coloneqq n_{\mathsf{Z},k} \pm\delta(N_\mathsf{Z}, \epsilon_{\mathsf{Z}, k}^{n,\pm}) \,, \label{eq:def_n_Z_k}
\end{equation}
and 
\begin{equation}
    \delta(N_\mathsf{Z}, \epsilon_{\mathsf{Z}, k}^{n,\pm})\coloneqq\sqrt{N_\mathsf{Z}\ln(1/\epsilon_{\mathsf{Z}, k}^{n,\pm})/2}
\end{equation}
is also called Hoeffding delta\footnote{Note that in a QKD protocol, we have an unknown state $\rho$, that produces random variables such as $n_{\mathsf{Z},k}$ and $s_{\mathsf{Z},m}$. The above expression (and other analogous expressions) should be interpreted such that the probability is over these random variables.}. In other words, $\epsilon_{\mathsf{Z}, k}^{n,\pm}$ is the probability that the number of photons $n_{\mathsf{Z},k}$ detected by Bob deviates by more than $\delta(N_\mathsf{Z}, \epsilon_{\mathsf{Z}, k}^{n,\pm})$ from the expectation value $n_{\mathsf{Z},k}^*$. For a formal derivation of the expression for the Hoeffding delta using Ref.~\cite[Eq.~(2.1)]{Hoeffding63} as a starting point, refer to App.~\ref{ap:hoeffding_delta}. Analogously,
\begin{empheq}[box=\fcolorbox{black}{equation_box}]{align}
    \Pr\left[c_{\mathsf{Z},k}^* > c_{\mathsf{Z},k}^+ \,|\, \Omega_\mathrm{EC}\right] &\leq \epsilon_{\mathsf{Z}, k}^{c,+} \, , \label{eq:def_c_plus} \\
    \Pr\left[c_{\mathsf{Z},k}^* < c_{\mathsf{Z},k}^- \,|\, \Omega_\mathrm{EC}\right] &\leq \epsilon_{\mathsf{Z}, k}^{c,-} \, . \label{eq:def_c_minus}
\end{empheq}
where
\begin{equation}
    c_{\mathsf{Z},k}^\pm \coloneqq c_{\mathsf{Z},k} \pm\delta(c_\mathsf{Z}, \epsilon_{\mathsf{Z}, k}^{c,\pm}) \,, \label{eq:def_c_Z_k}
\end{equation}
and we additionally condition Eqs.~\eqref{eq:def_c_plus} and \eqref{eq:def_c_minus} on error correction succeeding\footnote{Formally, this can be argued by noting that we can pretend that Alice assigns intensities to pulses only after error correction has succeeded.} as Bob needs to count the correct number of $\mathsf{Z}$-basis errors $c_{\mathsf{Z},k}$ when comparing the non-corrected key to the corrected key, cf. Fig.~\ref{fig:protocol_description}. The bounds $n_{\mathsf{X},k}^\pm$ and $c_{\mathsf{X},k}^\pm$ for the $\mathsf{X}$-basis are analogously defined but we don't require any conditioning on $\Omega_\mathrm{EC}$ in this case as the $\mathsf{X}$-basis detections are disclosed. We note that the use of Hoeffding's inequality for independent random variables is subtle in this context and we refer to Refs. \cite[Remark~13]{Tupkary24} and \cite{Curty2014} for a more thorough discussion. The goal of the following sections is to determine lower bounds for vacuum and single-photon events as well as an upper bound for the number of vacuum events and single-photon errors, which are required to perform the acceptance test, cf. Fig.~\ref{tab:acceptance_testing}.

\subsubsection{Lower bound on the number of vacuum events}
\label{sec:lower_bound_vacuum}
Using Bayes' theorem for the probability of choosing the intensity $k$ given that Bob detected an $m$-photon event,
\begin{equation}
    p_{k|m} = \frac{p_k}{\tau_m}p_{m|k} = \frac{p_k}{\tau_m}\frac{e^{-k} k^m}{m!}\, ,
    \label{eq:bayes_rule}
\end{equation}
where
\begin{equation}
    \tau_m = \sum_{k\in \{\mu_1, \mu_2\}} p_k \frac{e^{-k} k^m}{m!}
\end{equation}
is the average probability for Alice to transmit an $m$-photon state.
Now, using Eqs. \eqref{eq:def_n_star}, \eqref{eq:bayes_rule} and $p_{k|0} = \frac{p_k}{\tau_0}e^{-k}$, we find
\begin{equation}
    \frac{\mu_1 e^{\mu_2}n_{\mathsf{Z},\mu_2}^*}{p_{\mu_2}} - \frac{\mu_2 e^{\mu_1} n_{\mathsf{Z},\mu_1}^*}{p_{\mu_1}} = \frac{(\mu_1 - \mu_2)s_{\mathsf{Z},0}}{\tau_0} - \mu_1\mu_2\sum_{m=2}^\infty \frac{(\mu_1^{m-1}-\mu_2^{m-1})s_{\mathsf{Z},m}}{\tau_m m!}\, . 
\end{equation}
Now, solving for $s_{\mathsf{Z},0}$ and using $\mu_1 > \mu_2$ yields 
\begin{equation}
    s_{\mathsf{Z},0} \geq \frac{\tau_0}{(\mu_1 - \mu_2)}\left( \frac{\mu_1 e^{\mu_2} n_{\mathsf{Z},\mu_2}^*}{p_{\mu_2}} - \frac{\mu_2 e^{\mu_1} n_{\mathsf{Z},\mu_1}^*}{p_{\mu_1}} \right)\, .
\end{equation}
Finally, together with the bounds from Eqs. \eqref{eq:def_n_plus} and \eqref{eq:def_n_minus}, we have
\begin{empheq}[box=\fcolorbox{black}{equation_box}]{align}
    \Pr\left[s_{\mathsf{Z},0} < s_{\mathsf{Z},0}^- \coloneqq \frac{\tau_0}{(\mu_1 - \mu_2)}\left( \frac{\mu_1 e^{\mu_2} n_{\mathsf{Z},\mu_2}^-}{p_{\mu_2}} - \frac{\mu_2 e^{\mu_1} n_{\mathsf{Z},\mu_1}^+}{p_{\mu_1}} \right) \right] \leq \epsilon_{\mathsf{Z}, \mu_2}^{n,-} + \epsilon_{\mathsf{Z}, \mu_1}^{n,+}\, .
    \label{eq:lower_bound_vacuum}
\end{empheq}
The inequalities $n^*_{\mathsf{Z},\mu_2} \geq n_{\mathsf{Z},\mu_2}^-$ and $n^*_{\mathsf{Z},\mu_1} \leq n_{\mathsf{Z},\mu_1}^+$ hold with a probability of at least $1-\epsilon_{\mathsf{Z}, \mu_2}^{n,-}$ and $1-\epsilon_{\mathsf{Z}, \mu_1}^{n,+}$, respectively (see Eqs.~\eqref{eq:def_n_plus} and \eqref{eq:def_n_minus}). Hence, the probability that $s_{\mathsf{Z},0} \geq s_{\mathsf{Z},0}^-$ does not hold, i.e. at least one of the concentration inequalities does not hold, is upper-bounded by $\epsilon_{\mathsf{Z}, \mu_2}^{n,-}+ \epsilon_{\mathsf{Z}, \mu_1}^{n,+}$. This is a consequence of Boole's inequality.

\subsubsection{Upper bound on the number of vacuum events}
\label{sec:upper_bound_vacuum_events}
An upper bound on the number of vacuum events is derived in this section and will be needed to derive a lower bound on the number of single-photon events in Sec.~\ref{sec:lower_bound_single}. Trivially, following Eq.~\eqref{eq:total_error_detected}, the total number of errors detected in the $\mathsf{Z}$-basis can be lower bounded by the number of errors detected for vacuum events:
\begin{equation}
    c_\mathsf{Z} \geq v_{\mathsf{Z},0}\, .
\end{equation}
Due to the fact that vacuum events carry no information, the average number of vacuum errors $v_{\mathsf{Z},0}^*$ is half of the total number of vacuum events, namely\footnote{Here, we implicitly assume that Alice sends each bit with equal probability, both $\mathsf{Z}$-basis detectors have the same properties and no asymmetric losses are present.} \cite{Rusca18}
\begin{equation}
    \frac{v_{\mathsf{Z},0}^*}{s_{\mathsf{Z},0}} = \frac{1}{2}\, .
    \label{eq:value_v_x_0}
\end{equation}
By using Hoeffding's inequality, we can in turn bound $v_{\mathsf{Z},0}^*$ as a function of the observed number of errors for vacuum events $v_{\mathsf{Z},0}$ given a certain intensity,
\begin{equation}
    \Pr\left[v_{\mathsf{Z},0}^* > v_{\mathsf{Z},0} + \delta(N_\mathsf{Z}, \epsilon_{\mathsf{Z},0}^{v,+}) \right] \leq \epsilon_{\mathsf{Z},0}^{v,+}\,,
    \label{eq:bound_v_x_0}
\end{equation}
using $ v_{\mathsf{Z},0} + \delta(s_{\mathsf{Z},0}, \epsilon_{\mathsf{Z},0}^{v,+}) \leq v_{\mathsf{Z},0} + \delta(N_\mathsf{Z}, \epsilon_{\mathsf{Z},0}^{v,+})$. Additionally, we find an upper bound for $v_{\mathsf{Z},0}$ by rewriting
\begin{align}
    c_{\mathsf{Z},k}^* &=\sum_{m=0}^\infty p_{k|m} v_{\mathsf{Z},m} = \sum_{m=0}^\infty \frac{p_k}{\tau_m} \frac{e^{-k}k^m}{m!} v_{\mathsf{Z},m} \geq \frac{p_k}{\tau_0}e^{-k}v_{Z,0} \\
    \Leftrightarrow v_{\mathsf{Z},0} &\leq \frac{c_{\mathsf{Z},k}^*}{p_k}\tau_0e^k .
\end{align}
We can now plug in Eq.~\eqref{eq:def_c_plus} and write 
\begin{equation}
    \Pr\left[v_{\mathsf{Z},0} > \frac{c_{\mathsf{Z},k}^+}{p_k}\tau_0e^k \,\middle|\, \Omega_\mathrm{EC} \right] \leq \epsilon_{\mathsf{Z},k}^{c,+} \,,
    \label{eq:lower_bound_m_x_k}
\end{equation}
where any $k\in\{\mu_1, \mu_2\}$ can be chosen, e.g. to maximize the secure-key length. Now, using Eqs.~\eqref{eq:def_c_plus}, \eqref{eq:value_v_x_0} and the bounds from Eqs.~\eqref{eq:bound_v_x_0} and \eqref{eq:lower_bound_m_x_k}, we can determine an upper bound on the number of vacuum events,
\begin{empheq}[box=\fcolorbox{black}{equation_box}]{align}
    \Pr\left[s_{\mathsf{Z},0} > s_{\mathsf{Z},0}^+ \coloneqq 2\left(\frac{c_{\mathsf{Z},k}^+}{p_k}\tau_0e^k + \delta(N_\mathsf{Z}, \epsilon_{\mathsf{Z},0}^{v,+})\right) \, \middle| \, \Omega_\mathrm{EC}  \right] \leq \epsilon_{\mathsf{Z},0}^{v,+} + \epsilon_{\mathsf{Z},k}^{c,+}\, ,
    \label{eq:upper_bound_s_x_0}
\end{empheq}
where we condition on error correction succeeding, i.e. $\Omega_\mathrm{EC}$, as Bob requires the number of $\mathrm{Z}$-basis errors to compute $c_{\mathsf{Z},k}^+$, which he determines by comparing the verified key to the sifted key, as discussed in Sec.~\ref{sec:photon_event_statistics_and_bounds}.
In the following, we denote $k_\mathrm{min}^\mathsf{Z}\in\{\mu_1, \mu_2\}$ the intensity chosen to compute the bound in the inequality above, which can be chosen to maximize the secure-key length.

\subsubsection{Lower bound on the number of single-photon events}
\label{sec:lower_bound_single}
Analogously to Sec.~\ref{sec:lower_bound_vacuum}, we can write
\begin{align*}
    \frac{e^{\mu_2} n_{\mathsf{Z},\mu_2}^*}{p_{\mu_2}} - \frac{e^{\mu_1} n_{\mathsf{Z},\mu_1}^*}{p_{\mu_1}} &= \frac{(\mu_2 - \mu_1)s_{\mathsf{Z},1}}{\tau_1} + \sum_{m=2}^\infty \frac{(\mu_2^m - \mu_1^m)s_{\mathsf{Z},m}}{\tau_m m!} \\
    &\leq \frac{(\mu_2 - \mu_1)s_{\mathsf{Z},1}}{\tau_1} + \frac{\mu_2^2 - \mu_1^2}{\mu_1^2}\sum_{m=2}^\infty \frac{\mu_1^m s_{\mathsf{Z},m}}{\tau_m m!}\,,
\end{align*}
as
\begin{equation*}
    \mu_2^m - \mu_1^m = \mu_2^2\mu_2^{m-2} - \mu_1^2\mu_1^{m-2} = \mu_1^{m-2}\left(\mu_2^2 \frac{\mu_2^{m-2}}{\mu_1^{m-2}} - \mu_1^2\right) \leq (\mu_2^2 - \mu_1^2)\mu_1^{m-2}
\end{equation*}
for $m\geq 2$ and the last inequality holds for $\mu_1 > \mu_2$. Now, writing the sum of multi-photon events ($m\geq 2$) as 
\begin{equation}
    \sum_{m=2}^\infty \frac{\mu_1^m s_{\mathsf{Z},m}}{\tau_m m!} = \sum_{m=2}^\infty \frac{e^{\mu_1} e^{-\mu_1}p_{\mu_1} \mu_1^m s_{\mathsf{Z},m}}{p_{\mu_1}\tau_m m! } = \frac{e^{\mu_1}n_{\mathsf{Z},\mu_1}^*}{p_{\mu_1}} - \frac{s_{\mathsf{Z},0}}{\tau_0} - \frac{\mu_1 s_{\mathsf{Z},1}}{\tau_1}
\end{equation}
by using Equations \eqref{eq:def_n_star} and \eqref{eq:bayes_rule} yields
\begin{equation}
    \frac{e^{\mu_2} n_{\mathsf{Z},\mu_2}^*}{p_{\mu_2}} - \frac{e^{\mu_1} n_{\mathsf{Z},\mu_1}^*}{p_{\mu_1}} \leq \frac{(\mu_2 - \mu_1)s_{\mathsf{Z},1}}{\tau_1} + \frac{\mu_2^2 - \mu_1^2}{\mu_1^2}\left( \frac{e^{\mu_1}n_{\mathsf{Z},\mu_1}^*}{p_{\mu_1}} - \frac{s_{\mathsf{Z},0}}{\tau_0} - \frac{\mu_1 s_{\mathsf{Z},1}}{\tau_1} \right)\,.
\end{equation}
Now, solving for $s_{\mathsf{Z},1}$ yields
\begin{equation}
    s_{\mathsf{Z},1} \geq \frac{\mu_1 \tau_1}{\mu_2(\mu_1 - \mu_2)}\left( \frac{e^{\mu_2} n_{\mathsf{Z},\mu_2}^*}{p_{\mu_2}} - \frac{\mu_2^2}{\mu_1^2}\frac{e^{\mu_1}n_{\mathsf{Z},\mu_1}^*}{p_{\mu_1}} - \frac{(\mu_1^2 - \mu_2^2)}{\mu_1^2}\frac{s_{\mathsf{Z},0}}{\tau_0} \right)\,.
    \label{eq:lower_bound_single_photon_no_substitution}
\end{equation}
Finally, by substituting Eq.~\eqref{eq:upper_bound_s_x_0} into Eq.~\eqref{eq:lower_bound_single_photon_no_substitution} and using the lower and upper bounds for the number of detections $n_{\mathsf{Z},\mu_2}^*$ and $n_{\mathsf{Z},\mu_1}^*$ (see Eqs.~\eqref{eq:def_n_plus} and \eqref{eq:def_n_minus}), we can determine a lower bound for the number of single-photon events,\footnote{This bound is tight as $s_{\mathsf{Z},0}^+$ does not depend on $\epsilon_{\mathsf{Z},\mu_2}^{n,-}$ nor $\epsilon_{\mathsf{Z},\mu_1}^{n,+}$.}
\begin{empheq}[box=\fcolorbox{black}{equation_box}]{align}
    \Pr\left[s_{\mathsf{Z},1} < s_{\mathsf{Z},1}^- \,|\, \Omega_\mathrm{EC}\right] \leq \epsilon_{\mathsf{Z},\mu_2}^{n,-} + \epsilon_{\mathsf{Z},\mu_1}^{n,+} + \Pr\left[s_{\mathsf{Z},0} > s_{\mathsf{Z},0}^+\,|\,\Omega_\mathrm{EC}\right]\,,
    \label{eq:lower_bound_single}
\end{empheq}
where
\begin{equation}
    s_{\mathsf{Z},1}^- \coloneqq \frac{\mu_1 \tau_1}{\mu_2(\mu_1 - \mu_2)}\left( \frac{e^{\mu_2} n_{\mathsf{Z},\mu_2}^-}{p_{\mu_2}} - \frac{\mu_2^2}{\mu_1^2}\frac{e^{\mu_1}n_{\mathsf{Z},\mu_1}^+}{p_{\mu_1}} - \frac{(\mu_1^2 - \mu_2^2)}{\mu_1^2}\frac{s_{\mathsf{Z},0}^+}{\tau_0} \right) \,,
\end{equation}
and we condition on $\Omega_\mathrm{EC}$ for the same reason as in Sec.~\ref{sec:upper_bound_vacuum_events}.

\subsubsection{Upper bound on the number of single-photon errors}
\label{sec:upper_bound_single_photon_errors}
Finally, an upper bound for the number of single-photon errors is required to determine an upper bound on the single-photon QBER $\Lambda_\mathsf{X}$ (and thus the phase error rate $\Lambda_\mathsf{Z}$) in Sec.~\ref{sec:upper_bound_qber}. It can easily be derived, similarly to the other bounds, by rewriting
\begin{align*}
    \frac{e^{\mu_1} c_{\mathsf{X},\mu_1}^*}{p_{\mu_1}} - \frac{e^{\mu_2} c_{\mathsf{X},\mu_2}^*}{p_{\mu_2}} &= \frac{\mu_1-\mu_2}{\tau_1}v_{\mathsf{X},1} + \sum_{m=2}^\infty \frac{\mu_1^m}{\tau_m m!}v_{\mathsf{X},m} - \sum_{m=2}^\infty \frac{\mu_2^m}{\tau_m m!}v_{\mathsf{X},m}\, .
\end{align*}
Using $\mu_1 > \mu_2$, we have
\begin{align*}
    \frac{e^{\mu_1} c_{\mathsf{X},\mu_1}^*}{p_{\mu_1}} - \frac{e^{\mu_2} c_{\mathsf{X},\mu_2}^*}{p_{\mu_2}} \geq \frac{\mu_1-\mu_2}{\tau_1}v_{\mathsf{X},1}
\end{align*}
and thus
\begin{equation*}
     \frac{\tau_1}{\mu_1 - \mu_2}\left( \frac{e^{\mu_1}c_{\mathsf{X},\mu_1}^*}{p_{\mu_1}} - \frac{e^{\mu_2}c_{\mathsf{X},\mu_2}^*}{p_{\mu_2}}\right) \geq v_{\mathsf{X},1}\, .
\end{equation*}
Finally, using Eqs.~\eqref{eq:def_c_plus} and \eqref{eq:def_c_minus} yields an upper bound on the number of single-photon errors
\begin{empheq}[box=\fcolorbox{black}{equation_box}]{align}
    \Pr\left[v_{\mathsf{X},1} > v_{\mathsf{X},1}^+ \coloneqq \frac{\tau_1}{\mu_1 - \mu_2}\left( \frac{e^{\mu_1}c_{\mathsf{X},\mu_1}^+}{p_{\mu_1}} - \frac{e^{\mu_2}c_{\mathsf{X},\mu_2}^-}{p_{\mu_2}}\right) \right] \leq \epsilon_{\mathsf{X},\mu_1}^{c,+} + \epsilon_{\mathsf{X},\mu_2}^{c,-}
    \label{eq:upper_bound_error}
\end{empheq}
\begin{remark}
    Ineqs.~\eqref{eq:lower_bound_vacuum}, \eqref{eq:upper_bound_s_x_0}, \eqref{eq:lower_bound_single} and \eqref{eq:upper_bound_error} are equivalent to the bounds derived in Ref.~\cite{Rusca18}, except for the conditioning on $\Omega_\mathrm{EC}$ in Eqs.~\eqref{eq:upper_bound_s_x_0} and \eqref{eq:lower_bound_single}, which was not mentioned in Ref.~\cite{Rusca18}. For the derivation of the bounds for the 2-decoy protocol based on Ref.~\cite{LimCurtyWalenta14}, we refer to App.~\ref{ap:two_decoy_security_proof}.
\end{remark}
\begin{remark}
    All equations from this section remain valid if the bases used for the key distribution and parameter estimation are swapped.
    For example, although not explicitly shown, $s_{\mathsf{X},1}^-$ is given by replacing $\mathsf{Z}$ by $\mathsf{X}$ in Eq.~\eqref{eq:lower_bound_single}. This is needed to calculate the upper bound on the single-photon QBER $\Lambda_\mathsf{X}$ and thus the phase error rate $\Lambda_\mathsf{Z}$.
\end{remark}

\subsection{Upper bound on the phase error rate}
\label{sec:upper_bound_qber}
We recall that in Sec.~\ref{sec:decomposing_min_entropy} we have derived a bound on the phase error rate in terms of the single-photon QBER using Serfling's inequality, cf. Eq.~\eqref{eq:probability_upper_bound_phase_error}. The acceptance test requires to verify $\Lambda_\mathsf{X}^+ \leq \Lambda_\mathsf{X}^\mathrm{u}$. An upper bound on the single-photon QBER is explicitly given by\footnote{We note that this bound is not tight if $k_\mathrm{min}^\mathsf{X} = \mu_1$ as then $v_{\mathsf{X},1}^+$ is not independent of $s_{\mathsf{X},1}^-$. In this case, the expression can be optimized by removing one contribution $\epsilon_{\mathsf{X},\mu_1}^{c,+}$.} 
\begin{empheq}[box=\fcolorbox{black}{equation_box}]{align}
    \Pr\left[\Lambda_\mathsf{X} > \Lambda_\mathrm{X}^+ \coloneqq \frac{v_{\mathsf{X},1}^+}{s_{\mathsf{X},1}^-} \right] \leq \Pr\left[v_{\mathsf{X},1} > v_{\mathsf{X},1}^+\right] + \Pr\left[s_{\mathsf{X},1} < s_{\mathsf{X},1}^-\right]\, ,
    \label{eq:upper_bound_qber}
\end{empheq}
which corresponds to an upper bound on the probability that one of the bounds, $v_{\mathsf{X},1}^+$ or $s_{\mathsf{X},1}^-$, fails. We note that $s_{\mathsf{X},1}^-$ is given by replacing $\mathsf{Z}$ by $\mathsf{X}$ in Eq.~\eqref{eq:lower_bound_single}. Now that we have operational expressions for the decoy bounds $s_{\mathsf{Z},0}^-$, $s_{\mathsf{Z},1}^-$, $s_{\mathsf{X},1}^-$ and $\Lambda^+_\mathsf{X}$, the acceptance test described in Sec.~\ref{sec:long_protocol_description} and Fig~\ref{fig:protocol_description} can be performed on these bounds.

\section{Extractable secure-key length}
\label{sec:secret_key_length}
The first term in Eq.~\eqref{eq:expanded_security_def} has been treated in Sec.~\ref{sec:error_verification} by bounding branch (a) from Fig.~\ref{fig:prob_tree_protocol} using the properties of universal$_2$ hashing. In Sec.~\ref{sec:expansion_secrecy_ci}, we have split the second term in terms of the decoy concentration inequalities holding (branch (b)) and not holding (branch (c)). As seen in Sec.~\ref{sec:quantum_leftover_hash_lemma}, the quantum leftover hash lemma provides a bound on branch (b) in terms of the smooth min-entropy and the secure-key length, cf. Eq. \eqref{eq:branch_c_bound_distance_real_ideal}. Hence, the last term we need to bound is branch (c) and will be addressed in Sec.~\ref{sec:operational_expression_min_entropy} using the decoy bounds derived in last section. We then derive an operational expression on the secure-key length solely in terms of the acceptance parameters, which we further simplify in Sec.~\ref{sec:simplified_skl}.

\subsection{Operational expression for the secure-key length}
\label{sec:operational_expression_min_entropy}
 
\noindent After the error correction, error verification steps and acceptance test, Alice and Bob disclosed $\mathrm{leak}_{\mathrm{EC}} + \log_2\frac{2}{\epsilon_\mathrm{cor}}$ bits of information through the classical channel and possess a verified key pair $Z_A$ and $Z_B$, cf. Sec~\ref{sec:error_verification}. During the acceptance test, they check the observed statistics derived in Sec.~\ref{sec:decoy_state_bounds} against pre-defined parameters, namely they verify that $s_{\mathsf{Z},0}^- \geq s_{\mathsf{Z},0}^\mathrm{l}$, $s_{\mathsf{Z},1}^- \geq s_{\mathsf{Z}, 1}^\mathrm{l}$, $s_{\mathsf{X},1}^- \geq s_{\mathsf{X}, 1}^\mathrm{l}$ and $\Lambda_\mathsf{X}^+ \leq \Lambda_\mathsf{X}^\mathrm{u}$, cf. Sec.~\ref{sec:upper_bound_qber} and Table~\ref{tab:acceptance_testing}.
If at least one of the conditions does not hold, the protocol aborts, as discussed in Sec.~\ref{sec:preliminaries:protocol_description} and Fig.~\ref{fig:protocol_description}. 

In order to derive the decoy bounds, Hoeffding's inequalities, cf. Eqs.~\eqref{eq:def_n_plus} to \eqref{eq:def_c_minus}, have been used multiple times. Hence, as discussed in Sec.~\ref{sec:decoy_state_bounds}, the resulting bounds have a probability of not holding upper-bounded by Eqs.~\eqref{eq:lower_bound_vacuum}, \eqref{eq:lower_bound_single} and \eqref{eq:upper_bound_qber}. An illustration visualizing the different contributions is depicted in Fig.~\ref{fig:illustration_contribution_epsilon}. Using this, we can determine an upper-bound on branch (c), i.e. for the probability for at least one of the Hoeffding inequalities failing, cf. Eq.~\eqref{eq:reformulation_security_def}, by summing over all contributions, yielding
\begin{equation}
    \Pr[\neg \Omega_\mathrm{B}\land\Omega_\mathrm{EC}] \leq \epsilon_{\mathsf{Z},\mu_2}^{n,-} + \epsilon_{\mathsf{Z},\mu_1}^{n,+} + \epsilon_{\mathsf{Z},k^\mathsf{Z}_\mathrm{min}}^{c,+} + \epsilon_{\mathsf{Z},0}^{v,+} + \epsilon_{\mathsf{X},k^\mathsf{X}_\mathrm{min}}^{c,+} + \epsilon_{\mathsf{X},0}^{v,+} + \epsilon_{\mathsf{X},\mu_2}^{n,-} + \epsilon_{\mathsf{X},\mu_1}^{n,+} + \epsilon_{\mathsf{X},\mu_1}^{c,+} + \epsilon_{\mathsf{X},\mu_2}^{c,-}\eqqcolon \Delta_\mathrm{ci}\, ,
    \label{eq:def_delta_hi}
\end{equation}
where contributions appearing twice are only counted once because if a Hoeffding inequality holds, it holds for all occurences. We implicitly used that $\Pr[\neg \Omega_\mathrm{B}\land\Omega_\mathrm{EC}] \leq \Pr[\neg \Omega_\mathrm{B}\,|\,\Omega_\mathrm{EC}]$ and $\Pr[\neg \Omega_\mathrm{B}\land\Omega_\mathrm{EC}] \leq \Pr[\neg \Omega_\mathrm{B}]$. We thus directly have a bound on branch (c) as $\Pr[\tilde{\Omega} \land \neg \Omega_\mathrm{B}] \leq \Pr[\neg \Omega_\mathrm{B}\land\Omega_\mathrm{EC}]$. Substituting Eq.~\eqref{eq:def_delta_hi} in Eq.~\eqref{eq:branch_c_bound_distance_real_ideal}, we can write
\begin{equation}
   \Pr[\tilde{\Omega}] d_\mathrm{sec}(SEC)_{\rho|\tilde{\Omega}} \leq 2(\alpha_2 + \nu) + \Delta_\mathrm{pa}^\zeta + \Delta_\mathrm{ci} \leq \epsilon_\mathrm{sec}'\, ,
    \label{eq:final_bound_distance_real_ideal}
\end{equation}
and the parameter $\epsilon_\mathrm{sec}'$ is similarly defined to the secrecy parameter $\epsilon_\mathrm{sec}$, which corresponds to the second term in the expanded definition of security, cf. Eq.~\eqref{eq:expanded_security_def}, as discussed in  Remark~\ref{rem:secrec_correctness_not_seperated}. An operational expression for $\epsilon_\mathrm{sec}'$ is given by adding the contributions, yielding
\begin{equation}
    \epsilon_\mathrm{sec}' = \underbrace{2\nu}_\text{Serfling's inequality} + \,\,\,
    \underbrace{2\alpha_2}_\text{Chain rule}\,\,\, +
    \underbrace{\Delta_\mathrm{pa}^\zeta}_\text{Privacy amplification} + \underbrace{\Delta_\mathrm{ci} \vphantom{(\Delta_\mathrm{pa}^\zeta)}}_\text{Concentration inequalities}.
    \label{eq:sum_of_epsilons}
\end{equation}
The term $\nu$ results from the use of Serfling's inequality to bound the single-photon phase error rate using the single-photon QBER, cf. Sec.~\ref{sec:entropic_uncertainty_relation}. The terms $\alpha_2$ and $\nu$ result from optimizing the min-entropy in an $\zeta$-ball around $\rho_{ZE}$ (see Sec.~\ref{sec:decomposing_min_entropy}) and thus represent the probability that the optimized density operator $\bar{\rho}_{ZE}$ can be distinguished from $\rho_{ZE}$. The term $\Delta_\mathrm{pa}^\zeta$ results from the privacy amplification itself and originates from the quantum leftover hash lemma for min-entropies, cf. Eq.~\eqref{eq:def_delta_pa_zeta}. Finally, $\Delta_\mathrm{ci}$ represents the probability that the decoy bounds, as derived in Sec.~\ref{sec:decoy_state_bounds} and discussed above, fail.

Substituting the definition of $\Delta_\mathrm{pa}^\zeta$, cf. Eq.~\eqref{eq:def_delta_pa_zeta}, into Eq.~\eqref{eq:final_bound_distance_real_ideal}, using the bound on the min-entropy, cf. Eq.~\eqref{eq:H_min_operational_hypothetical}, and isolating $l$ yields an expression for the maximum extractable secure-key length, as given in Theorem~\ref{th:final_skr}. \added{We have thus bounded both terms in the expanded definition in security, cf. Eq.~\eqref{eq:expanded_security_def}, i.e. all branches in Fig.~\ref{fig:prob_tree_protocol}, such that the protocol described in Sec.~\ref{sec:long_protocol_description}, which either aborts or generates a key of length $l$, is $\epsilon$-secure. We recall that the length $l$ of the key should be fixed before the protocol run, during the parameter agreement step. 
\begin{theorem}
    The protocol described in Fig.~\ref{fig:protocol_description} is $\epsilon$-secure, where $\epsilon_\mathrm{cor} + \epsilon_\mathrm{sec}'\leq \epsilon$, if the length of the secure key output when the protocol does not abort is
    \begin{empheq}[box=\fcolorbox{black}{equation_box}]{align}
    l= \left\lfloor{s_{\mathsf{Z},0}^\mathrm{l} + s_{\mathsf{Z}, 1}^\mathrm{l}(1 - h(\Lambda_\mathsf{X}^\mathrm{u} + \gamma)) - \mathrm{leak}_{\mathrm{EC}} - \log_2{\frac{4}{\epsilon_\mathrm{cor}\alpha_2^2}} - 2\log_2{\frac{1}{2(\epsilon_\mathrm{sec}'-2\zeta' - \Delta_\mathrm{ci})}}} \right\rfloor\, ,
    \label{eq:max_secret_key_formula}
    \end{empheq}
    where $\zeta' = \alpha_2 + \nu$, the acceptance parameters $s_{\mathsf{Z},0}^\mathrm{l}$, $s_{\mathsf{Z}, 1}^\mathrm{l}$ and $\Lambda_\mathsf{X}^\mathrm{u}$ are fixed before the protocol run, cf. Fig.~\ref{fig:protocol_description}, $\gamma = \gamma(\nu, s^\mathrm{l}_{\mathsf{Z},1}, s^\mathrm{l}_{\mathsf{X},1})$ is given by Eq.~\eqref{eq:def_gamma}, and $\mathrm{leak}_\mathrm{EC}$ is the number of bits leaked for error correction. The terms appearing in the above expression can be set to arbitrary values (e.g. to the maximize the secure-key length) such that $2\zeta' + \Delta_\mathrm{pa}^\zeta + \Delta_\mathrm{ci} \leq \epsilon_\mathrm{sec}'$ and $\epsilon_\mathrm{cor} + \epsilon_\mathrm{sec}' \leq \epsilon$. 
    \label{th:final_skr}
\end{theorem}
\begin{proof}
    The proof for $\epsilon_\mathrm{sec}'$-secrecy is given above, cf. Eq.~\eqref{eq:final_bound_distance_real_ideal}, while the $\epsilon_\mathrm{cor}$-correctness follows from Theorem~\ref{th:epsilon_cor}. Then the protocol is $\epsilon$-secure following Eq.~\eqref{eq:expanded_security_def}.
\end{proof}}

\noindent For each term appearing in the above equation, we have plugged in the value minimizing the secure-key length that satisfies the conditions imposed by the acceptance set $Q$. In our case, this can be done analytically as the secure-key length is monotonously dependent on the observed statistics. As such, the value minimizing the secure-key length is either a lower or upper bound on the observed statistics. This also explains why the acceptance set has been specifically chosen as described in Fig.~\ref{fig:protocol_description}. We can see that the smooth min-entropy has an operational meaning in the sense that it is an upper bound on the number of secret bits Alice can extract from $Z$. 

\added{Note that the expression for secure-key length in Eq.~\eqref{eq:max_secret_key_formula} may in principle be negative for certain parameters. However, without loss of generality, we assume that the parameters defining the protocol are chosen such that the resulting secure-key length is strictly positive (which can always be evaluated in advance using Eq.~\eqref{eq:max_secret_key_formula} as we are considering a fixed-length protocol).}

\subsection{Simplified expression for the secure-key length}
\label{sec:simplified_skl}
We may simplify Eq.~\eqref{eq:max_secret_key_formula} by setting the terms $\alpha_2$, $\nu$, $\Delta_\mathrm{pa}^\zeta$ and error terms in $\Delta_\mathrm{ci}$ to a common value $\epsilon_0$. This reduces Eq.~\eqref{eq:sum_of_epsilons} to $\epsilon_\mathrm{sec}' = 15\epsilon_0$. This expression can be plugged back into Eq.~\eqref{eq:max_secret_key_formula}, where
\begin{equation}
    \epsilon_\mathrm{sec}' - 2\zeta' - \Delta_\mathrm{ci} = \epsilon_0 = \frac{\epsilon_\mathrm{sec}'}{15}\, ,
    \label{eq:def_epsilon_0}
\end{equation}
yielding
\begin{empheq}[box=\fcolorbox{black}{equation_box}]{align}
    l = s_{\mathsf{Z},0}^\mathrm{l} + s_{\mathsf{Z}, 1}^\mathrm{l}(1 - h(\Lambda_\mathsf{X}^\mathrm{u} + \gamma)) - \mathrm{leak}_{\mathrm{EC}} - \log_2{\frac{2}{\epsilon_\mathrm{cor}}} - 4\log_2{\frac{15}{\epsilon_\mathrm{sec}' \sqrt[4]{2}}} \, ,
    \label{eq:l_max_final_equation}
\end{empheq}
where we have set $\zeta' = 2\epsilon_0$ and $\Delta_\mathrm{ci} = 10\epsilon_0$. We note that the bound on the secure-key length is slightly tighter than in the original works \cite{Rusca18, LimCurtyWalenta14}, as we have directly removed the multi-photon event register in the min-entropy in Eq.~\eqref{eq:removed_multi_photon_events} without using the chain rule, leading to $\epsilon_\mathrm{sec}' = 15\epsilon_0$ instead of $\epsilon_\mathrm{sec}' = 19\epsilon_0$ \cite{Rusca18} and the last term in Eq.~\eqref{eq:def_epsilon_0} instead of $6\log_2(19/\epsilon_\mathrm{sec}')$ \cite{Rusca18}. The same optimization is done for the 2-decoy state protocol and an expression for the secure-key length of the 2-decoy variant is derived in App.~\ref{ap:two_decoy_security_proof}. Essentially, the security proof for the 2-decoy state protocol only differs from the 1-decoy state protocol in the decoy bounds used to perform the acceptance test.

\begin{figure}
    \centering
    \includegraphics[]{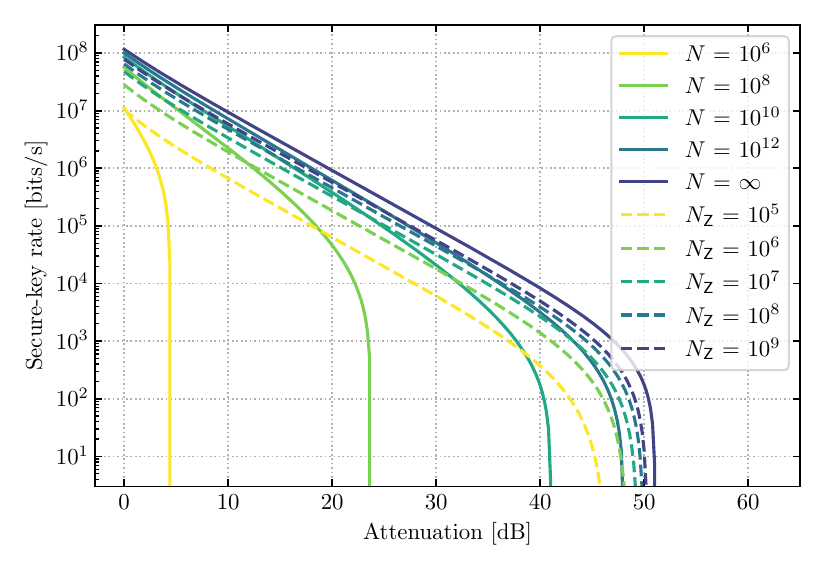}
    \caption{\added{Secure-key rate as a function of the attenuation for various numbers of signals sent $N$ and block sizes $N_\mathsf{Z}$, computed using Eq.~\eqref{eq:l_max_final_equation}. At each point, the parameters $\mu_1$, $\mu_2$, $p_{\mu_1}$ and $p_\mathsf{X}^{A}$ are optimized to maximize the secure-key rate. The values for the various protocol parameters are listed in the main text.}}
    \label{fig:skr_plot}
\end{figure}

\added{The secure-key rate is simulated using the simplified expression, Eq.~\eqref{eq:l_max_final_equation}, and depicted as a function of the channel attenuation for various numbers of signals sent $N$ and for various block sizes $N_\mathsf{Z}$, cf. Fig.~\ref{fig:skr_plot}. We recall that the block size is defined as the number of $\mathsf{Z}$-basis detections remaining after sifting, i.e. the size of the block used for post-processing. In this case, either $N$ is fixed and $N_\mathsf{Z}$ variable or the other way around}\footnote{\added{We recall that in practice $N$ and $N_\mathsf{Z}$ must be fixed prior to running the protocol.}}. \added{At each point, the free parameters $\mu_1$, $\mu_2$, $p_{\mu_1}$ and $p_\mathsf{X}^{A}$ are optimized to maximize the secure-key rate, cf. Fig.~\ref{fig:protocol_description}. For simplicity, we set Alice's and Bob's basis choice probabilities to be equal. We assume a $625\,$MHz repetition rate for Alice, $200\,$Hz dark-count rate for Bob and a probability of error (i.e. misalignment) of $2$\,\%. We use Eq.~\eqref{eq:bound_epsilon_cor} to simulate the error correction cost and set the error correction inefficiency to $f_{\mathrm{EC}}=1.16$. Finally, we set the security parameters to $\epsilon_\mathrm{sec}=\epsilon_\mathrm{cor}=10^{-12}$. 
Under these conditions, we observe that we can reach about $50$\,dB attenuation for a block size of $N_\mathsf{Z}=10^{7}$, which corresponds to about $250$\,km at $0.2$\,dB/km. The asymptotic case, $N=\infty$, directly follows from Eq.~\eqref{eq:l_max_final_equation} as all correction terms which do not scale with $N$ disappear, cf. Appendix~\ref{ap:asymptotic_key_rate}.}
\begin{remark}
     We can observe that for a given $\epsilon_\mathrm{sec}'$ and $\epsilon_\mathrm{cor}$, multiple variables can be optimized in Eqs.~\eqref{eq:sum_of_epsilons} and \eqref{eq:max_secret_key_formula}, namely $\nu$, $\alpha_2$ and the different $\epsilon$-terms. They can be adjusted individually to maximize $l$ as long as $2\zeta' + \Delta_\mathrm{pa}^\zeta + \Delta_\mathrm{ci} \leq \epsilon_\mathrm{sec}'$.
\end{remark}

\added{
\begin{remark}
    A comparison of the mathematical tools underlying different proof techniques, i.e. entropic uncertainty relation, entropy-accumulation-theorem-based and post-selection technique, is provided in Ref.~\cite{tupkary25a}. A comparison of the performance of our work with the recently published marginal-constrained entropy accumulation theorem (MEAT) framework \cite{arqand25} can be found in Ref.~\cite{kamin25}, while a comparison with both the MEAT and the post-selection technique can be found in Ref.~\cite{kaminthesis}.
\end{remark}
}

\tikzstyle{level 1}=[level distance=2cm, sibling distance=2.3cm]
\tikzstyle{level 2}=[level distance=2cm, sibling distance=2cm]
\tikzstyle{level 3}=[level distance=2cm, sibling distance=2cm]

\tikzstyle{bag} = [text width=8em, text centered]
\tikzstyle{end} = [circle, minimum width=3pt,fill, inner sep=0pt]
\begin{figure}[h]
    \centering
    \makebox[\textwidth][c]{\begin{tikzpicture}[grow=down, sloped, edge from parent fork down]
        \node[draw, anchor = north, fill = black!8, very thick] (c1) {\textbf{Term} $\mathbf{s_{\mathsf{Z},0}^-}$}
            child [anchor = north] {
                {node [draw,fill = black!8] (n1) {$n_{\mathsf{Z},\mu_2}^-$}}
                }
            child [anchor = north] {
                {node [draw,fill = black!8] (n2){$n_{\mathsf{Z},\mu_1}^+$}}
            };

            \node[draw, fill = black!8, very thick] (c2) at ([xshift=5cm]c1) {\textbf{Term} $\mathbf{s_{\mathsf{Z},1}^-}$}
            child[anchor = north] {
                node[draw,fill = black!8, very thick] {$\mathbf{s_{\mathsf{Z},0}^+}$}
                child [anchor = north] {
                        {node [draw,fill = black!8] (nxmu2) {$c_{\mathsf{Z},k_\mathrm{min}^\mathsf{Z}}^+$}}
                        }
                child [anchor = north] {
                    {node [draw,fill = black!8] (nxmu1)  {$\delta(N_\mathsf{Z},\epsilon_{\mathsf{Z},0}^{v,+})$}}
                    }
                edge from parent
            }
            child [anchor = north] {
                {node [draw,fill = black!8] (n3) {$n_{\mathsf{Z},\mu_2}^-$}}
                }
            child [anchor = north] {
                    {node [draw,fill = black!8] (n4) {$n_{\mathsf{Z},\mu_1}^+$}}
            };

            \node[draw, fill = black!8, very thick] (c3) at ([xshift=6.5cm]c2) {\textbf{Term} $\mathbf{\Lambda_\mathsf{X}^+ + \gamma}$}
            child[anchor = north] {
                node[draw,fill = black!8, very thick] {$\mathbf{s_{\mathsf{X},1}^-}$}
                child {
                    node[draw,fill = black!8, very thick] {$\mathbf{s_{\mathsf{X},0}^+}$}
                    child{
                        node[draw, fill = black!8] (n5) {$c_{\mathsf{X},k_\mathrm{min}^\mathsf{X}}^+$}
                    }
                    child{
                        node[draw, fill = black!8] (n6) {$\delta(N_\mathsf{X},\epsilon_{\mathsf{X},0}^{v,+})$}
                    }
                }
                child {
                            node[draw, fill = black!8] (n7) {$n_{\mathsf{X},\mu_2}^-$}
                    }
                    child {
                            node[draw, fill = black!8] (n8) {$n_{\mathsf{X},\mu_1}^+$}
                    }
                edge from parent
            }
            child[anchor = north]  {
                node[draw,fill = black!8, very thick] (sx1) {$\mathbf{s_{\mathsf{Z},1}^-}$}
                edge from parent
            }
            child[anchor = north] {
                node[draw,fill = black!8, very thick] {$\mathbf{v_{\mathsf{X},1}^+}$}
                child {
                    node[draw, fill = black!8] (n9) {$c_{\mathsf{X},\mu_1}^+$}
                }
                child {
                    node[draw, fill = black!8] (n10) {$c_{\mathsf{X},\mu_2}^-$}
                }
                edge from parent
            };
            
            \tikzset{point/.style={
            thick,
            draw=black,
            cross out,
            inner sep=0pt,
            minimum width=26pt,
            minimum height=21pt,
            }}
            \node[point] at (sx1) {};

            \tikzset{point/.style={
            thick,
            draw=black,
            cross out,
            inner sep=0pt,
            minimum width=32pt,
            minimum height=21pt,
            }}

            \node[point] at (n3) {};

            \node[point] at (n4) {};

            \node[shift=({0,-0.8})] at (n1) {$\epsilon_{\mathsf{Z},\mu_2}^{n,-}$};
            \node[shift=({0,-0.8})] at (n2) {$\epsilon_{\mathsf{Z},\mu_1}^{n,+}$};
            \node[shift=({0,-0.9})] at (nxmu2) {$\epsilon_{\mathsf{Z},k_\mathrm{min}^\mathsf{Z}}^{c,+}$};
            \node[shift=({0,-0.8})] at (nxmu1) {$\epsilon_{\mathsf{Z},0}^{v,+}$};
            \node[shift=({0,-0.9})] at (n5) {$\epsilon_{\mathsf{X},k_\mathrm{min}^\mathsf{X}}^{c,+}$};
            \node[shift=({0,-0.8})] at (n6) {$\epsilon_{\mathsf{X},0}^{v,+}$};
            \node[shift=({0,-0.8})] at (n7) {$\epsilon_{\mathsf{X},\mu_2}^{n,-}$};
            \node[shift=({0,-0.8})] at (n8) {$\epsilon_{\mathsf{X},\mu_1}^{n,+}$};
            \node[shift=({0,-0.8})] at (n9) {$\epsilon_{\mathsf{X},\mu_1}^{c,+}$};
            \node[shift=({0,-0.8})] at (n10) {$\epsilon_{\mathsf{X},\mu_2}^{c,-}$};
            
        \end{tikzpicture}}
    \caption{Illustration of finite-size effects appearing in the bound on the min-entropy (Eq.~\eqref{eq:H_min_operational_hypothetical}) and their contribution to Eq.~\eqref{eq:def_delta_hi}. The crossed-out boxes correspond to terms that appear twice and are thus only counted once.}
    \label{fig:illustration_contribution_epsilon}
\end{figure}

\section{\label{sec:discussion}Discussion}
This section serves as a general discussion about the 1-decoy and 2-decoy state BB84 protocols as well as Renner's EUR framework and dwelves into insights often omitted from security proofs.

\paragraph{Impact of noise on performance.} One notable difference between the 1-decoy and 2-decoy state protocols, which was not discussed in Ref.~\cite{Rusca18}, is their resilience in regards to noise. In fact, for the 1-decoy state protocol, the upper bound on the number of vacuum events, i.e. Eq.~\eqref{eq:upper_bound_s_x_0}, directly depends on the number of errors in the basis. This means that the advantage of this approach, which avoids sending the vacuum state itself, is best achieved in experiments with low noise. In fact, an increase in noise impacts this protocol twice; first in the single-photon QBER, Eq.~\eqref{eq:upper_bound_qber}, and second in the lower bound on the number of single-photon events, Eq.~\eqref{eq:lower_bound_single}. In contrast, for the 2-decoy state protocol, the number of errors only appears once, namely in the bound for the single-photon QBER, cf. App.~\ref{ap:two_decoy_security_proof}. 

\paragraph{Numerical analysis.} When performing numerical simulations, particular caution should be taken regarding the accepted range for the bounds on the number of photon events and errors as well as the phase error rate. In fact, the bounds should be clipped when exceeding their allowed range. The lower bounds, Eqs.~\eqref{eq:def_n_minus}, \eqref{eq:def_c_minus}, \eqref{eq:lower_bound_vacuum}, \eqref{eq:lower_bound_single}, can, in theory, be negative and should, in this case, be clipped to zero. 
Additionally, in numerical simulations, one usually optimized the parameters from the parameter agreement step to maximize the secure-key rate, cf. Sec.~\ref{sec:long_protocol_description}. For this optimization, the error terms fom Eq.~\eqref{eq:def_delta_hi} are often set to a common value, which significantly reduced the degrees of freedom and results in Eq.~\eqref{eq:l_max_final_equation}. However, in principle, one can use the more general expression, cf. Eq.~\eqref{eq:max_secret_key_formula}, and optimize the full set of error terms to yield higher secure-key rates, albeit increasing the complexity of the numerical optimizations.

\paragraph{EUR framework limitations.} Two important considerations must be kept in mind when applying the EUR in Sec.~\ref{sec:entropic_uncertainty_relation}. First, in order to write down the EUR statement, we require an active basis choice on Alice's side. Second, we assume that Eve's attack strategy does not depend on Alice's and Bob's basis choice to estimate the phase error rate in the $\mathsf{Z}$-basis from the single-photon QBER in the $\mathsf{X}$-basis using statistical sampling. This requires that Alice prepares perfect states, i.e. her basis choice is not leaked to Eve via the state she prepares. This also requires that the detection probabilities are equal on Bob's side. Various implementation designs and device imperfections lead to exploitable asymmetric losses between the bases, which conflicts with this assumption. Examples include detection-efficiency mismatch or dark count rate mismatch. We note that a recent work resolves this long-standing limitation of the EUR framework on the detector side for active basis choices \cite{Tupkary24}.

\paragraph{Reverse error correction.} Finally, we note that reverse error correction, a commonly used post-processing variant where Alice corrects her key to match Bob's, is not straightforward to incorporate in this security proof framework. Indeed, Eve can influence Bob's detection outcomes and thus directly affect the sifted key if reverse error correction is used. Throughout the analysis, the number of $m$-photon events refers to the photon number distribution leaving Alice's system, cf. Sec.~\ref{sec:photon_event_statistics_and_bounds}. Now, if reverse error correction is used, this discussion does not hold anymore. For example, in this context it is not possible to assume that Eve has no information about the vacuum events, which was used in Sec.~\ref{sec:chain_rule_smooth_min_entropies}, as she may influence Bob's detections and Alice corrects her key to match Bob's. Even if the vacuum events are assumed to be unsafe, and thus the vacuum term left out, the use of the EUR is not directly justified in this context. In contrast, with forward error correction, the verified key is effectively Alice's sifted key, as Bob corrects his key to match hers, simplifying the theoretical analysis as Eve cannot influence Alice's state preparation under the assumptions listed in App.~\ref{ap:assumptions}. Therefore, we advise practical implementations based on this framework (which also include the works \cite{Tomamichel12, LimCurtyWalenta14, Rusca18}) to employ forward error correction.

\section{Conclusion}
In this work, we consolidate the security proof for the 1-decoy and 2-decoy state BB84 protocol by formulating it in a rigorous yet accessible manner in Renner's entropic uncertainty relation framework. By addressing technical inconsistencies and unclarities in existing works, we refine the proof with a unified language, rigorously handling the conditioning on states and thoroughly discussing the error terms. Additionally, we provide a rigorous treatment for fixed-length protocols, an aspect overlooked in previous analyses \cite{ LimCurtyWalenta14, Rusca18}. An important distinction, previously unaddressed, concerns the 1-decoy state protocol's acceptance test, which is performed after error correction. We rigorously discuss and resolve these issues in our analysis. By providing a constructive approach, beginning with the general definition of security, and, bounding each term separately, we offer a clear outline and step-by-step framework for the proof. We unify the ideas developed in various works to form a robust reference for practical implementations of the protocol, and a basis for further discussions of the underlying assumptions and potential vulnerabilities. 

While this security proof serves as a solid starting point for any practical implementation wishing to use the decoy-state BB84 protocol, it is important to note that side-channel attacks, resulting from device imperfections, are not modeled in the security proof. Prominent attacks such as the Trojan-horse attack \cite{Lucamarini15, Wang18} or detector-efficiency mismatch attack \cite{Zhang21, Tupkary24} can be incorporated into the security proof by relaxing the assumptions about the devices. On the hardware side, QKD systems often either require adjustments or additional components, such as optical isolators and filters, that need to be precisely characterized\footnote{Alternatively, one can opt for a different protocol, such as measurement-device-independent QKD \cite{Lo12} or device-independent QKD \cite{Pironio09}.}. A recent study by the German Federal Office for Information Security (BSI) \cite{BSI24} lists most known side-channel attacks and their countermeasures. Unfortunately, these attacks are often neglected \cite{Sajeed21, Makarov23} and thus compromise the security claims. Incorporating the most critical side-channel attacks into a complete security proof would form the natural next step towards bridging the gap between theory and experiment. While significant effort has been devoted to combining different countermeasures, developing a security proof encompassing most of today's known attacks is far from trivial and remains an open task. 

\section*{Acknowledgements}

We would like to thank Shlok Nahar, Víctor Zapatero, Davide Orsucci, Jonas Treplin and Benedikt Leible for valuable comments on the first version of the manuscript. We thank Ernest Y.-Z. Tan for pointing out that reverse error correction may not be trivial to apply in this framework. We also thank Andy Schreier and Hermann Kampermann for engaging in fruitful discussions.  For semantical polishing of the introduction and conclusion, large language models were used. This research was conducted within the scope of the QuNET-initiative, funded by the German Federal Ministry of Education and Research (BMBF). D.R. thanks the Galician Regional Government (consolidation of Research Units: AtlantTIC), MICIN with funding from the European Union NextGenerationEU (PRTR-C17.I1) and the Galician Regional Government with own funding through the “Planes Complementarios de I+D+I con las Comunidades Autónomas” in Quantum Communication and The European Union’s Horizon Europe Framework Programme under the project “Quantum Security Networks Partnership” (QSNP, grant agreement No 101114043). This work was also funded by the NSERC Discovery Grant, and was partially conducted at the Institute for Quantum Computing, University of Waterloo, which is funded by Government of Canada through ISED. D.T. is partially funded by the Mike and Ophelia Lazaridis Fellowship.

\section*{Author contributions}
J.K. and N.W. proposed and supervised the research.
J.W. and J.K. performed the research for the first version of the manuscript.
J.W. wrote the first draft of the manuscript.
J.W. and J.K. edited the manuscript.
J.W., J.K., D.R. and N.W. discussed and reviewed the first version of the manuscript. N.L. and D.T. were brought onto the project during the second version of the manuscript to address subtleties and enhance its rigor. J.W. and D.T. led the editing of the second version, implementing a series of improvements that clarified and strengthened the arguments. All authors reviewed and approved the updated manuscript for finalization.

\section*{Software availability}
\added{
The code used to generate Fig.~\ref{fig:skr_plot} is openly available at \cite{Wiesemann_QKD_simulation_2025}.
}

\addcontentsline{toc}{section}{List of symbols}
\nomenclature{$k \in \{\mu_1, \mu_2, \mu_3 \}$}{Levels of intensities used to transmit photons, where $mu_3$ is only used for the 2-decoy state protocol}
\nomenclature{$Q$}{Acceptance set}
\nomenclature{$S_A$, $S_B$}{Alice's and Bob's secure keys}
\nomenclature{$p_b^A, p_b^A$}{Probability for Alice and Bob to choose a basis $b$}
\nomenclature{$p_k$}{Probability for Alice to choose an intensity $k$}
\nomenclature{$\epsilon_\mathrm{sec}'$}{Parameter similarly defined to the secrecy parameter but additionally conditioned on $\Omega_\mathrm{EC}$}
\nomenclature{$\epsilon_\mathrm{sec}$}{Secrecy parameter}
\nomenclature{$\epsilon_\mathrm{cor}$}{Correctness parameter}
\nomenclature{$\epsilon$}{Security parameter}
\nomenclature{$\mathrm{leak_{EC}}$}{Number of bits leaked during error correction}
\nomenclature{$N$}{Total number of signals sent by Alice during a protocol run}
\nomenclature{$s_{\mathsf{Z},0}^\mathrm{l}, s_{\mathsf{Z},1}^\mathrm{l}, s_{\mathsf{X},0}^\mathrm{l}, \Lambda_\mathsf{X}^\mathrm{u}$}{Acceptance set parameters for the decoy bounds}
\nomenclature{$s_{\mathsf{Z},0}^-, s_{\mathsf{Z},1}^-, s_{\mathsf{X},0}^-, \Lambda_\mathsf{X}^+$}{Decoy bounds on the number of photon events and the single-photon QBER}
\nomenclature{$N_\mathrm{X}, N_\mathrm{Z}$}{Number of bits used for acceptance testing and key generation}
\nomenclature{$l$}{Secure-key length}
\nomenclature{$\tilde{Z}_A, \tilde{Z}_B$}{Alice and Bob's sifted keys}
\nomenclature{$Z_A, Z_B$}{Alice and Bob's corrected keys, $\tilde{Z}_A = Z_A$}
\nomenclature{$\tilde{X}_A, \tilde{X}_B$}{Alice and Bob's sifted bit strings in the monitoring $\mathsf{X}$-basis}
\nomenclature{$\Omega_\mathrm{EC}, \Omega_\mathrm{EV}, \Omega_\mathrm{AT}, \Omega_\mathrm{B}$}{Events describing the error correction succeeding, error verification passing, acceptance test passing and decoy bounds holding}
\nomenclature{$\Omega_\top$}{Event describing the protocol not abort, i.e. $\Omega_\top = \Omega_\mathrm{EV} \land \Omega_\mathrm{AT}$}
\nomenclature{$\tilde{\Omega}$}{Defined as $\tilde{\Omega} = \Omega_\top \land \Omega_\mathrm{EC}$}
\nomenclature{$\Omega_\circ$}{Virtual event where the true values of the statistics are in the acceptance set bounds}
\nomenclature{$\rho_A$}{Density operator representing a quantum state  $A$}
\nomenclature{$\rho_{AB}$}{Bipartite system of quantum systems $A$ and $B$}
\nomenclature{$\rho_{A|\Omega}$}{Density operator describing of a quantum system $A$ conditioned on an event $\Omega$}
\nomenclature{$U_S$, $U_Y$}{Density operator representation of the uniform distribution of all possible keys $S$ and hash functions $Y$}
\nomenclature{$\norm{\cdot}_1$}{Trace norm}
\nomenclature{$D(\rho_A, \rho_B)$}{Trace distance between $\rho_A$ and $\rho_B$}
\nomenclature{$A$, $B$, $E$}{Alice's, Bob's and Eve's quantum systems}
\nomenclature{$\mathsf{X}, \mathsf{Z}$}{Basis in which the qubits are transmitted}
\nomenclature{$d_\mathrm{sec}(\cdot)$}{Distance to an ideal protocol}
\nomenclature{$Z$}{Describes $Z_A$ and $Z_B$ interchangeably when conditioning on $\Omega_\mathrm{EC}$}
\nomenclature{$S$}{Describes $S_A$ and $S_B$ interchangeably when conditioning on $\Omega_\mathrm{EC}$}
\nomenclature{$C$}{Classical register storing the classical communications}
\nomenclature{$\tilde{C}^N, C_\mathrm{EC}, C_\mathrm{EV}, C_\mathrm{AT}, C_\mathrm{PA}$}{Various classical registers listed in Table~\ref{tab:classical_registers_comm_description}}
\nomenclature{$f_y(\cdot)$}{Hash function taken from a universal$_2$ family of hash functions $\{f_y\}_y$}
\nomenclature{$H_\text{min}(\rho)$}{Min-entropy of $\rho$}
\nomenclature{$H_\text{min}(Z\vert E), H_\text{min}^\zeta(Z\vert E)$}{Min-entropy and smooth min-entropy of $Z$ given side information $E$}
\nomenclature{$\alpha_i$}{Smoothing parameters for min-entropies}
\nomenclature{$\nu^2$}{Probability for Serfling's inequality not holding}
\nomenclature{$\Delta_\text{ci}$}{Upper bound on the probability that one of the decoy concentration inequalities fails}
\nomenclature{$h(\cdot)$}{Truncated binary entropy function}
\nomenclature{$c_b$}{Total number of errors in the basis $b$}
\nomenclature{$s_{b,m}$}{Number of detections of $m$-photon events in the basis $b$}
\nomenclature{$v_{b,m}$}{Number of errors associated with $s_{b,m}$}
\nomenclature{$\Lambda_\mathsf{X}$}{Single photon QBER in the $\mathsf{X}$ basis}
\nomenclature{$\Lambda_\mathsf{Z}$}{Phase error rate in the $\mathsf{Z}$ basis}
\nomenclature{$\zeta$}{Smoothing parameter for min-entropies, $\zeta = 2\alpha_1 + \alpha_2 + \alpha_3$}
\nomenclature{$P(\rho_A, \rho_B)$}{Purified distance between $\rho_A$ and $\rho_B$}
\nomenclature{$\mathcal{B}^\zeta(\rho)$}{$\zeta$-ball around $\rho$}
\nomenclature{$\delta(\cdot, \cdot)$}{Hoeffding-delta}
\nomenclature{$k$}{As a subscript, denotes an intensity}
\nomenclature{$m$}{As a subscript, denotes a number of photons}
\nomenclature{$\gamma$}{Deviation term for the phase error rate, following from Serfling's inequality}

\printnomenclature

\appendix

\section{Appendix}
\subsection{Model assumptions}
\label{ap:assumptions}

This section provides a list of assumptions that have been used in the security proof. References to further readings are provided for most items. We recall the fundamental assumptions of quantum key distribution \cite[Sec.~IV]{Portmann22}:
\begin{enumerate}
    \item Quantum physics is a correct and complete theory.
    \item The classical channel used by Alice and Bob is authenticated.
    \item The devices used during the protocol behave in a fully predictable and controllable manner.
\end{enumerate}
Following the last point, we assume that no device imperfections and thus no side-channels are present \cite{Sajeed21, Makarov23, BSI24}. Specifically, we make the following adversarial and model assumptions:
\begin{enumerate}
    \item Eve has only access to the classical and quantum channel, but not to the sender and receiver devices, nor to their environment. She may perform any attack allowed by quantum mechanics, i.e. including coherent attacks.
    \item The transmitted signals are uniformly phase-randomized and coherent \cite{Lo05b}. Eve has no a priori information about Alice's intensity choice. Thus, we can write Alice's signals as a coherent superposition of Fock states \cite{Lo07}
    \begin{equation}
        \rho_k = \sum_{m=0}^\infty e^{-k}\frac{k^m}{m!}\ketbra{m}{m}\,.
    \end{equation}
    \item The $\mathsf{X}$ and $\mathsf{Z}$ bases are perfectly diagonal, i.e. the quality factor is $q = 1$, cf. \cite{Tomamichel12}.
    \item Eve's attack strategy does not depend on Alice and Bob's basis choices. Eve has no a priori knowledge about the basis choices. This implies basis-independent losses, i.e. no detection-efficiency mismatch nor dark-count rate mismatch \cite{Tupkary24}.
    \item Alice has access to a true randomness source (e.g. when choosing the error verification and privacy amplification hash functions or for state encoding).
\end{enumerate}

\subsection{Bounds for the 2-decoy state protocol}
\label{ap:two_decoy_security_proof}
In the case where two decoy states are used, also known as \textit{2-decoy state protocol}, Alice can choose between three intensities when preparing her states, namely $\mu_1$, $\mu_2$ and $\mu_3$, for which $\mu_1 > \mu_2 + \mu_3$ and $\mu_2 > \mu_3 \geq 0$\cite{Lim14, LimCurtyWalenta14}. The security proof is done analogously to the 1-decoy security proof with the exception that the bounds on photon events and errors differ, leading to different terms for $\Delta_\mathrm{ci}$ in Eq.~\eqref{eq:def_delta_hi}. In this section, we will first derive the relevant bounds and finally determine $\Delta_\mathrm{ci}$ in the case of two decoy states. The derivation of the bounds is based on Ref.~\cite[Sec.~3.1]{Lim14}.
\subsubsection{Lower bound for the vacuum events}
In full analogy to Sec.~\ref{sec:lower_bound_vacuum}, by replacing $\mu_1$ by $\mu_2$ and $\mu_2$ by $\mu_3$, we can derive a lower bound for the number of vacuum event
\begin{equation}
    \frac{\mu_2 e^{\mu_3}n_{\mathsf{Z},\mu_3}^*}{p_{\mu_3}} - \frac{\mu_3 e^{\mu_2} n_{\mathsf{Z},\mu_2}^*}{p_{\mu_2}} = \frac{(\mu_2 - \mu_3)s_{\mathsf{Z},0}}{\tau_0} - \mu_2\mu_3\sum_{m=2}^\infty \frac{(\mu_2^{m-1}-\mu_3^{m-1})s_{\mathsf{Z},m}}{\tau_m m!}\, ,
\end{equation}
yielding
\begin{empheq}[box=\fcolorbox{black}{equation_box}]{align}
    \Pr\left[s_{\mathsf{Z},0} < s_{\mathsf{Z},0}^-\right] \leq \epsilon_{\mathsf{Z}, \mu_3}^{n,-}+ \epsilon_{\mathsf{Z}, \mu_2}^{n,+}\, ,
\end{empheq}
where 
\begin{equation}
     s_{\mathsf{Z},0}^- \coloneqq \frac{\tau_0}{(\mu_2 - \mu_3)}\left( \frac{\mu_2 e^{\mu_3} n_{\mathsf{Z},\mu_3}^-}{p_{\mu_3}} - \frac{\mu_3 e^{\mu_2} n_{\mathsf{Z},\mu_2}^+}{p_{\mu_2}} \right)\,.
\end{equation}

\subsubsection{Lower bound for the vacuum events}
Analogously to Sec.~\ref{sec:lower_bound_single}, we can write
\begin{align*}
    \frac{e^{\mu_2} n_{\mathsf{Z},\mu_2}^*}{p_{\mu_2}} - \frac{e^{\mu_3} n_{\mathsf{Z},\mu_3}^*}{p_{\mu_3}} &= \frac{(\mu_2 - \mu_3)s_{\mathsf{Z},1}}{\tau_1} + \sum_{m=2}^\infty \frac{(\mu_2^m - \mu_3^m)s_{\mathsf{Z},m}}{\tau_m m!} \\
    &\leq \frac{(\mu_2 - \mu_3)s_{\mathsf{Z},1}}{\tau_1} + \frac{\mu_2^2 - \mu_3^2}{\mu_1^2}\sum_{m=2}^\infty \frac{\mu_1^m s_{\mathsf{Z},m}}{\tau_m m!}
\end{align*}
as we can write
\begin{equation*}
    \mu_2^m - \mu_3^m = \frac{\mu_2^2 - \mu_3^2}{\mu_2 + \mu_3}\sum_{i=0}^{m-1}\mu_2^{m-i-1}\mu_3^{i} \leq (\mu_2^2-\mu_3^2)(\mu_2 + \mu_3)^{m-2}\leq (\mu_2^2-\mu_3^2)\mu_1^{m-2}\, ,
\end{equation*}
for $m\geq 2$ where the last inequality holds for $\mu_2 + \mu_3 \leq \mu_1$ and $\sum_{i=0}^{m-1}\mu_2^{m-i-1}\mu_3^{i}\leq(\mu_2+\mu_3)^{m-1}$ for $m\geq 2$ has been used, which directly follows from the binomial theorem. This can easily be proven by induction. Now, writing the sum of multi-photon events ($m\geq 2$) as 
\begin{equation}
    \sum_{m=2}^\infty \frac{\mu_1^m s_{\mathsf{Z},m}}{\tau_m m!} = \sum_{m=2}^\infty \frac{e^{\mu_1} e^{-\mu_1}p_{\mu_1} \mu_1^m s_{\mathsf{Z},m}}{p_{\mu_1}\tau_m m! } = \frac{e^{\mu_1}n_{\mathsf{Z},\mu_1}^*}{p_{\mu_1}} - \frac{s_{\mathsf{Z},0}}{\tau_0} - \frac{\mu_1 s_{\mathsf{Z},1}}{\tau_1}
\end{equation}
by using Eqs.~\eqref{eq:bayes_rule} and \eqref{eq:def_n_star} yields
\begin{equation}
    \frac{e^{\mu_2} n_{\mathsf{Z},\mu_2}^*}{p_{\mu_2}} - \frac{e^{\mu_3} n_{\mathsf{Z},\mu_3}^*}{p_{\mu_3}} \leq \frac{(\mu_2 - \mu_3)s_{\mathsf{Z},1}}{\tau_1} + \frac{\mu_2^2 - \mu_3^2}{\mu_1^2}\left( \frac{e^{\mu_1}n_{\mathsf{Z},\mu_1}^*}{p_{\mu_1}} - \frac{s_{\mathsf{Z},0}}{\tau_0} - \frac{\mu_1 s_{\mathsf{Z},1}}{\tau_1} \right)\, .
\end{equation}
Finally, solving for $s_{\mathsf{Z},1}$ yields
\begin{equation}
    s_{\mathsf{Z},1} \geq \frac{\mu_1 \tau_1}{\mu_1(\mu_2 - \mu_3)-(\mu_2^2-\mu_3^2)}\left( \frac{e^{\mu_2} n_{\mathsf{Z},\mu_2}^*}{p_{\mu_2}} - \frac{e^{\mu_3}n_{\mathsf{Z},\mu_3}^*}{p_{\mu_3}} + \frac{\mu_2^2-\mu_3^2}{\mu_1^2}\left(\frac{s_{\mathsf{Z},0}}{\tau_0}-\frac{e^{\mu_1}n_{\mathsf{Z},\mu_1}^*}{p_{\mu_1}}\right) \right)
\end{equation}
and using $\mu_1(\mu_2-\mu_3)-(\mu_2^2-\mu_3^2) = (\mu_1 - \mu_3 + \mu_2)(\mu_2-\mu_3) > 0$ we find
\begin{empheq}[box=\fcolorbox{black}{equation_box}]{align}
\label{eq:2decoy_s1}
    \Pr\left[s_{\mathsf{Z},1} < s_{\mathsf{Z},1}^-\right] \leq \epsilon_{\mathsf{Z}, \mu_2}^{n,-}+ \epsilon_{\mathsf{Z}, \mu_3}^{n,+} + \epsilon_{\mathsf{Z}, \mu_1}^{n,+}+ \epsilon_{\mathsf{Z}, \mu_3}^{n,-}+ \epsilon_{\mathsf{Z}, \mu_2}^{n,+} \, ,
\end{empheq}
where
\begin{equation}
    s_{\mathsf{Z},1}^- \coloneqq \frac{\mu_1 \tau_1}{\mu_1(\mu_2 - \mu_3)-(\mu_2^2-\mu_3^2)}\left( \frac{e^{\mu_2} n_{\mathsf{Z},\mu_2}^-}{p_{\mu_2}} - \frac{e^{\mu_3}n_{\mathsf{Z},\mu_3}^+}{p_{\mu_3}} + \frac{\mu_2^2-\mu_3^2}{\mu_1^2}\left(\frac{s_{\mathsf{Z},0}^-}{\tau_0}-\frac{e^{\mu_1}n_{\mathsf{Z},\mu_1}^+}{p_{\mu_1}}\right) \right) \,.
\end{equation}

\subsubsection{Upper bound on the number of single-photon events}
In complete analogy to Sec.~\ref{sec:upper_bound_single_photon_errors}, we can derive an upper bound on the number of single-photon events by replacing $\mu_1\rightarrow\mu_2$ and $\mu_2\rightarrow\mu_3$, leading to
\begin{empheq}[box=\fcolorbox{black}{equation_box}]{align}
    \Pr\left[v_{\mathsf{X},1} > v_{\mathsf{X},1}^+\right] \leq \epsilon_{\mathsf{X}, \mu_2}^{c,+}+ \epsilon_{\mathsf{X}, \mu_3}^{c,-}\,,
\end{empheq}
where
\begin{equation}
    v_{\mathsf{X},1}^+ \coloneqq \frac{\tau_1}{\mu_2 - \mu_3}\left( \frac{e^{\mu_2}c_{\mathsf{X},\mu_2}^+}{p_{\mu_2}} - \frac{e^{\mu_3}c_{\mathsf{X},\mu_3}^-}{p_{\mu_3}}\right) \,.
\end{equation}

\subsubsection{Security analysis}
As discussed above, the security proof of the 2-decoy protocol solely differs from that of the 1-decoy regarding the bounds on the photon events and errors. As such, for the 2-decoy state protocol, the protocol description from Fig.~\ref{fig:protocol_description} can be used with an additional intensity $\mu_3$ and where the bounds derived above are used for the acceptance test. The proof can be performed analogously to the 1-decoy state protocol yielding
\begin{equation}
   \Pr[\tilde{\Omega}] d_\mathrm{sec}(SEC)_{\rho|\tilde{\Omega}} \leq 2(\alpha_2 + \nu) + \Delta_\mathrm{pa}^\zeta + \Delta_\mathrm{ci} \leq \epsilon_\mathrm{sec}'\, ,
\end{equation}
where we adjust the concentration inequality term using the bounds derived above,
\begin{align}
    \Delta_\mathrm{ci} &= \epsilon_{\mathsf{Z}, \mu_2}^{n,-}+ \epsilon_{\mathsf{Z}, \mu_3}^{n,+} + \epsilon_{\mathsf{Z}, \mu_1}^{n,+}+ \epsilon_{\mathsf{Z}, \mu_3}^{n,-}+ \epsilon_{\mathsf{Z}, \mu_2}^{n,+} + \epsilon_{\mathsf{X}, \mu_2}^{n,-}+ \epsilon_{\mathsf{X}, \mu_3}^{n,+} \nonumber \\&+ \epsilon_{\mathsf{X}, \mu_1}^{n,+}+ \epsilon_{\mathsf{X}, \mu_3}^{n,-}+ \epsilon_{\mathsf{X}, \mu_2}^{n,+} + \epsilon_{\mathsf{X}, \mu_2}^{c,+}+ \epsilon_{\mathsf{X}, \mu_3}^{c,-}\, .
    \label{eq:sum_epsilons_two_decoy}
\end{align}
Now, in analogy to Sec.~\ref{sec:secret_key_length}, using Eq.~\eqref{eq:sum_epsilons_two_decoy} and setting all error terms to a common value yields an expression for the maximum extractable secure-key length in terms of $\epsilon_\mathrm{cor}$ and $\epsilon_\mathrm{sec}'$,
\begin{empheq}[box=\fcolorbox{black}{equation_box}]{align}
    l = s_{\mathsf{Z},0}^\mathrm{l} + s_{\mathsf{Z}, 1}^\mathrm{l}(1 - h(\Lambda_\mathsf{X}^\mathrm{u} + \gamma)) - \mathrm{leak}_{\mathrm{EC}} - \log_2{\frac{2}{\epsilon_\mathrm{cor}}} - 4\log_2{\frac{17}{\epsilon_\mathrm{sec}'\sqrt[4]{2}}}\, .
\end{empheq}

\subsection{Intuition on the trace distance}
\label{ap:intuition_trace_norm}
Given two probability distributions $P_Z$ and $Q_Z$, the probability of an event $z \in \mathcal{Z}$ occurring is given by $P_Z(z)$ and $Q_Z(z)$ respectively. The goal is to distinguish both probability distributions by correctly assigning a random sample $s\in\{z\}$ to the probability distribution it originated from. For this purpose, we introduce the \textit{total variation distance} (called trace distance when generalized to density operators), defined as
\begin{equation}
    TV(P_Z, Q_Z) \coloneqq \frac{1}{2}\sum_z |P_Z(z) - Q_Z(z)| \le 1\, .
\end{equation} 
The definition above can be rewritten as (see Ref.~\cite[Eq.~(9.3)]{ChuangNielsen10})
\begin{equation}
    TV(P_Z, Q_Z) = \max_\Omega \left| \sum_{z\in \Omega} P_Z(z) - \sum_{z\in \Omega} Q_Z(z) \right|\, ,
\end{equation}
where the maximization is taken over all subsets $\Omega\subseteq \{x\}$. In other words, the $\Omega$ for which the difference is maximized is the optimal event to consider when one wants to distinguish $P_Z$ and $Q_Z$ as it describes the event for which the probabilities of both probability distributions differ the most.

If the total variation distance is upper-bounded by $\zeta$, i.e. $TV(P_Z, Q_Z) \leq \zeta$, it can be shown that the probability of assigning a sample $s$ to the correct probability distribution is upper-bounded by (see Refs.~\cite[Sec.~9.1.4]{Wilde16}, \cite[Sec.~2.4.2]{Tsybakov2009} and \cite[Sec.~3.2.1]{TomamichelThesis12})
\begin{equation}
    P_\mathrm{distinguish}(P_Z, Q_Z) \leq \frac{1}{2} + \frac{\zeta}{2}\,.
\end{equation}
In other words, $\zeta$ describes the \textit{distinguishing advantage} when trying to distinguish two probability distributions. In the special case where $\zeta=0$, and thus both probability distributions yield equal results, one can only randomly assign $s$ with a success rate of 50\,\%. In this case, both probability distributions are indistinguishable.

This concept also applies when density operators $\rho$ and $\sigma$ are considered instead of classical probability distributions. The generalization of the total variation distance to quantum states is discussed in Refs.~\cite[Sec.~9.2]{ChuangNielsen10} and \cite[Sec.~3.2]{TomamichelThesis12}.

\subsection{Comparison between the von Neumann entropy and min-entropy}
\label{ap:comparison_neumann_min_entropy}
As a more conservative measure of entropy than the von Neumann entropy, the min-entropy outputs the entropy in the case where Eve is the most likely to guess the secure key. As an example, let $n$ be the number of possible bit string combinations, $n = \dim{\mathcal{S}}$. Now, assuming that the $j$-th combination occurs $50\%$ of the time and that the remaining $n-1$ strings are uniformly distributed with $p_i = \frac{1/2}{n-1}$, $i\neq j$. The von Neumann entropy yields
\begin{align*}
    S(\rho) &= -\Tr{\rho \log_2 \rho}\\
    &= -\sum_{i=1}^{n} p_i \log_2{p_i}\\
    &= -\frac{1}{2} \log_2 \frac{1}{2} -\sum_{i=1, i\neq j} \frac{1}{2(n-1)} \log_2{\frac{1}{2(n-1)}} \\
    &= \frac{1}{2} \log_2{(2(n-1))} + \frac{1}{2}\, .
\end{align*}
For $n=2^{64}$, i.e. a $64$-bit string, we find $S(\rho) \approx 33$.
In contrast, the min-entropy Eq.~\eqref{eq:min_entropy} yields
\begin{equation*}
    H_\mathrm{min}(\rho)=-\log_2{p_\mathrm{guess}} = 1\, ,
\end{equation*}
independently of $n$, where $p_\mathrm{guess} = \frac{1}{2}$. The von Neumann entropy describes the average uncertainty of a system equivalently to the Shannon entropy, while the min-entropy solely assumes the outcome with highest probability, i.e. the case where the observer has the smallest uncertainty about the system and thus accounts for a more conservative entropy measure, which is assumed in the realm of quantum cryptography.

\subsection{Derivation of the expression for the Hoeffding-delta}
\label{ap:hoeffding_delta}
Hoeffding's inequality \cite[Eq.~(2.1)]{Hoeffding63} states that for independent and identically distributed random variables $V_1, \ldots, V_n$, such that $0\leq V_i \leq 1$, 
\begin{equation}
    P\left(\frac{1}{n}\sum_{i=1}^n V_i - E[V] > t\right) \le \exp(-2nt^2)\, ,
\end{equation}
where $E[V]$ is the common expectation value of each variables $V_i$ and $t > 0$. In other words, Hoeffding's inequality states that, for a finite number of events $n$, the measured mean $\frac{1}{n}\sum_{i=1}^n V_i$ deviates by more than $t$ from the common expectation value with a probability no more than $\exp(-2nt^2)$. The expectation value is common to all variables $V_i$ as they are independent and identically distributed. The expectation value for the sum of all random variables is then $n E[V]$, where $n$ is the number of events considered. Now, if we choose $t = \tilde{\delta}(n, \epsilon) \coloneqq \sqrt{\frac{1}{2n}\ln\frac{1}{\epsilon}}$, then Hoeffding's inequality holds since
\begin{align}
    P\left(\frac{1}{n}\sum_{i=1}^n V_i - E[V] > \tilde\delta(n, \epsilon)\right) &\leq \exp(-2n\tilde{\delta}(n, \epsilon)^2) \\
    &= \exp(-2n\frac{1}{2n}\ln\frac{1}{\epsilon}) \\
    &= \epsilon\, .
\end{align}
Note, however, that we are considering the expectation value of the sum of all variables and not the expectation value of each variable in Sec.~\ref{sec:photon_event_statistics_and_bounds}. As the expectation value is common to all variables, we can use Ref.~\cite[Eq.~(1.3)]{Hoeffding63}, effectively multiplying each term with $n$, yielding
\begin{equation}
    P\left(\sum_{i=1}^n V_i - n E[V] > n \tilde{\delta}(n, \epsilon)\right) \leq \epsilon\, .
\end{equation}
We can then define a new Hoeffding-delta
\begin{equation}
    \delta(n, \epsilon) \coloneqq n \tilde{\delta}(n, \epsilon) = \sqrt{\frac{n}{2}\ln\frac{1}{\epsilon}} \, ,
\end{equation}
which fulfills the above equation. This choice of $\delta(n, \epsilon)$ is precisely the one introduced in Sec.~\ref{sec:photon_event_statistics_and_bounds}. Note that from Ref.~\cite[Eq.~(1.4)]{Hoeffding63}, using the symmetry of the bounds, we can state that
\begin{equation}
    P\left(-\sum_{i=1}^n V_i + n E[V] > \delta(n, \epsilon)\right) \leq \epsilon
\end{equation}
As an example, this corresponds to the cases
\begin{equation}
    P\left(n_{\mathsf{Z},\mu_1} - n_{\mathsf{Z},\mu_1}^* > \delta(N_\mathsf{Z}, \epsilon)\right) \leq \epsilon
\end{equation}
and
\begin{equation}
    P\left( n_{\mathsf{Z},\mu_1}^* - n_{\mathsf{Z},\mu_1} > \delta(N_\mathsf{Z}, \epsilon)\right) \leq \epsilon
\end{equation}
or, equivalently,
\begin{equation}
    P\left( \left| n_{\mathsf{Z},\mu_1}^* - n_{\mathsf{Z},\mu_1} \right| > \delta(N_\mathsf{Z}, \epsilon)\right) \leq 2\epsilon \,.
\end{equation}

\subsection{Scenario where Alice chooses the intensity after detection}
\label{sec:scenario_alice_intensity_before_detection}
If Alice sends $m$ photons, then Eve's and Bob's detections are independent of the intensity $\mu_1$ or $\mu_2$ the $m$-photon event originates from.
To illustrate this, consider the probability tree diagram depicted in Fig.~\ref{fig:prob_tree_trace_detections}. Let $p_e(m)$ represent Eve's influence on Bob's photon detections. Intuitively, as Eve cannot base her attack strategy on the intensity chosen, this term only depends on the number of photons sent $m$. Note that this is a simplification as $p_e(m)$ can also depend on other variables that are however not relevant for this discussion.

\tikzstyle{level 1}=[level distance=5cm, sibling distance=4cm]
\tikzstyle{level 2}=[level distance=5cm, sibling distance=1.5cm]

\tikzstyle{bag} = [text width=8em, text centered]
\tikzstyle{end} = [circle, minimum width=3pt,fill, inner sep=0pt]
\begin{figure}[H]
    \centering
    \begin{tikzpicture}[grow=right, sloped]
        \node[draw, fill=black!8] {Choosing the intensity $k$}
            child {
                node[draw, fill=black!8] {$k = \mu_2$}
                child {
                        node[end, label=right:
                        {$\ldots$}] {}
                        edge from parent
                        node[above] {$p_{m|\mu_2} p_e(m)$}
                    }
                    child {
                        node[end, label=right:
                        {$m = 1$}] {}
                        edge from parent
                        node[above] {$p_{1|\mu_2} p_e(m=1)$}
                    }
                    child {
                        node[end, label=right:
                        {$m=0$}] {}
                        edge from parent
                        node[above] {$p_{0|\mu_2} p_e(m=0)$}
                    }
                edge from parent
                node[above] {$p_{\mu_2}$}
            }
            child {
                node[draw, fill=black!8] {$k= \mu_1$}
                child {
                    node[end, label=right:
                    {$\ldots$}] {}
                    edge from parent
                    node[above] {$p_{m|\mu_1} p_e(m)$}
                }
                child {
                    node[end, label=right:
                    {$m = 1$}] {}
                    edge from parent
                    node[above] {$p_{1|\mu_1} p_e(m=1)$}
                }
                child {
                    node[end, label=right:
                    {$m=0$}] {}
                    edge from parent
                    node[above] {$p_{0|\mu_1} p_e(m=0)$}
                }
                edge from parent
                node[above] {$p_{\mu_1}$}
            };
        \end{tikzpicture}
    \caption{Probability tree diagram representing the different paths leading to an $m$-photon detection on Bob's side given the intensities $\mu_1$ and $\mu_2$. Here, $p_e(m)$ represents Eve's influence on Bob's detections due to her attack strategy.}
    \label{fig:prob_tree_trace_detections}
    \end{figure}

\noindent In the case where Eve cannot base her attack strategy on the intensity chosen, the probability of the intensity $\mu_1$ being chosen given that Bob detects $m$ photons is given by using Bayes' theorem, cf. Eq.~\eqref{eq:bayes_rule},
\begin{align}
    p_{\mu_1|m}&=\frac{p_e(m)p_{\mu_1} p_{m|\mu_1}}{p_e(m)p_{\mu_1} p_{m|\mu_1} + p_e(m)p_{\mu_2} p_{m|\mu_2}} \\
    &= \frac{p_{\mu_1} p_{m|\mu_1}}{p_{\mu_1} p_{m|\mu_1} + p_{\mu_2} p_{m|\mu_2}}\, .
\end{align}
Intuitively, this represents the fraction of the paths originating from an intensity $\mu_1$, given that $m$ photons where detected, compared to all paths leading to the detection of an $m$-photon event. 

This shows that assuming that Alice chooses the intensity $k$ after an $m$-photon event was detected is mathematically equivalent to choosing the intensity before measuring an $m$-photon event. Now, if we were to assume that Eve could base her attack strategy on the level of intensity $k$ chosen, then her influence $p_e(m, k)$ would depend on the level of intensity $k$ in additional to $m$, yielding a new expression for $p_{\mu_1|m}$, namely
\begin{equation}
    p_{\mu_1|m}=\frac{p_e(m, \mu_1)p_{\mu_1} p_{m|\mu_1}}{p_e(m, \mu_1)p_{\mu_1} p_{m|\mu_1} + p_e(m, \mu_2)p_{\mu_2} p_{m|\mu_2}}\, .
\end{equation}
Here, the $p_e(m,k)$ terms do not cancel out and we could not determine $p_{k|m}$ solely as a function of $p_{m|k}$ and $p_k$ using Bayes' theorem effectively meaning that both scenarios would not be equivalent.

\subsection{Asymptotic secure-key rate}
\label{ap:asymptotic_key_rate}
\added{An expression for the asymptotic secure-key rate can be found based on the finite-size case. In the following, we use the simplified finite-size expression, cf. Eq.~\eqref{eq:l_max_final_equation}, but the same argument holds for Eq.~\eqref{eq:max_secret_key_formula}. For a given channel and detection model, we denote the probability, each round, to detect a click in a basis $a$ (which is not discarded) with intensity choice $k$ as $p_{a, k}$. Similarly, we denote the probability for an error in a basis $a$ with intensity choice $k$ as $e_{a, k}$. As we are considering the asymptotic case, and therefore directly dealing with probabilities, concentration inequalities are not needed and the deviation terms from Eqs.~\eqref{eq:def_n_Z_k} and \eqref{eq:def_c_Z_k} do not appear. Then, analogously to Sec.~\ref{sec:decoy_state_bounds}, we can derive bounds on the probabilities for vacuum and single-photon events
\begin{align}
    Y_{\mathsf{Z},0}^- &\coloneqq \frac{\tau_0}{(\mu_1 - \mu_2)}\left( \frac{\mu_1 e^{\mu_2} p_{\mathsf{Z},\mu_2}}{p_{\mu_2}} - \frac{\mu_2 e^{\mu_1} p_{\mathsf{Z},\mu_1}}{p_{\mu_1}} \right) \,,\\
    Y_{a,0}^+ &\coloneqq 2\frac{e_{a,k}}{p_k}\tau_0e^k \,,\\
    Y_{a,1}^- &\coloneqq \frac{\mu_1 \tau_1}{\mu_2(\mu_1 - \mu_2)}\left( \frac{e^{\mu_2} p_{a,\mu_2}}{p_{\mu_2}} - \frac{\mu_2^2}{\mu_1^2}\frac{e^{\mu_1}p_{a,\mu_1}}{p_{\mu_1}} - \frac{(\mu_1^2 - \mu_2^2)}{\mu_1^2}\frac{Y_{a,0}^+}{\tau_0} \right)\,,
\end{align}
where any $k\in\{\mu_1, \mu_2\}$ can be chosen, e.g. to maximize the secure-key rate. Analogously to the discussion in Sec.~\ref{sec:decoy_state_bounds}, the probability for a single-photon error is given by
\begin{equation}
    E_{\mathsf{X}, 1}^+ = \frac{\tau_1}{\mu_1 - \mu_2}\left( \frac{e^{\mu_1}e_{\mathsf{X},\mu_1}}{p_{\mu_1}} - \frac{e^{\mu_2}e_{\mathsf{X},\mu_2}}{p_{\mu_2}}\right)\,.
\end{equation}
Then, the asymptotic secure-key rate is given by
\begin{equation}
    r=Y_{\mathsf{Z},0}^- + Y_{\mathsf{Z},1}^-\left(1 - h\left(\frac{E_{\mathsf{X}, 1}^+}{Y_{\mathsf{X},1}^-}\right)\right) - h\left(e_\mathsf{Z}\right) \,,
\end{equation}
where the last term follows from Eq.~\eqref{eq:approx_leak_EC} and $e_\mathsf{Z}$ is the probability of an error in the $\mathsf{Z}$-basis, which is given by the channel and detection model. The asymptotic secure-key rate is plotted in Fig.~\ref{fig:skr_plot} and compared to the finite-size regime.}

\bibliographystyle{unsrtdoi}
\bibliography{literature}{}

@misc{inprep_authentication,
	title = {Authentication in {Security} {Proofs} for {Quantum} {Key} {Distribution}},
	url = {http://arxiv.org/abs/2601.17960},
	doi = {10.48550/arXiv.2601.17960},
	urldate = {2026-03-16},
	publisher = {arXiv},
	author = {Tupkary, D. and Nahar, S. and Tan, E. Y.-Z.},
	year = {2026}
}

@misc{ferradini2025definingsecurityquantumkey,
      title={Defining Security in Quantum Key Distribution}, 
      author={Carla Ferradini and Martin Sandfuchs and Ramona Wolf and Renato Renner},
      year={2025},
      eprint={2509.13405},
      archivePrefix={arXiv},
      primaryClass={quant-ph},
      url={https://arxiv.org/abs/2509.13405}, 
      doi={10.48550/arXiv.2509.13405}
}

@article{kamin2024improveddecoystate,
	title = {Improved decoy-state and flag-state squashing methods},
	volume = {6},
	url = {https://link.aps.org/doi/10.1103/PhysRevResearch.6.043223},
	doi = {10.1103/PhysRevResearch.6.043223},
	number = {4},
	urldate = {2026-03-10},
	journal = {Physical Review Research},
	publisher = {American Physical Society},
	author = {Kamin, L. and Lütkenhaus, N.},
	year = {2024},
	pages = {043223}
}

@article{kamin2024decoygeat,
	title = {Finite-{Size} {Analysis} of {Prepare}-and-{Measure} and {Decoy}-{State} {Quantum} {Key} {Distribution} via {Entropy} {Accumulation}},
	volume = {6},
	url = {https://link.aps.org/doi/10.1103/PRXQuantum.6.020342},
	doi = {10.1103/PRXQuantum.6.020342},
	number = {2},
	urldate = {2026-03-16},
	journal = {PRX Quantum},
	publisher = {American Physical Society},
	author = {Kamin, L. and Arqand, A. and George, I. and Lütkenhaus, N. and Tan, E. Y.-Z.},
	year = {2025},
	pages = {020342}
}

@article{Curty2014,
  title = {Finite-key analysis for measurement-device-independent quantum key distribution},
  volume = {5},
  ISSN = {2041-1723},
  url = {http://dx.doi.org/10.1038/ncomms4732},
  DOI = {10.1038/ncomms4732},
  number = {1},
  journal = {Nature Communications},
  publisher = {Springer Science and Business Media LLC},
  author = {Curty,  M. and Xu,  F. and Cui,  W. and Lim,  C. and Tamaki,  K. and Lo,  H.-K.},
  year = {2014},
}

@phdthesis{Lim14,
       author = "Lim, C. C. W.",
       title = "Tight security bounds for quantum key distribution",
       year = "2014",
       school= {Université de Genève}}

@misc{Renner05,
  doi = {10.48550/arXiv.quant-ph/0512258},
  
  author = {Renner, R.},
  
  title = {Security of Quantum Key Distribution},
  
  publisher = {arXiv},
  
  year = {2005},
  
  copyright = {Assumed arXiv.org perpetual, non-exclusive license to distribute this article for submissions made before January 2004}
}

@book{ChuangNielsen10, place={Cambridge}, title={Quantum Computation and Quantum Information: 10th Anniversary Edition}, DOI={10.1017/CBO9780511976667}, publisher={Cambridge University Press}, author={Nielsen, M. and Chuang, I.}, year={2010}}

@book{Audretsch07,
    author= "J. Audretsch",
    publisher = {John Wiley \& Sons, Ltd},
    isbn = {9783527619153},
    title = {Entangled Systems: New Directions in Quantum Physics},
    doi = {10.1002/9783527619153},
    year = {2007}
}

@article{Rusca18,
	title = {Finite-key analysis for the 1-decoy state {QKD} protocol},
	volume = {112},
	issn = {0003-6951},
	doi = {10.1063/1.5023340},
	number = {17},
	journal = {Applied Physics Letters},
	author = {Rusca, D. and Boaron, A. and Grünenfelder, F. and Martin, A. and Zbinden, H.},
	year = {2018},
	pages = {171104},
}

@article{wang2005,
  title={Beating the photon-number-splitting attack in practical quantum cryptography},
  author={Wang, X.-B.},
  journal={Physical Review Letters},
  volume={94},
  number={23},
  pages={230503},
  year={2005},
  publisher={APS},
doi = {10.1103/PhysRevLett.94.230503}
}

@article{Tomamichel12,
	doi = {10.1038/ncomms1631},
  
	year = 2012,
  
	publisher = {Springer Science and Business Media {LLC}},
  
	volume = {3},
  
	number = {1},
  
	author = {M. Tomamichel and C. C. W. Lim and N. Gisin and R. Renner},
  
	title = {Tight finite-key analysis for quantum cryptography},
  
	journal = {Nature Communications}
}

@article{Vitanov13,
	doi = {10.1109/tit.2013.2238656},
  
	url = {https://doi.org/10.1109%2Ftit.2013.2238656},
  
	year = 2013,
  
	publisher = {Institute of Electrical and Electronics Engineers ({IEEE})},
  
	volume = {59},
  
	number = {5},
  
	pages = {2603--2612},
  
	author = {A. Vitanov and F. Dupuis and M. Tomamichel and R. Renner},
  
	title = {Chain Rules for Smooth Min- and Max-Entropies},
  
	journal = {{IEEE} Transactions on Information Theory}
}

@article{Konig09,
	doi = {10.1109/tit.2009.2025545},
  
	url = {https://doi.org/10.1109%2Ftit.2009.2025545},
  
	year = 2009,
  
	publisher = {Institute of Electrical and Electronics Engineers ({IEEE})},
  
	volume = {55},
  
	number = {9},
  
	pages = {4337--4347},
  
	author = {R. Konig and R. Renner and C. Schaffner},
  
	title = {The Operational Meaning of Min- and Max-Entropy},
  
	journal = {{IEEE} Transactions on Information Theory}
}

@book{Tsybakov2009,
      author        = "Tsybakov, A.",
      title         = "{Introduction to Nonparametric Estimation}",
      publisher     = "Springer",
      address       = "Dordrecht",
      series        = "Springer series in statistics",
      year          = "2009",
      url           = "https://cds.cern.ch/record/1315296",
      doi           = "10.1007/b13794",
}

@article{Bennett84,
	doi = {10.1016/j.tcs.2014.05.025},
  
	url = {https://doi.org/10.1016%2Fj.tcs.2014.05.025},
  
	year = 1984,
  
	publisher = {Elsevier {BV}},
  
	volume = {560},
  
	pages = {7--11},
  
	author = {C. H. Bennett and G. Brassard},
  
	title = {Quantum cryptography: Public key distribution and coin tossing},
  
	journal = {Theoretical Computer Science}
}

@article{Carter79,
	title = {Universal classes of hash functions},
	volume = {18},
	issn = {0022-0000},
	url = {https://www.sciencedirect.com/science/article/pii/0022000079900448},
	doi = {10.1016/0022-0000(79)90044-8},
	language = {en},
	number = {2},
	urldate = {2023-03-15},
	journal = {Journal of Computer and System Sciences},
	author = {Carter, J. and Wegman, M.},
	year = {1979},
	pages = {143--154}
}

@inproceedings{RennerKoenig05,
    author="Renner, R.
    and K{\"o}nig, R.",
    editor="Kilian, Joe",
    title="Universally Composable Privacy Amplification Against Quantum Adversaries",
    booktitle="Theory of Cryptography",
    year="2005",
    publisher="Springer Berlin Heidelberg",
    address="Berlin, Heidelberg",
    pages="407--425",
    doi= "10.1007/978-3-540-30576-7_22"
}

@misc{TomamichelThesis12,
  doi = {10.48550/arxiv.1203.2142},
  url = {https://arxiv.org/abs/1203.2142},
  author = {Tomamichel, M.},
  title = {{A Framework for Non-Asymptotic Quantum Information Theory}},
  publisher = {arXiv},
  year = {2012},
  copyright = {arXiv.org perpetual, non-exclusive license}
}

@article{Tomamichel11,
	doi = {10.1109/tit.2011.2158473},
	url = {https://doi.org/10.1109%2Ftit.2011.2158473},
	year = 2011,
	publisher = {Institute of Electrical and Electronics Engineers ({IEEE})},
	volume = {57},
	number = {8},
	pages = {5524--5535},
	author = {M. Tomamichel and C. Schaffner and A. Smith and R. Renner},
	title = {Leftover Hashing Against Quantum Side Information},
	journal = {{IEEE} Transactions on Information Theory}
}

@article{LimCurtyWalenta14,
	title = {Concise security bounds for practical decoy-state quantum key distribution},
	volume = {89},
	url = {https://link.aps.org/doi/10.1103/PhysRevA.89.022307},
	doi = {10.1103/PhysRevA.89.022307},
	number = {2},
	urldate = {2024-01-19},
	journal = {Physical Review A},
	author = {Lim, C. C. W. and Curty, M. and Walenta, N. and Xu, F. and Zbinden, H.},
	year = {2014},
	pages = {022307}
}

@article{Hoeffding63,
	title = {Probability {Inequalities} for {Sums} of {Bounded} {Random} {Variables}},
	volume = {58},
	issn = {0162-1459},
	url = {https://www.jstor.org/stable/2282952},
	doi = {10.2307/2282952},
	number = {301},
	urldate = {2023-04-21},
	journal = {Journal of the American Statistical Association},
	author = {Hoeffding, W.},
	year = {1963},
	pages = {13--30}
}

@book{Wilde16,
	doi = {10.1017/9781316809976.001},
	url = {https://doi.org/10.1017%2F9781316809976.001},
	year = 2016,
	publisher = {Cambridge University Press},
	title = {{From} {Classical} to {Quantum} {Shannon} {Theory}},
	booktitle = {{Quantum Information Theory}},
	author = {M. Wilde}
}

@article{Tomamichel17,
	doi = {10.22331/q-2017-07-14-14},
	url = {https://doi.org/10.22331%2Fq-2017-07-14-14},
	year = 2017,
	publisher = {Verein zur Forderung des Open Access Publizierens in den Quantenwissenschaften},
	volume = {1},
	pages = {14},
	author = {M. Tomamichel and A. Leverrier},
	title = {A largely self-contained and complete security proof for quantum key distribution},
	journal = {Quantum}
}

@article{TomamichelRenner11,
	title = {The {Uncertainty} {Relation} for {Smooth} {Entropies}},
	volume = {106},
	issn = {0031-9007, 1079-7114},
	url = {http://arxiv.org/abs/1009.2015},
	doi = {10.1103/PhysRevLett.106.110506},
	number = {11},
	urldate = {2024-03-12},
	journal = {Physical Review Letters},
	author = {Tomamichel, M. and Renner, R.},
	year = {2011},
	pages = {110506}
}

@article{Mueller-Quade09,
	title = {Composability in quantum cryptography},
	volume = {11},
	issn = {1367-2630},
	url = {https://iopscience.iop.org/article/10.1088/1367-2630/11/8/085006},
	doi = {10.1088/1367-2630/11/8/085006},
	language = {en},
	number = {8},
	urldate = {2023-10-19},
	journal = {New Journal of Physics},
	author = {Müller-Quade, J. and Renner, R.},
	year = {2009},
	pages = {085006},
}

@book{Smart04,
  author         = {N. Smart},
  publisher      = {Mcgraw-Hill College},
  title          = {Cryptography: An Introduction, 3rd Edition},
  year           = {2004}
}

@article{Portmann22,
	title = {Security in {Quantum} {Cryptography}},
	volume = {94},
	issn = {0034-6861, 1539-0756},
	url = {http://arxiv.org/abs/2102.00021},
	doi = {10.1103/RevModPhys.94.025008},
	number = {2},
	urldate = {2023-11-05},
	journal = {Reviews of Modern Physics},
	author = {Portmann, C. and Renner, R.},
	year = {2022},
	pages = {025008},
}

@book{Tomamichel16,
   title={Quantum Information Processing with Finite Resources},
   ISBN={9783319218915},
   ISSN={2197-1765},
   url={http://dx.doi.org/10.1007/978-3-319-21891-5},
   doi={10.1007/978-3-319-21891-5},
   journal={SpringerBriefs in Mathematical Physics},
   publisher={Springer International Publishing},
   author={Tomamichel, Marco},
   year={2016} }

@article{Jain16,
	title = {Attacks on practical quantum key distribution systems (and how to prevent them)},
	volume = {57},
	issn = {0010-7514, 1366-5812},
	url = {http://arxiv.org/abs/1512.07990},
	doi = {10.1080/00107514.2016.1148333},
	number = {3},
	urldate = {2023-10-19},
	journal = {Contemporary Physics},
	author = {Jain, N. and Stiller, B. and Khan, I. and Elser, D. and Marquardt, C. and Leuchs, G.},
	year = {2016},
	pages = {366--387},
}

@misc{BSI24,
  author       = {German Federal Office for Information Security},
  title        = {{Implementation Attacks against QKD Systems}},
  year         = {2024},
note = "\url{https://www.bsi.bund.de/EN/Service-Navi/Publikationen/Studien/QKD-Systems/Implementation_Attacks_QKD_Systems_node.html} Last accessed on 18-05-24."
}

@book{Wolf21,
	address = {Cham},
	title = {Quantum {Key} {Distribution}: {An} {Introduction} with {Exercises}},
	volume = {988},
	shorttitle = {Quantum {Key} {Distribution}},
	url = {https://link.springer.com/10.1007/978-3-030-73991-1},
	language = {en},
	urldate = {2024-01-06},
	publisher = {Springer International Publishing},
	author = {Wolf, R.},
	year = {2021},
	doi = {10.1007/978-3-030-73991-1},
}

@book{Grasselli21,
	address = {Cham},
	series = {Quantum {Science} and {Technology}},
	title = {Quantum {Cryptography}: {From} {Key} {Distribution} to {Conference} {Key} {Agreement}},
	copyright = {http://www.springer.com/tdm},
	isbn = {978-3-030-64359-1 978-3-030-64360-7},
	shorttitle = {Quantum {Cryptography}},
	url = {http://link.springer.com/10.1007/978-3-030-64360-7},
	language = {en},
	urldate = {2024-05-18},
	publisher = {Springer International Publishing},
	author = {Grasselli, F.},
	year = {2021},
	doi = {10.1007/978-3-030-64360-7},
}

@article{Makarov23,
	title = {Preparing a commercial quantum key distribution system for certification against implementation loopholes},
	volume = {22},
	url = {https://link.aps.org/doi/10.1103/PhysRevApplied.22.044076},
	doi = {10.1103/PhysRevApplied.22.044076},
	number = {4},
	urldate = {2026-03-16},
	journal = {Physical Review Applied},
	publisher = {American Physical Society},
	author = {Makarov, V. and Abrikosov, A. and Chaiwongkhot, P. and Fedorov, A. K. and Huang, A. and Kiktenko, E. and Petrov, M. and Ponosova, A. and Ruzhitskaya, D. and Tayduganov, A. and Trefilov, D. and Zaitsev, K.},
	year = {2024},
	pages = {044076}
}

@article{Sajeed21,
	title = {An approach for security evaluation and certification of a complete quantum communication system},
	volume = {11},
	copyright = {2021 The Author(s)},
	issn = {2045-2322},
	url = {https://www.nature.com/articles/s41598-021-84139-3},
	doi = {10.1038/s41598-021-84139-3},
	language = {en},
	number = {1},
	urldate = {2023-12-07},
	journal = {Scientific Reports},
	author = {Sajeed, S. and Chaiwongkhot, P. and Huang, A. and Qin, H. and Egorov, V. and Kozubov, A. and Gaidash, A. and Chistiakov, V. and Vasiliev, A. and Gleim, A. and Makarov, V.},
	year = {2021},
	pages = {5110}
}

@inproceedings{Canetti2001,
    author={Canetti, R.},
    booktitle={Proceedings 42nd IEEE Symposium on Foundations of Computer Science}, 
    title={Universally composable security: a new paradigm for cryptographic protocols}, 
    year={2001},
    volume={},
    number={},
    pages={136-145},
    doi={10.1109/SFCS.2001.959888}
}

@article{Rivest78,
	title = {A method for obtaining digital signatures and public-key cryptosystems},
	volume = {21},
	issn = {0001-0782},
	url = {https://dl.acm.org/doi/10.1145/359340.359342},
	doi = {10.1145/359340.359342},
	number = {2},
	urldate = {2023-12-18},
	journal = {Communications of the ACM},
	author = {Rivest, R. L. and Shamir, A. and Adleman, L.},
	year = {1978},
	pages = {120--126}
}

@article{Gisin02,
	title = {Quantum {Cryptography}},
	volume = {74},
	issn = {0034-6861, 1539-0756},
	url = {http://arxiv.org/abs/quant-ph/0101098},
	doi = {10.1103/RevModPhys.74.145},
	number = {1},
	urldate = {2023-10-14},
	journal = {Reviews of Modern Physics},
	author = {Gisin, N. and Ribordy, G. and Tittel, W. and Zbinden, H.},
	year = {2002},
	pages = {145--195}
}

@article{Scarani09,
	title = {The {Security} of {Practical} {Quantum} {Key} {Distribution}},
	volume = {81},
	issn = {0034-6861, 1539-0756},
	url = {http://arxiv.org/abs/0802.4155},
	doi = {10.1103/RevModPhys.81.1301},
	number = {3},
	urldate = {2023-10-18},
	journal = {Reviews of Modern Physics},
	author = {Scarani, V. and Bechmann-Pasquinucci, H. and Cerf, N. J. and Dusek, M. and L{\"u}tkenhaus, N. and Peev, M.},
	year = {2009},
	pages = {1301--1350}
}

@article{Pirandola20,
	title = {Advances in {Quantum} {Cryptography}},
	volume = {12},
	issn = {1943-8206},
	url = {http://arxiv.org/abs/1906.01645},
	doi = {10.1364/AOP.361502},
	number = {4},
	urldate = {2023-10-19},
	journal = {Advances in Optics and Photonics},
	author = {Pirandola, S. and Andersen, U. L. and Banchi, L. and Berta, M. and Bunandar, D. and Colbeck, R. and Englund, D. and Gehring, T. and Lupo, C. and Ottaviani, C. and Pereira, J. and Razavi, M. and Shaari, J. S. and Tomamichel, M. and Usenko, V. C. and Vallone, G. and Villoresi, P. and Wallden, P.},
	year = {2020},
	pages = {1012}
}

@article{Hwang03,
	title = {Quantum {Key} {Distribution} with {High} {Loss}: {Toward} {Global} {Secure} {Communication}},
	volume = {91},
	issn = {0031-9007, 1079-7114},
	shorttitle = {Quantum {Key} {Distribution} with {High} {Loss}},
	url = {https://link.aps.org/doi/10.1103/PhysRevLett.91.057901},
	doi = {10.1103/PhysRevLett.91.057901},
	language = {en},
	number = {5},
	urldate = {2023-10-18},
	journal = {Physical Review Letters},
	author = {Hwang, W.-Y.},
	year = {2003},
	pages = {057901}
}

@article{Huttner95,
	title = {Quantum cryptography with coherent states},
	volume = {51},
	url = {https://link.aps.org/doi/10.1103/PhysRevA.51.1863},
	doi = {10.1103/PhysRevA.51.1863},
	number = {3},
	urldate = {2023-10-21},
	journal = {Physical Review A},
	author = {Huttner, B. and Imoto, N. and Gisin, N. and Mor, T.},
	year = {1995},
	pages = {1863--1869}
}

@article{Lutkenhaus02,
	title = {Quantum key distribution with realistic states: photon-number statistics in the photon-number splitting attack},
	volume = {4},
	issn = {13672630},
	shorttitle = {Quantum key distribution with realistic states},
	doi = {10.1088/1367-2630/4/1/344},
	urldate = {2023-10-21},
	journal = {New Journal of Physics},
	author = {Lütkenhaus, N. and Jahma, M.},
	year = {2002},
	pages = {44--44}
}

@article{Lo05,
	title = {Decoy {State} {Quantum} {Key} {Distribution}},
	volume = {94},
	issn = {0031-9007, 1079-7114},
	url = {https://link.aps.org/doi/10.1103/PhysRevLett.94.230504},
	doi = {10.1103/PhysRevLett.94.230504},
	language = {en},
	number = {23},
	urldate = {2023-10-18},
	journal = {Physical Review Letters},
	author = {Lo, H.-K. and Ma, X. and Chen, K.},
	year = {2005},
	pages = {230504}
}

@article{Ekert91,
  title = {{Quantum cryptography based on Bell's theorem}},
  author = {Ekert, A.},
  journal = {Physical Review Letters},
  volume = {67},
  issue = {6},
  pages = {661--663},
  numpages = {0},
  year = {1991},
  publisher = {American Physical Society},
  doi = {10.1103/PhysRevLett.67.661},
  url = {https://link.aps.org/doi/10.1103/PhysRevLett.67.661}
}

@article{Scarani04,
   title={Quantum Cryptography Protocols Robust against Photon Number Splitting Attacks for Weak Laser Pulse Implementations},
   volume={92},
   ISSN={1079-7114},
   url={http://dx.doi.org/10.1103/PhysRevLett.92.057901},
   DOI={10.1103/physrevlett.92.057901},
   number={5},
   journal={Physical Review Letters},
   publisher={American Physical Society (APS)},
   author={Scarani, V. and Acín, A. and Ribordy, G. and Gisin, N.},
   year={2004}
}

@article{Lo12,
   title={Measurement-Device-Independent Quantum Key Distribution},
   volume={108},
   ISSN={1079-7114},
   url={http://dx.doi.org/10.1103/PhysRevLett.108.130503},
   DOI={10.1103/physrevlett.108.130503},
   number={13},
   journal={Physical Review Letters},
   publisher={American Physical Society (APS)},
   author={Lo, H.-K. and Curty, M. and Qi, B.},
   year={2012}
}

@article{Shor00,
  title = {{{Simple Proof of Security of the BB84 Quantum Key Distribution Protocol}}},
  author = {Shor, P. and Preskill, J.},
  journal = {Physical Review Letters},
  volume = {85},
  issue = {2},
  pages = {441--444},
  numpages = {0},
  year = {2000},
  publisher = {American Physical Society},
  doi = {10.1103/PhysRevLett.85.441},
  url = {https://link.aps.org/doi/10.1103/PhysRevLett.85.441}
}

@article{Mayers01, author = {Mayers, D.}, title = {Unconditional Security in Quantum Cryptography}, year = {2001}, issue_date = {May 2001}, publisher = {Association for Computing Machinery}, address = {New York, NY, USA}, volume = {48}, number = {3}, url = {https://doi.org/10.1145/382780.382781}, doi = {10.1145/382780.382781}, journal = {J. ACM}, pages = {351–406}, numpages = {56} }

@article{Tomamichel10,
	title = {Duality {Between} {Smooth} {Min}- and {Max}-{Entropies}},
	volume = {56},
	issn = {1557-9654},
	url = {https://ieeexplore.ieee.org/document/5550419},
	doi = {10.1109/TIT.2010.2054130},
	number = {9},
	urldate = {2024-01-18},
	journal = {IEEE Transactions on Information Theory},
	author = {Tomamichel, M. and Colbeck, R. and Renner, R.},
	year = {2010},
	pages = {4674--4681}
}

@article{Tomamichel17b,
	title = {Fundamental {Finite} {Key} {Limits} for {One}-{Way} {Information} {Reconciliation} in {Quantum} {Key} {Distribution}},
	volume = {16},
	issn = {1570-0755, 1573-1332},
	url = {http://arxiv.org/abs/1401.5194},
	doi = {10.1007/s11128-017-1709-5},
	number = {11},
	urldate = {2024-01-18},
	journal = {Quantum Information Processing},
	author = {Tomamichel, M. and Martinez-Mateo, J. and Pacher, C. and Elkouss, D.},
	year = {2017},
	pages = {280}
}

@article{Ekert14,
	title = {The ultimate physical limits of privacy},
	volume = {507},
	copyright = {2014 Springer Nature Limited},
	issn = {1476-4687},
	url = {https://www.nature.com/articles/nature13132},
	doi = {10.1038/nature13132},
	language = {en},
	number = {7493},
	urldate = {2023-11-07},
	journal = {Nature},
	author = {Ekert, A. and Renner, R.},
	year = {2014},
	pages = {443--447}
}

@article{Vernam26,
	title = {Cipher printing telegraph systems: {For} secret wire and radio telegraphic communications},
	volume = {45},
	issn = {2376-5976},
	shorttitle = {Cipher printing telegraph systems},
	url = {https://ieeexplore.ieee.org/document/6534724},
	doi = {10.1109/JAIEE.1926.6534724},
	number = {2},
	urldate = {2024-01-18},
	journal = {Journal of the A.I.E.E.},
	author = {Vernam, G. S.},
	year = {1926},
	pages = {109--115}
}

@article{Shannon49,
	title = {Communication theory of secrecy systems},
	volume = {28},
	issn = {0005-8580},
	url = {https://ieeexplore.ieee.org/document/6769090},
	doi = {10.1002/j.1538-7305.1949.tb00928.x},
	number = {4},
	urldate = {2024-01-18},
	journal = {The Bell System Technical Journal},
	author = {Shannon, C. E.},
	year = {1949},
	pages = {656--715}
}

@article{Herrero17,
	title = {Quantum random number generators},
	volume = {89},
	url = {https://link.aps.org/doi/10.1103/RevModPhys.89.015004},
	doi = {10.1103/RevModPhys.89.015004},
	number = {1},
	urldate = {2023-11-12},
	journal = {Reviews of Modern Physics},
	author = {Herrero-Collantes, M. and Garcia-Escartin, J. C.},
	year = {2017},
	pages = {015004}
}

@article{Lucamarini15,
	title = {Practical {Security} {Bounds} {Against} the {Trojan}-{Horse} {Attack} in {Quantum} {Key} {Distribution}},
	volume = {5},
	issn = {2160-3308},
	doi = {10.1103/PhysRevX.5.031030},
	language = {en},
	number = {3},
	urldate = {2023-10-20},
	journal = {Physical Review X},
	author = {Lucamarini, M. and Choi, I. and Ward, M. B. and Dynes, J. F. and Yuan, Z. L. and Shields, A. J.},
	year = {2015},
	pages = {031030}
}

@article{Wang18,
	title = {Finite-key security analysis for quantum key distribution with leaky sources},
	volume = {20},
	issn = {1367-2630},
	doi = {10.1088/1367-2630/aad839},
	language = {en},
	number = {8},
	urldate = {2024-01-16},
	journal = {New Journal of Physics},
	author = {Wang, W. and Tamaki, K. and Curty, M.},
	year = {2018},
	pages = {083027},
}

@article{Zhang21,
	title = {Security proof of practical quantum key distribution with detection-efficiency mismatch},
	volume = {3},
	url = {https://link.aps.org/doi/10.1103/PhysRevResearch.3.013076},
	doi = {10.1103/PhysRevResearch.3.013076},
	number = {1},
	urldate = {2023-11-09},
	journal = {Physical Review Research},
	author = {Zhang, Y. and Coles, P. J. and Winick, A. and Lin, J. and Lütkenhaus, N.},
	year = {2021},
	pages = {013076}
}

@misc{Lo05b,
	title = {Phase randomization improves the security of quantum key distribution},
	doi = {10.48550/arXiv.quant-ph/0504209},
	urldate = {2024-01-23},
	publisher = {arXiv},
	author = {Lo, H.-K. and Preskill, J.},
	year = {2005}
}

@article{boaronSecureQuantumKey2018,
  title = {Secure {{Quantum Key Distribution}} over 421 Km of {{Optical Fiber}}},
  author = {Boaron, A. and Boso, G. and Rusca, D. and Vulliez, C. and Autebert, C. and Caloz, M. and Perrenoud, M. and Gras, G. and Bussi{\`e}res, F. and Li, M.-J. and Nolan, D. and Martin, A. and Zbinden, H.},
  year = {2018},
  journal = {Physical Review Letters},
  volume = {121},
  number = {19},
  pages = {190502},
  issn = {0031-9007, 1079-7114},
  doi = {10.1103/PhysRevLett.121.190502},
  urldate = {2022-10-31},
  langid = {english}
}

@article{lucamariniOvercomingRateDistance2018,
  title = {Overcoming the Rate{\textendash}Distance Limit of Quantum Key Distribution without Quantum Repeaters},
  author = {Lucamarini, M. and Yuan, Z. L. and Dynes, J. F. and Shields, A. J.},
  year = {2018},
  journal = {Nature},
  volume = {557},
  number = {7705},
  pages = {400--403},
  issn = {0028-0836, 1476-4687},
  doi = {10.1038/s41586-018-0066-6},
  urldate = {2022-10-31},
  langid = {english}
}

@article{zengModepairingQuantumKey2022,
  title = {Mode-Pairing Quantum Key Distribution},
  author = {Zeng, P. and Zhou, H. and Wu, W. and Ma, X.},
  year = {2022},
  journal = {Nature Communications},
  volume = {13},
  number = {1},
  pages = {3903},
  issn = {2041-1723},
  doi = {10.1038/s41467-022-31534-7},
  urldate = {2023-01-24}
}

@article{colbeckNoExtensionQuantum2011,
  title = {No Extension of Quantum Theory Can Have Improved Predictive Power},
  author = {Colbeck, R. and Renner, R.},
  year = {2011},
  journal = {Nature Communications},
  volume = {2},
  number = {1},
  pages = {411},
  issn = {2041-1723},
  doi = {10.1038/ncomms1416},
  urldate = {2024-02-17},
  langid = {english}
}

@article{Coles16,
	title = {Numerical approach for unstructured quantum key distribution},
	volume = {7},
	copyright = {2016 The Author(s)},
	issn = {2041-1723},
	url = {https://www.nature.com/articles/ncomms11712},
	doi = {10.1038/ncomms11712},
	language = {en},
	number = {1},
	urldate = {2024-03-11},
	journal = {Nature Communications},
	author = {Coles, P. J. and Metodiev, R. M. and Lütkenhaus, N.},
	year = {2016},
	pages = {11712}
}

@article{George21,
	title = {Numerical calculations of the finite key rate for general quantum key distribution protocols},
	volume = {3},
	url = {https://link.aps.org/doi/10.1103/PhysRevResearch.3.013274},
	doi = {10.1103/PhysRevResearch.3.013274},
	number = {1},
	urldate = {2024-03-11},
	journal = {Physical Review Research},
	author = {George, I. and Lin, J. and Lütkenhaus, N.},
	year = {2021},
	pages = {013274}
}

@article{Tupkary23,
  title = {Security proof for variable-length quantum key distribution},
  author = {Tupkary, D. and Tan, E. Y.-Z. and L\"utkenhaus, N.},
  journal = {Phys. Rev. Res.},
  volume = {6},
  issue = {2},
  pages = {023002},
  numpages = {20},
  year = {2024},
  publisher = {American Physical Society},
  doi = {10.1103/PhysRevResearch.6.023002},
  url = {https://link.aps.org/doi/10.1103/PhysRevResearch.6.023002}
}

@article{Portmann14,
	title = {Key recycling in authentication},
	volume = {60},
	issn = {0018-9448, 1557-9654},
	url = {http://arxiv.org/abs/1202.1229},
	doi = {10.1109/TIT.2014.2317312},
	number = {7},
	urldate = {2024-03-14},
	journal = {IEEE Transactions on Information Theory},
	author = {Portmann, C.},
	year = {2014},
	pages = {4383--4396},
}

@article{metgerSecurityQuantumKey2023,
  title = {Security of Quantum Key Distribution from Generalised Entropy Accumulation},
  author = {Metger, T. and Renner, R.},
  year = {2023},
  journal = {Nature Communications},
  volume = {14},
  number = {1},
  pages = {5272},
  issn = {2041-1723},
  doi = {10.1038/s41467-023-40920-8},
  urldate = {2024-03-21},
  langid = {english},
}

@article{maPracticalDecoyState2005,
  title = {Practical Decoy State for Quantum Key Distribution},
  author = {Ma, X. and Qi, B. and Zhao, Y. and Lo, H.-K.},
  year = {2005},
  journal = {Physical Review A},
  volume = {72},
  number = {1},
  pages = {012326},
  issn = {1050-2947, 1094-1622},
  doi = {10.1103/PhysRevA.72.012326},
  urldate = {2022-10-27},
  langid = {english}
}

@article{Nahar24,
  title = {Postselection Technique for Optical Quantum Key Distribution with Improved de Finetti Reductions},
  author = {Nahar, S. and Tupkary, D. and Zhao, Y. and L\"utkenhaus, N. and Tan, E. Y.-Z.},
  journal = {PRX Quantum},
  volume = {5},
  issue = {4},
  pages = {040315},
  numpages = {35},
  year = {2024},
  publisher = {American Physical Society},
  doi = {10.1103/PRXQuantum.5.040315},
  url = {https://link.aps.org/doi/10.1103/PRXQuantum.5.040315}
}

@misc{Biham05,
	title = {A {Proof} of the {Security} of {Quantum} {Key} {Distribution}},
	url = {http://arxiv.org/abs/quant-ph/0511175},
	doi = {10.48550/arXiv.quant-ph/0511175},
	urldate = {2024-04-01},
	author = {Biham, E. and Boyer, M. and Boykin, P. O. and Mor, T. and Roychowdhury, V.},
	year = {2005}
}

@article{Wootters82,
  title = {A Single Quantum Cannot Be Cloned},
  author = {Wootters, W. K. and Zurek, W. H.},
  year = {1982},
  journal = {Nature},
  volume = {299},
  number = {5886},
  pages = {802--803},
  publisher = {Nature Publishing Group},
  issn = {1476-4687},
  doi = {10.1038/299802a0},
}

@article{Christandl09,
	title = {Postselection {Technique} for {Quantum} {Channels} with {Applications} to {Quantum} {Cryptography}},
	volume = {102},
	url = {https://link.aps.org/doi/10.1103/PhysRevLett.102.020504},
	doi = {10.1103/PhysRevLett.102.020504},
	number = {2},
	urldate = {2024-03-23},
	journal = {Physical Review Letters},
	author = {Christandl, M. and König, R. and Renner, R.},
	year = {2009},
	pages = {020504}
}

@article{Stinson94,
	title = {Universal hashing and authentication codes},
	volume = {4},
	issn = {1573-7586},
	url = {https://doi.org/10.1007/BF01388651},
	doi = {10.1007/BF01388651},
	language = {en},
	number = {3},
	urldate = {2024-05-17},
	journal = {Designs, Codes and Cryptography},
	author = {Stinson, D. R.},
	year = {1994},
	pages = {369--380},
}

@article{Wegman81,
	title = {New hash functions and their use in authentication and set equality},
	volume = {22},
	issn = {0022-0000},
	url = {https://www.sciencedirect.com/science/article/pii/0022000081900337},
	doi = {10.1016/0022-0000(81)90033-7},
	number = {3},
	urldate = {2024-05-13},
	journal = {Journal of Computer and System Sciences},
	author = {Wegman, M. N. and Carter, J. L.},
	year = {1981},
	pages = {265--279},
}

@article{lutkenhausSecurityIndividualAttacks2000,
  title = {Security against Individual Attacks for Realistic Quantum Key Distribution},
  author = {L{\"u}tkenhaus, N.},
  year = {2000},
  journal = {Physical Review A},
  volume = {61},
  number = {5},
  pages = {052304},
  doi = {10.1103/PhysRevA.61.052304},
}

@article{tyagiUniversalHashingInformationTheoretic2015,
  title = {Universal {{Hashing}} for {{Information-Theoretic Security}}},
  author = {Tyagi, H. and Vardy, A.},
  year = {2015},
  journal = {Proceedings of the IEEE},
  volume = {103},
  number = {10},
  pages = {1781--1795},
  doi = {10.1109/JPROC.2015.2462774},
}

@article{Pfister16,
	title = {Sifting attacks in finite-size quantum key distribution},
	volume = {18},
	issn = {1367-2630},
	url = {https://dx.doi.org/10.1088/1367-2630/18/5/053001},
	doi = {10.1088/1367-2630/18/5/053001},
	language = {en},
	number = {5},
	urldate = {2024-08-02},
	journal = {New Journal of Physics},
	author = {Pfister, C. and Lütkenhaus, N. and Wehner, S. and Coles, P. J.},
	year = {2016},
	pages = {053001},
}

@article{Serfling74,
	title = {Probability {Inequalities} for the {Sum} in {Sampling} without {Replacement}},
	volume = {2},
	issn = {0090-5364, 2168-8966},
	url = {https://projecteuclid.org/journals/annals-of-statistics/volume-2/issue-1/Probability-Inequalities-for-the-Sum-in-Sampling-without-Replacement/10.1214/aos/1176342611.full},
	doi = {10.1214/aos/1176342611},
	number = {1},
	urldate = {2024-07-30},
	journal = {The Annals of Statistics},
	author = {Serfling, R. J.},
	year = {1974},
	pages = {39--48},
}

@article{Tamaki18,
	title = {Security of quantum key distribution with iterative sifting},
	volume = {3},
	issn = {2058-9565},
	url = {http://arxiv.org/abs/1610.06499},
	doi = {10.1088/2058-9565/aa89bd},
	number = {1},
	urldate = {2024-09-23},
	journal = {Quantum Science and Technology},
	author = {Tamaki, K. and Lo, H.-K. and Mizutani, A. and Kato, G. and Lim, C. C. W. and Azuma, K. and Curty, M.},
	year = {2018},
	pages = {014002},
}

@article{Curras24,
	title = {Security framework for quantum key distribution with imperfect sources},
	volume = {3},
	copyright = {© 2025 Optica Publishing Group},
	issn = {2837-6714},
	url = {https://opg.optica.org/opticaq/abstract.cfm?uri=opticaq-3-6-525},
	doi = {10.1364/OPTICAQ.569424},
	language = {EN},
	number = {6},
	urldate = {2026-03-16},
	journal = {Optica Quantum},
	publisher = {Optica Publishing Group},
	author = {Currás-Lorenzo, G. and Pereira, M. and Kato, G. and Curty, M. and Tamaki, K.},
	year = {2025},
	pages = {525--534}
}

@article{Zapatero23,
author = {Zapatero, V. and Navarrete, Á. and Curty, M.},
title = {Implementation Security in Quantum Key Distribution},
journal = {Advanced Quantum Technologies},
volume = {8},
number = {2},
pages = {2300380},
doi = {https://doi.org/10.1002/qute.202300380},
url = {https://advanced.onlinelibrary.wiley.com/doi/abs/10.1002/qute.202300380},
eprint = {https://advanced.onlinelibrary.wiley.com/doi/pdf/10.1002/qute.202300380},
year = {2025}
}

@article{Tupkary24,
  doi = {10.22331/q-2025-12-11-1937},
  url = {https://doi.org/10.22331/q-2025-12-11-1937},
  title = {Phase error rate estimation in {QKD} with imperfect detectors},
  author = {Tupkary, D. and Nahar, S. and Sinha, P. and L{\"{u}}tkenhaus, N.},
  journal = {{Quantum}},
  issn = {2521-327X},
  publisher = {{Verein zur F{\"{o}}rderung des Open Access Publizierens in den Quantenwissenschaften}},
  volume = {9},
  pages = {1937},
  year = {2025}
}

@article{Curras21,
	title = {Tight finite-key security for twin-field quantum key distribution},
	volume = {7},
	copyright = {2021 The Author(s)},
	issn = {2056-6387},
	url = {https://www.nature.com/articles/s41534-020-00345-3},
	doi = {10.1038/s41534-020-00345-3},
	language = {en},
	number = {1},
	urldate = {2024-09-26},
	journal = {npj Quantum Information},
	author = {Currás-Lorenzo, G. and Navarrete, A. and Azuma, K. and Kato, G. and Curty, M. and Razavi, M.},
	year = {2021},
	pages = {1--9},
}

@article{Hayashi12,
   title={Concise and tight security analysis of the {Bennett–Brassard} 1984 protocol with finite key lengths},
   volume={14},
   ISSN={1367-2630},
   url={http://dx.doi.org/10.1088/1367-2630/14/9/093014},
   DOI={10.1088/1367-2630/14/9/093014},
   number={9},
   journal={New Journal of Physics},
   publisher={IOP Publishing},
   author={Hayashi, M. and Tsurumaru, T.},
   year={2012},
pages={093014} }

@article{Kawakami17,
	title = {Finite-key analysis for quantum key distribution with weak coherent pulses based on {Bernoulli} sampling},
	volume = {96},
	url = {https://link.aps.org/doi/10.1103/PhysRevA.96.012305},
	doi = {10.1103/PhysRevA.96.012305},
	number = {1},
	urldate = {2026-03-16},
	journal = {Physical Review A},
	publisher = {American Physical Society},
	author = {Kawakami, S. and Sasaki, T. and Koashi, M.},
	year = {2017},
	pages = {012305}
}

@article{Mannalath24,
	title = {Sharp {Finite} {Statistics} for {Quantum} {Key} {Distribution}},
	volume = {135},
	url = {https://link.aps.org/doi/10.1103/l735-x48g},
	doi = {10.1103/l735-x48g},
	number = {2},
	urldate = {2026-03-16},
	journal = {Physical Review Letters},
	publisher = {American Physical Society},
	author = {Mannalath, V. and Zapatero, V. and Curty, M.},
	year = {2025},
	pages = {020803}
}

@book{Schneier96,
    author = {B. Schneier},
    title = {Applied Cryptography: Protocols, Algorithms, and Source Code in C},
    publisher = {John Wiley \& Sons Inc},
    year = 1996
}

@misc{Lo07,
      title={Security of quantum key distribution using weak coherent states with nonrandom phases}, 
      author={H.-K. Lo and J. Preskill},
      year={2007},
      eprint={quant-ph/0610203},
      archivePrefix={arXiv},
      primaryClass={quant-ph},
      url={https://arxiv.org/abs/quant-ph/0610203}, 
      doi = {10.48550/arXiv.quant-ph/0610203}
}

@article{Curras24b,
	title = {Security of quantum key distribution with imperfect phase randomisation},
	volume = {9},
	issn = {2058-9565},
	url = {http://arxiv.org/abs/2210.08183},
	doi = {10.1088/2058-9565/ad141c},
	number = {1},
	urldate = {2024-06-13},
	journal = {Quantum Science and Technology},
	author = {Currás-Lorenzo, G. and Nahar, S. and Lütkenhaus, N. and Tamaki, K. and Curty, M.},
	year = {2024},
	pages = {015025},
}

@article{Sixto23,
	title = {Secret key rate bounds for quantum key distribution with faulty active phase randomization},
	volume = {10},
	issn = {2196-0763},
	url = {https://doi.org/10.1140/epjqt/s40507-023-00210-0},
	doi = {10.1140/epjqt/s40507-023-00210-0},
	language = {en},
	number = {1},
	urldate = {2026-03-16},
	journal = {EPJ Quantum Technology},
	author = {Sixto, X. and Currás-Lorenzo, G. and Tamaki, K. and Curty, M.},
	year = {2023},
	pages = {53}
}

@article{Nahar23,
  title = {Imperfect phase randomization and generalized decoy-state quantum key distribution},
  author = {Nahar, S. and Upadhyaya, T. and L\"utkenhaus, N.},
  journal = {Phys. Rev. Appl.},
  volume = {20},
  issue = {6},
  pages = {064031},
  numpages = {24},
  year = {2023},
  publisher = {American Physical Society},
  doi = {10.1103/PhysRevApplied.20.064031},
  url = {https://link.aps.org/doi/10.1103/PhysRevApplied.20.064031}
}

@article{Koashi09,
doi = {10.1088/1367-2630/11/4/045018},
url = {https://dx.doi.org/10.1088/1367-2630/11/4/045018},
year = {2009},
publisher = {},
volume = {11},
number = {4},
pages = {045018},
author = {M. Koashi},
title = {Simple security proof of quantum key distribution based on complementarity},
journal = {New Journal of Physics}
}

@article{Curty04,
  title = {Entanglement as a Precondition for Secure Quantum Key Distribution},
  author = {Curty, M. and Lewenstein, M. and L\"utkenhaus, N.},
  journal = {Phys. Rev. Lett.},
  volume = {92},
  issue = {21},
  pages = {217903},
  numpages = {4},
  year = {2004},
  publisher = {American Physical Society},
  doi = {10.1103/PhysRevLett.92.217903},
  url = {https://link.aps.org/doi/10.1103/PhysRevLett.92.217903}
}

@article{Ferenczi12,
  title = {Symmetries in quantum key distribution and the connection between optimal attacks and optimal cloning},
  author = {Ferenczi, A. and L\"utkenhaus, N.},
  journal = {Phys. Rev. A},
  volume = {85},
  issue = {5},
  pages = {052310},
  numpages = {17},
  year = {2012},
  publisher = {American Physical Society},
  doi = {10.1103/PhysRevA.85.052310},
  url = {https://link.aps.org/doi/10.1103/PhysRevA.85.052310}
}

@article{Diffie76,
  author={Diffie, W. and Hellman, M.},
  journal={IEEE Transactions on Information Theory}, 
  title={New directions in cryptography}, 
  year={1976},
  volume={22},
  number={6},
  pages={644-654},
  doi={10.1109/TIT.1976.1055638}
}

@article{Koblitz87,
  author = {Koblitz, N.},
  issn = {0025-5718},
  journal = {Mathematics of Computation},
  mrclass = {94A60 (11T71 11Y16 68P25)},
  mrnumber = {88b:94017},
  mrreviewer = {Harald Niederreiter},
  number = 177,
  pages = {203--209},
  title = {Elliptic Curve Cryptosystems},
  volume = 48,
  year = 1987,
  doi={10.1090/S0025-5718-1987-0866109-5}
}

@InProceedings{Miller85,
author="Miller, V. S.",
editor="Williams, H. C.",
title="Use of Elliptic Curves in Cryptography",
booktitle="Advances in Cryptology --- CRYPTO '85 Proceedings",
year="1986",
publisher="Springer Berlin Heidelberg",
address="Berlin, Heidelberg",
pages="417--426",
isbn="978-3-540-39799-1",
doi={10.1007/3-540-39799-X_31}
}

@inproceedings{Brassard05,
   title={Brief history of quantum cryptography: a personal perspective},
   url={http://dx.doi.org/10.1109/ITWTPI.2005.1543949},
   DOI={10.1109/itwtpi.2005.1543949},
   booktitle={IEEE Information Theory Workshop on Theory and Practice in Information-Theoretic Security},
   publisher={IEEE},
   author={Brassard, G.},
   pages={19–23}, year = {2005} }

@article{Bennet92,
    author = {C. Bennett and F. Bessette and G. Brassard and L. Salvail and J. Smolin},
    title = {Experimental quantum cryptography},
    journal = {Journal of Cryptology},
    year = {1992},
    doi = {10.1007/BF00191318}
}

@misc{Mosca12,
      title={Quantum Key Distribution in the Classical Authenticated Key Exchange Framework}, 
      author={N. Mosca and D. Stebila and B. Ustaoglu},
      year={2012},
      eprint={1206.6150},
      archivePrefix={arXiv},
      primaryClass={quant-ph},
      url={https://arxiv.org/abs/1206.6150}, 
        doi = {10.48550/arXiv.1206.6150}
}

@book{Scully97, place={Cambridge}, title={Quantum Optics}, publisher={Cambridge University Press}, author={Scully, M. O. and Zubairy, M. S.}, year={1997}}

@article{Pironio09,
doi = {10.1088/1367-2630/11/4/045021},
url = {https://doi.org/10.1088/1367-2630/11/4/045021},
year = {2009},
publisher = {},
volume = {11},
number = {4},
pages = {045021},
author = {Pironio, S. and Acín, A. and Brunner, N. and Gisin, N. and Massar, S. and Scarani, V.},
title = {Device-independent quantum key distribution secure against collective attacks},
journal = {New Journal of Physics}
}

@misc{Wiesemann_QKD_simulation_2025,
  author       = {Wiesemann, G.},
  title        = {Quantum key distribution secure-key rate simulation (1-decoy {BB84})},
  year         = {2025},
  howpublished = {\url{https://github.com/JeromeWiesemann/Quantum-key-distribution-secure-key-rate-simulation-1-decoy-BB84}},
  note         = {Accessed: 2025-12-03}
}

@misc{arqand25,
	title = {Marginal-constrained entropy accumulation theorem},
	url = {http://arxiv.org/abs/2502.02563},
	doi = {10.48550/arXiv.2502.02563},
	urldate = {2025-11-24},
	publisher = {arXiv},
	author = {Arqand, A. and Tan, E. Y.-Z.},
	year = {2025},
}

@misc{kamin25,
	title = {Rényi security framework against coherent attacks applied to decoy-state {QKD}},
	url = {http://arxiv.org/abs/2504.12248},
	doi = {10.48550/arXiv.2504.12248},
	urldate = {2025-05-16},
	publisher = {arXiv},
	author = {Kamin, L. and Burniston, J. and Tan, E. Y.-Z.},
	year = {2025}
}

@misc{tupkary25a,
	title = {{QKD} security proofs for decoy-state {BB84}: protocol variations, proof techniques, gaps and limitations},
	shorttitle = {{QKD} security proofs for decoy-state {BB84}},
	url = {http://arxiv.org/abs/2502.10340},
	doi = {10.48550/arXiv.2502.10340},
	urldate = {2025-03-05},
	publisher = {arXiv},
	author = {Tupkary, D. and Tan, E. Y.-Z. and Nahar, S. and Kamin, L. and Lütkenhaus, N.},
	year = {2025},
}

@phdthesis{kaminthesis,
    title = {From Asymptotic to Finite-Size Security in Decoy-State Quantum Key Distribution}, author = {Kamin, L.}, school = {University of Waterloo}, year = {2026}, note = {In preparation}
}

\end{document}